\documentclass[]{aa}  

\usepackage{natbib}
\usepackage{graphicx}
\usepackage{siunitx}
\usepackage{txfonts}
\usepackage[svgnames,dvipsnames]{xcolor} 

\usepackage{wasysym}

\newcommand{\jp}{\ensuremath{J^{\prime}}}

\newcommand{\bp}{\ensuremath{B^{\prime}}}

\begin{document}
\title{The EDIBLES survey \\V: Line profile variations in the $\lambda\lambda$5797, 6379, and 6614 diffuse interstellar bands as a tool to constrain carrier sizes.}
\titlerunning{EDIBLES V: DIB Line Profile Variations}
\authorrunning{MacIsaac et al.}

   \author{
Heather MacIsaac \inst{1,2} \and 
Jan Cami \inst{1,2,3} \and
Nick L.J.~Cox\inst{4} \and
Amin Farhang \inst{5,1} \and
Jonathan Smoker \inst{6} \and 
Meriem Elyajouri \inst{7} \and 
Rosine Lallement \inst{7} \and 
Peter J.~Sarre \inst{8} \and 
Martin A.~Cordiner \inst{9,10} \and
Haoyu Fan \inst{1} \and 
Klay Kulik \inst{1} \and
Harold Linnartz \inst{11} \and
Bernard H. Foing \inst{12} \and 
Jacco Th. van Loon \inst{13} \and
Giacomo Mulas \inst{14} \and
Keith T. Smith \inst{15}
}

   \institute{
{Department of Physics and Astronomy, The University of Western Ontario,
 London, ON N6A 3K7, Canada}
\and
{Institute for Earth and Space Exploration, The University of Western Ontario, London, ON N6A 3K7, Canada}
\and
{SETI Institute, 189 Bernardo Ave, Suite 100, Mountain View, CA
  94043, USA}
\and
{ACRI-ST, 260 route du Pin Montard, 06904, Sophia Antipolis, France }
\and
{School of Astronomy, Institute for Research in Fundamental
  Sciences, 19395-5531 Tehran, Iran}
\and 
{European Southern Observatory, Alonso de Cordova 3107, Vitacura,
  Santiago, Chile}
\and 
 {GEPI, Observatoire de Paris, PSL Research University, CNRS,
  Universit\'e Paris-Diderot, Sorbonne Paris Cit\'e, Place Jules
  Janssen, 92195 Meudon, France}
\and 
 {School of Chemistry, The University of Nottingham, University
  Park, Nottingham NG7 2RD, UK}
  \and 
{Astrochemistry Laboratory, NASA Goddard Space Flight Center,
Code 691, 8800 Greenbelt Road, Greenbelt, MD 20771, USA}
\and 
{Department of Physics, The Catholic University of America,
Washington, DC 20064, USA}
  \and 
{Laboratory for Astrophysics, Leiden Observatory, Leiden
University, PO Box 9513, 2300 RA Leiden, The Netherlands}
\and 
{ESTEC, ESA, Keplerlaan 1, 2201 AZ Noordwijk, The Netherlands}
\and 
{Lennard-Jones Laboratories, Keele University, ST5 5BG, UK}
 \and 
 {INAF–Osservatorio Astronomico di Cagliari, via della Scienza 5,
09047 Selargius, Italy}
\and 
{AAAS Science International, Clarendon House, Clarendon Road,
  Cambridge CB2 8FH, UK}}

   \date{}

 
  \abstract
   {Several diffuse interstellar bands (DIBs) have profiles with resolved sub-peaks that resemble rotational bands of large molecules. Analysis of these profiles can constrain the sizes and geometries of the DIB carriers, especially if the profiles exhibit clear variations along lines of sight probing different physical conditions. }
   {Using the extensive data set from the ESO Diffuse Interstellar Bands Large Exploration Survey (EDIBLES) we searched for systematic variations in the peak-to-peak separation of these sub-peaks for three well-known DIBs in lines of sight with a single dominant interstellar cloud. 
   }
   {We used the spectra of twelve single-cloud sight lines to examine 
   the $\lambda\lambda$5797, 6379, and 6614 DIB profiles. We measured the peak-to-peak separation in the band profile substructures for these DIBs. We adopted the rotational contour formalism for linear or spherical top molecules to infer the rotational constant for each DIB carrier and the rotational excitation temperature in the sight lines. We compared these to experimentally or theoretically obtained rotational constants for linear and spherical molecules to estimate the DIB carrier sizes.}
   {All three DIBs have peak separations that vary systematically between lines of sight, indicating correlated changes in the rotational excitation temperatures. The rotational constant $B$ of the $\lambda$6614 DIB was determined independently of the rotational excitation temperature; we derived $B_{6614}$=$(22.2\pm8.9)\times 10^{-3}$~cm$^{-1}$, consistent with previous estimates. Assuming a similar rotational temperature for the $\lambda$6614 DIB carrier and assuming a linear carrier, we found B$_{5797}^{\rm linear}=(5.1\pm2.0)\times10^{-3}~{\rm cm}^{-1}$ and B$_{6379}^{\rm linear} =(2.3\pm0.9)\times10^{-3}~{\rm cm}^{-1}$.  If the carriers of those DIBs are spherical species, on the other hand, their rotational constants are half that value, $B_{5797}^{\rm spherical} = (2.6\pm1.0)\times10^{-3}~{\rm cm}^{-1}$ and $B_{6379}^{\rm spherical} = (1.1\pm0.4)\times10^{-3}~{\rm cm}^{-1}$.}
   {Systematic variations in the DIB profiles provide the means to constrain the molecular properties. We estimate molecule sizes that range from 7--9 carbon atoms ($\lambda$6614 carrier, linear) to 77--114 carbon atoms ($\lambda$6379, spherical). }


   \maketitle
%


\section{Introduction}
\label{Sec:introduction}
The diffuse interstellar bands (DIBs) are a set of hundreds of unidentified optical absorption features that arise from the interstellar medium (ISM); readers can refer to \cite{Herbig1995}, \cite{Sarre2006}, and \cite{Snow2013} for reviews and \citet{DIBconf2014} for information on further progress in the field.  \citet{Hobbs:HD204827, Hobbs:HD183143} and \citet{Fan_2019} provide catalogues of known DIBs. \citet{1922LicOB..10..146H} established that DIBs are of interstellar origin by showing that they are stationary in the spectra of spectroscopic binary stars. Further evidence was provided by the rough correlation between DIB absorption strength and interstellar reddening $E(B-V)$ \citep{Merrill38}. DIBs are widespread throughout the ISM: nearly any sightline with non-negligible reddening also shows DIBs in its spectrum, both in the Milky Way and other galaxies \citep[see for example,][]{2002ApJ...576L.117E, 2011ApJ...726...39C, 2005A&A...429..559S,2018A&A...615A..33M}. The carriers that cause the DIBs must be abundant and survive the harsh conditions in the ISM, such as the ultraviolet (UV) radiation field. The current consensus is that the DIB carriers are most likely large, stable, carbonaceous molecules such as carbon chains, polycyclic aromatic hydrocarbons \citep[PAHs, see for example,][]{Salama:1996}, fullerenes, or related species \citep[see, for example, various contributions in][]{DIBconf2014}.

Support for this hypothesis has grown with the first convincing identification of a DIB carrier. \citet{Foing:C60_1} discovered two near-infrared DIBs close to the expected wavelengths of the electronic bands of C$_{60}^+$. Improved laboratory techniques confirmed that these two DIBs -- and a few weaker optical bands -- are indeed due to C$_{60}^+$, making  C$_{60}^+$ the only widely accepted DIB carrier (\citealp{campbell:c60+dibs, walker, 2017ApJ...843...56W, 2017ApJ...843L...2C, 2017ApJ...846..168S, EDIBLES2, 2019ApJ...875L..28C};   see \citealp{Harold:C60plusreview} for a review). 

Comparing DIBs to experimental data and theoretical models can constrain the nature of the unknown carriers. DIBs have a wide range of strengths and band profile shapes. Band widths vary from tens of \AA\ to a fraction of an \AA, with the broader DIBs mostly having smooth, often symmetric profiles while the narrower bands often have a resolved substructure. The strongest DIB, $\lambda$4428, has a broad Lorentzian profile \citep{Snow:4430profile}. 
This shape is expected for lifetime broadening of a short-lived upper energy level, implying femtosecond lifetimes for the excited states of the carrier molecule -- similar to laboratory measurements of PAHs \citep{Snow:4430profile}. DIB profiles with resolved substructures have shapes similar to the P, Q, and R branches of rotational bands \citep{Sarre1995,Kerr98}. Modelling these band shapes shows they can be explained by rotational bands of large molecules and have been used to estimate the rotational constants of the carrier molecules (\citealt{Cossart-Magos:contours, Kerr96,Ehrenfreund_rotation,2015MolPh.113.2159H}). Measuring the separations between the absorption peaks of the DIB profile substructures provides an estimate of the moment of inertia of the carrier, and thus the molecule size. Adopting rotational excitation temperatures of $\sim$50~K, expected for PAHs in diffuse clouds, this yields typical carrier sizes of 40--60 carbon atoms for the $\lambda\lambda$ 5797, 6379, and 6614 DIBs \citep{Ehrenfreund_rotation}.

It is well established that DIBs exhibit an environmental dependence: some of their properties vary between sightlines with similar reddening but different physical conditions. A well-known example of this environmental behaviour is offered by the relative strengths of the $\lambda\lambda$5780 and 5797 DIBs. Their ratio can change by a factor of four between sightlines similar to HD~147165 ($\sigma$~Sco; environments more exposed to UV radiation) and others more similar to HD~149757 \citep[$\zeta$~Oph; more shielded environments; see, for example,][]{1995ASSL..202...13K, Cami:correlations}. Such variations find their origin in different physical conditions \citep[for example, exposure to the interstellar radiation field][]{Cami:correlations,2005A&A...432..515R,2017ApJ...850..194F} and can then be used to estimate molecular properties of DIB carriers, such as their ionization potential \citep[][]{Sonnentrucker2013}.

 Environmental variations have also been established for DIB line profiles. There are systematic variations in the substructure of the $\lambda$6614 DIB between single-cloud sightlines with different physical conditions \citep{cami2004rotational}. These variations were interpreted as changes in the rotational excitation temperature which then allows the carrier's rotational constant to be determined independently of the carrier's rotational temperature in each line of sight. \citet{cami2004rotational} found lower rotational temperatures ($\sim$20--25~K) for $\lambda$6614 and consequently smaller carrier sizes than proposed by \citet{Ehrenfreund_rotation}. 

In this paper, we expand on the work presented by \citet{cami2004rotational}. We use the higher-quality spectra from the ESO Diffuse Interstellar Bands Large Exploration Survey \citep[EDIBLES; ][]{EDIBLES1} to select a sample of single-cloud lines of sight, then search for systematic line profile variations in three DIBs with resolved substructures ($\lambda\lambda$5797, 6379 and 6614). 
Our goals are to verify the results of \citet{cami2004rotational} using a larger sample
and to determine the carrier rotational constants and sizes for the two additional DIBs. In Sect.~\ref{sect:obsanddata} we describe the observations, target selection, and data processing. Sect.~\ref{sect:measurements} details the measurement of DIB substructure peak positions. In Sect.~\ref{Sec:Analysis} we determine the rotational constants and rotational excitation temperatures for our selected DIBs and sightlines. Sect.~\ref{sect:estimates} presents a comparison to experimentally obtained or theoretically calculated rotational constants to estimate DIB carrier sizes and Sect.~\ref{sect:discuss} discusses the astronomical implications for the DIB carriers. Appendix~\ref{Sect:App:formalism} provides details about the rotational contour formalism that we use in this paper while appendices~\ref{Sect:Na_Doublet_plots}--\ref{Sect:App_6379} provide supplementary figures and data tables. 


\section{Observations and target selection}
\label{sect:obsanddata}

\subsection{The EDIBLES survey}
Our data are taken from the ESO Diffuse Interstellar Bands Large Exploration Survey \citep[EDIBLES; ][]{EDIBLES1}, which has collected spectra at a high signal-to-noise ratio (S/N$\sim$1000) and high spectral resolution (resolving power $R \sim$80,000--110,000) over a wide spectral range ($\sim$300~nm -- 1$\mu$m). The EDIBLES sample comprises 123 O- and B-type stars, chosen to sample a range of interstellar conditions. \citet{EDIBLES1} provide details on the target selection and data reduction procedures. 

\subsection{Target selection: single cloud sightlines}
\label{sec:targets}

\begin{table*}
\centering     
\caption{Single-cloud lines of sight we selected from the EDIBLES dataset. We list the target names, coordinates (J2000 equinox), main ISM cloud radial velocity ($v_\text{ISM}$), stellar spectral type, reddening ($E(B-V)$), and fraction of molecular hydrogen $f(\text{H}_2)$. The values of $v_\text{ISM}$ are measured in Sect.~\ref{sec:targets}, all other parameters are taken from \citet{EDIBLES1}.} 
\label{target_data}
\begin{tabular}{cccr@{.}l@{$\pm$}lcccS} 
\hline \hline
Target & Right    & Declination    & \multicolumn{3}{c}{$\varv_\text{ ISM}$} & Spectral type & $E(B-V)$ & $f(\text{H}_2)$\\ 
       & ascension &  & \multicolumn{3}{c}{[km~s$^{-1}$]} &      & [mag] & \\ 

\hline
HD23180 & 03:44:19.1 & +32:17:17.7 & 13 &3 &0.2 & B1 III & 0.28 & 0.51  \\  
HD24398 & 03:54:07.9 & +31:53:01.1 & 13&8 &0.1 & B1 Ib  & 0.29 & 0.60 \\ 
HD144470 & 16:06.48.4 & $-$20:40:09.1 & $-$10&1 & 0.1 & B1 V & 0.21 & 0.132 \\  
HD147165 & 16:21:11.3 & $-$25:35:34.1 & $-$6&4  & 0.3 & B1 III & 0.37 & 0.053 \\  
HD147683 & 16:24:42.7 & $-$34:53:37.5 & $-$0&8  & 0.2 & B3: Vn (SB2) & 0.29 & 0.377\\  
HD149757 & 16:37:09.5 & $-$10:34:01.5 & $-$13&9 & 0.1 & O9.2 IVnn  & 0.32 & 0.630 \\  
HD166937 & 18:13:45.8 & $-$21:03:31.8 & $-$6&4  & 0.2 & B8 Iab(e) & 0.22 & --- \\  
HD170740 & 18:31:25.7 & $-$10:47:45.0 & $-$10&1 & 0.2 & B2 V & 0.45 & 0.575 \\  
HD184915 & 19:36:53.5 & $-$07:01:38.9 & $-$12&0 & 0.2 & B0.5 IIIn  & 0.22  & 0.366\\  
HD185418 & 19:38:27.5 & +17:15:26.1 & $-$10&1  & 0.2 & B0.5 V & 0.42 & 0.398 \\  
HD185859 & 19:40:28.3 & +20:38:37.5 & $-$8&2   & 0.1 & B0.5 Ia & 0.56 & --- \\  
HD203532 & 21:33:54.6 & $-$82:40:59.1 & 14&2  & 0.1 & B3 IV  & 0.30 & 0.84 \\  
\hline
\end{tabular}
\end{table*}

To isolate variations in the DIB profiles due to changes in the physical conditions, we restricted our targets to single-cloud sightlines. Ideally, these have only one intervening interstellar cloud with sufficiently uniform properties to be considered a single environment. In practice, almost all sightlines show multiple cloud components in strong atomic interstellar lines, for example, \ion{Na}{i}~D, so there are very few truly single cloud sightlines known. We consider a sightline effectively a single cloud if, by eye, there is a single dominant component to interstellar UV \ion{Na}{i} lines at 3302~\AA~(see Figs.~\ref{Fig:Na1}-\ref{Fig:Na2}). These UV lines are far less saturated than the Na D lines, so should more accurately reflect the relative column densities in each cloud component.

We found that twelve of the EDIBLES targets fulfil this single-cloud requirement, listed in Table~\ref{target_data}. Six of these targets overlap with those analysed in \cite{EDIBLES3}. We exclude HD147889, which was included in their analysis because we found that the \ion{Na}{i}~D lines are composed of two, roughly equally strong and overlapping components in this sightline. Although most of our selected targets show weaker cloud components as well, those have much lower column densities than the main cloud in each sightline (see Appendix~\ref{Sect:Na_Doublet_plots}). We derived cloud velocities from the strongest UV \ion{Na}{i} line, which has laboratory wavelength 3302.368~\AA\ \citep{NIST_ASD}. We compared these to velocities for the same lines of sight determined from \ion{K}{i} lines \citep{2001ApJS..133..345W} and found them to be consistent.


\subsection{Target selection: DIBs}
\label{Sect:DIBselection}
For this study, we need to first and foremost select DIBs that show a clearly resolved substructure in their profile at the resolution of our EDIBLES observations. High-resolution studies of DIB profiles \citep[for example,][]{Sarre1995,1997ApJ...477..209K,Kerr98,Galazutdinov2002,gala08,2006A&A...448..221S} have indeed revealed several DIBs with clearly resolved substructures. However, the peak separation for some DIBs (for example, the $\lambda6196$ DIB) is too small, while other DIBs \citep[for example, the C$_2$-DIBs,][]{EDIBLES3} are too weak for their peak separation to be reliably measured in our EDIBLES data. Only the $\lambda\lambda$5797, 6379 and 6614 DIBs were found to be suitable for our purposes here.  

For all three DIBs in our study, there is furthermore some evidence for profile variability that could be the result of changes in the rotational excitation. 
\citet{cami2004rotational} found variations in substructure separation of no more than 0.07~\AA\ for the $\lambda$6614 DIB, which corresponds to about a resolution element at the EDIBLES resolving power. This DIB thus allows us to directly compare our results to \cite{cami2004rotational} and is the only one of these three DIBs with a triple-peak substructure (with an additional red wing). For this particular DIB, it has been suggested that the profile includes contributions from vibrational hot bands \citep{2015MNRAS.453.3912M}; including these hot bands in our analysis greatly improves the overall fit of the observed profiles but should not greatly affect the peak positions we measure. 

All three DIBs also show an extended tail to the red (ETR) when observed toward Herschel~36 \citep{2013ApJ...773...41D,2013ApJ...773...42O}, likely due to radiative pumping of the rotational states, coupled with a slightly smaller rotational constant in the excited state. The redward tails are much more pronounced for the $\lambda\lambda$5797 and 6614 DIBs than for the $\lambda$6379 DIB.


\subsection{Co-adding observations}
\begin{figure} 
\includegraphics[width=\columnwidth]{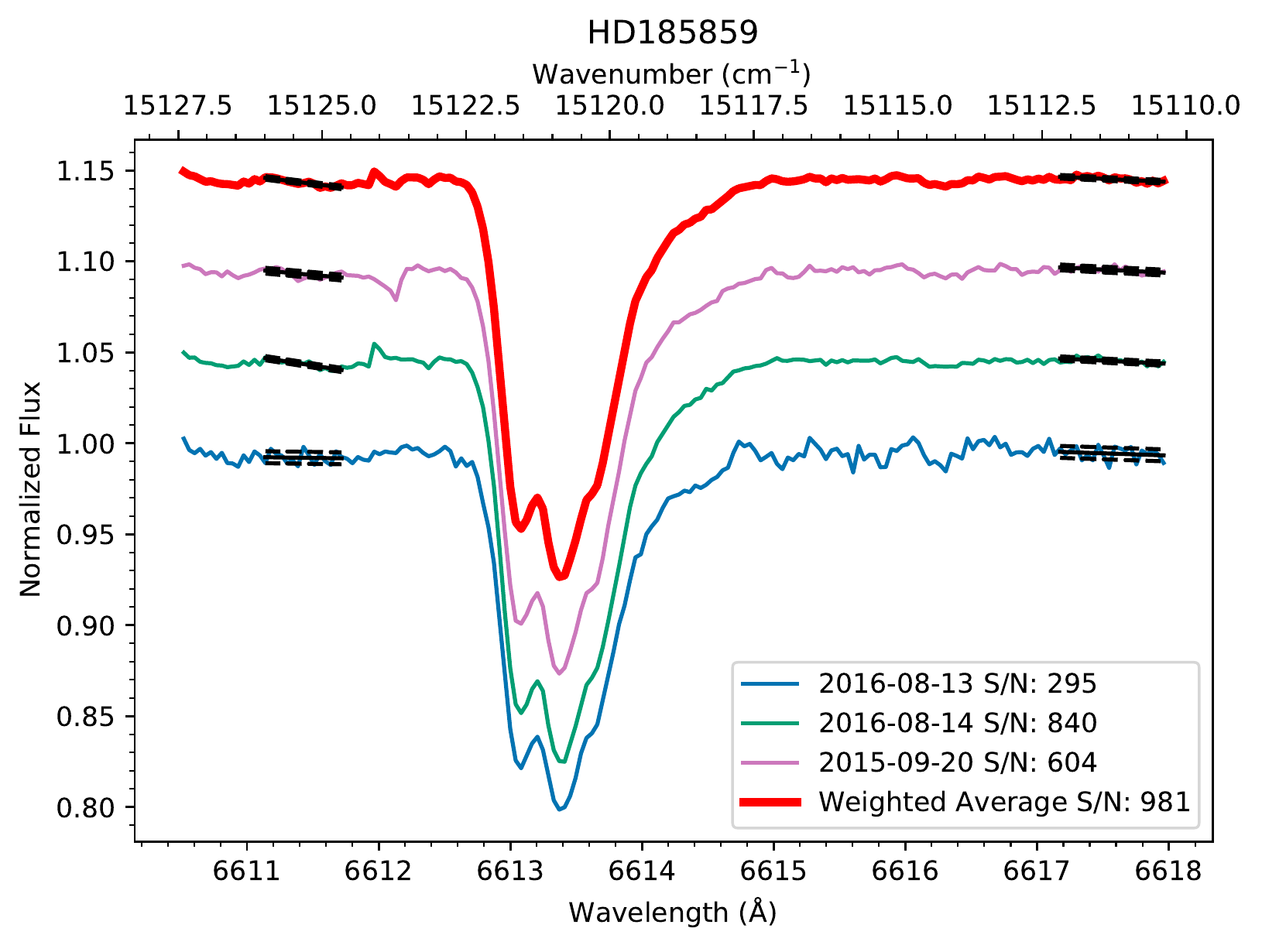}
\caption{All available EDIBLES observations of the $\lambda$6614 DIB toward HD 185859 (thin lines), and our weighted average spectrum (thick red line). Observation dates and S/N are indicated in the legend. The S/N was estimated from the standard deviations in the regions marked in black. A vertical offset is added for clarity.}
\label{Fig:6614_weighted_average}
\end{figure}

\begin{figure}
    \includegraphics[width=\columnwidth]{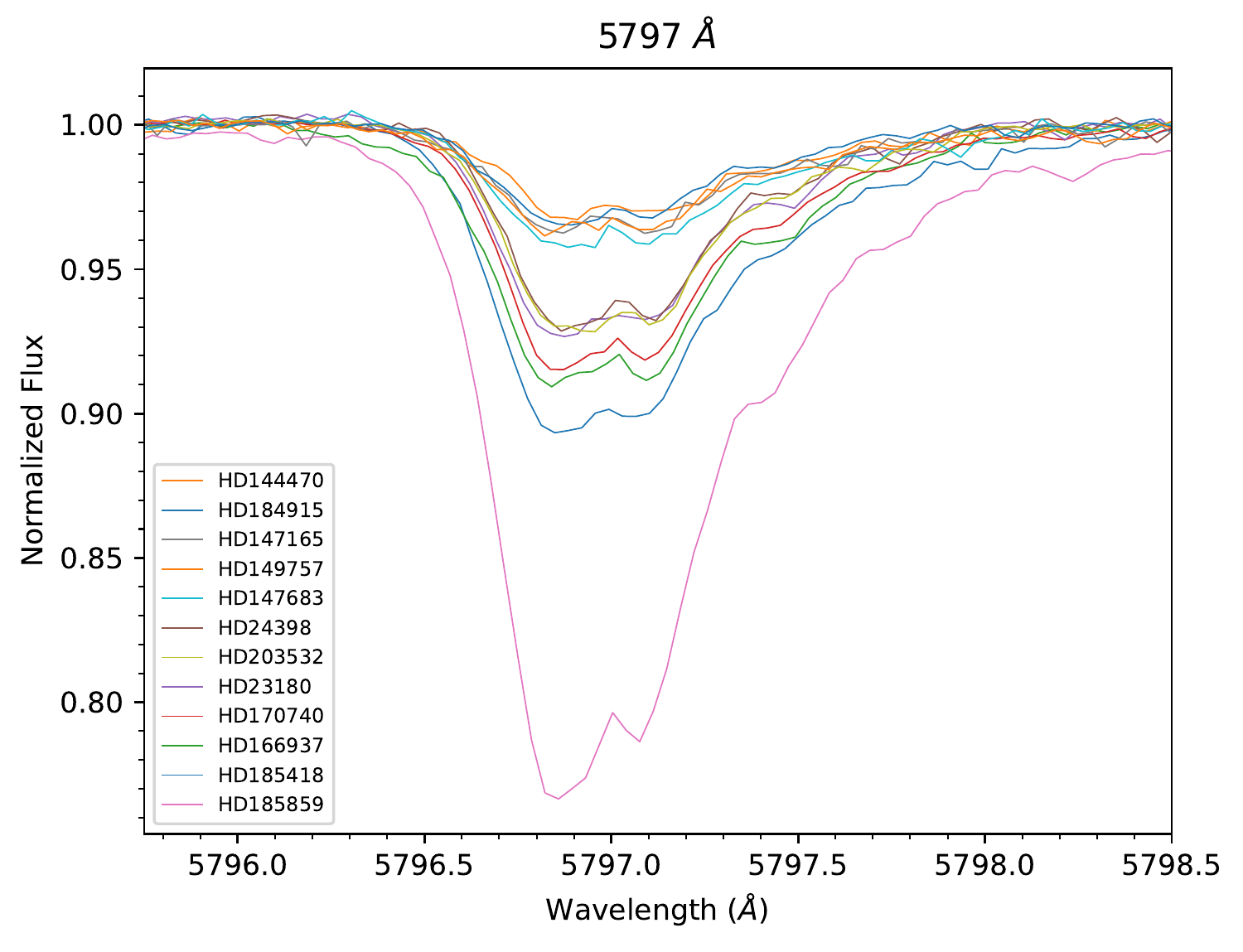}\\
    \includegraphics[width=\columnwidth]{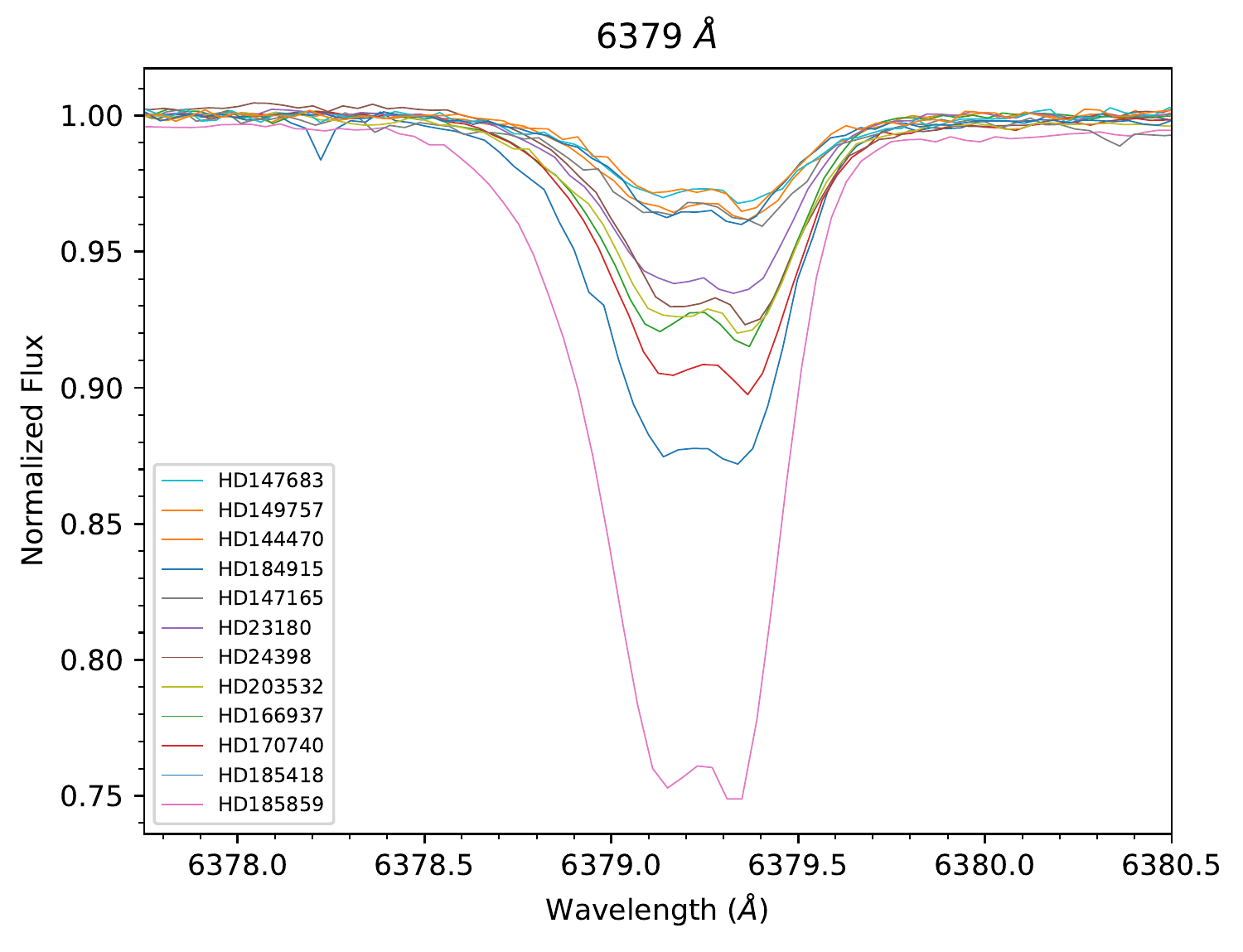}
    \includegraphics[width=\columnwidth]{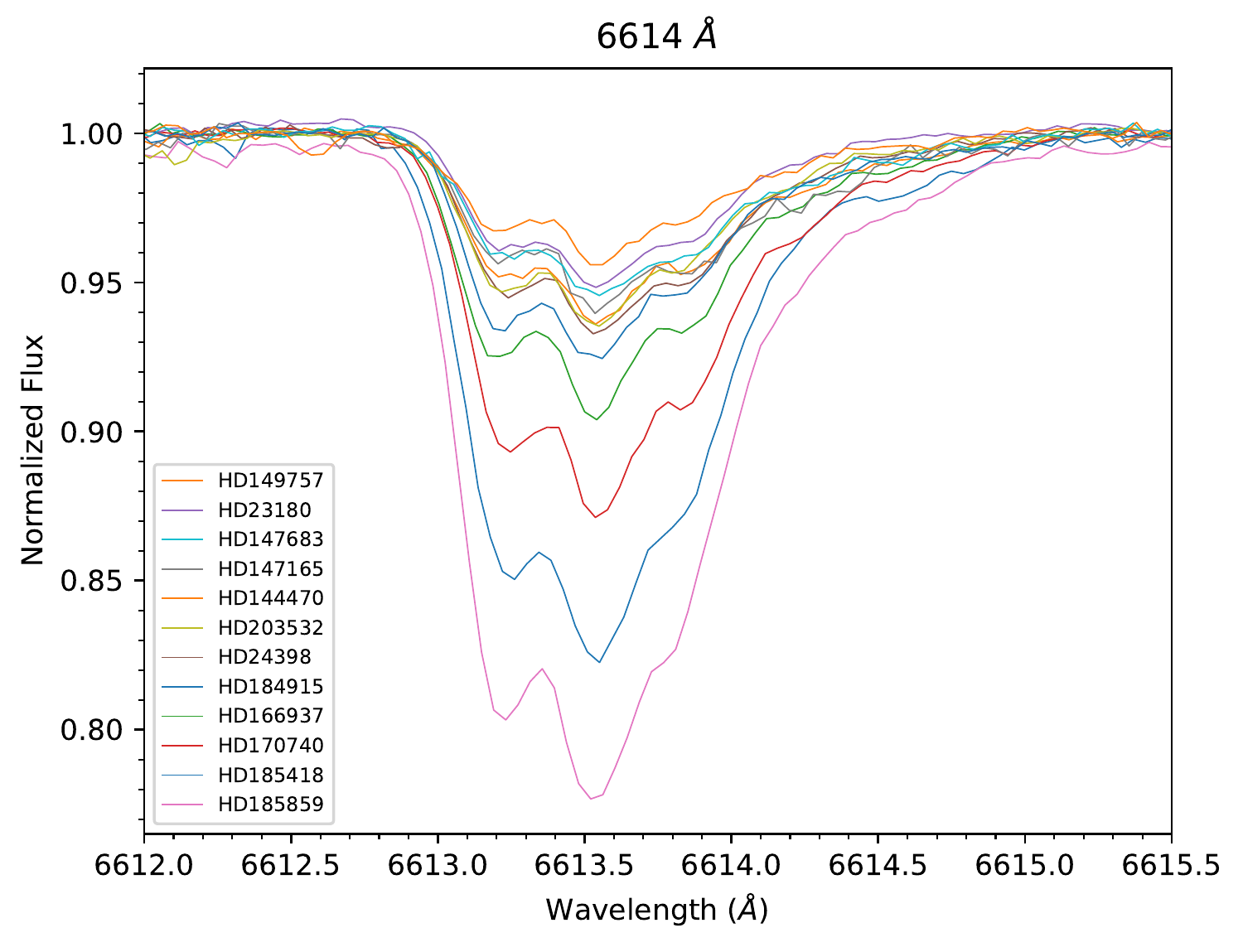}
    \caption{The co-added observations of our sample sight lines, shifted to the interstellar rest frame for respectively the (a)$\lambda\lambda$5797 (top panel),  (b) 6379 (middle panel) and (c) 6614 (bottom panel) DIBs. }
    \label{fig:co_added}
\end{figure}

Many of our target sightlines were observed by EDIBLES on multiple nights. We co-added these separate observations to increase the S/N. Before co-addition, we first identified continuum regions on either side of the DIBs, then fitted a cubic spline model to those continuum regions. The spectra were normalised by dividing by this continuum model. We then co-added the normalised spectra using an inverse-variance weighted average of all available observations (that is, using a weight $w_i\propto1/\sigma_i^2$ with $\sigma_i$ the uncertainty on observation $i$). As an example, Fig.~\ref{Fig:6614_weighted_average} shows the three available spectra for HD~185859 as well as the weighted average co-added spectrum of these three observations shown in red. In Fig.~\ref{fig:co_added}, these averaged spectra are presented for the three DIBs studied and for the sightlines listed in Table~\ref{target_data}; thus, the red spectrum in Fig.~\ref{Fig:6614_weighted_average} corresponds to the pink spectrum in Fig.~\ref{fig:co_added}c. All measurements and analyses were performed on these averaged spectra. 

\section{Measuring peak positions}
\label{sect:measurements}

Our analysis requires measurements of the precise locations of the sub-peaks in the profiles. \citet{Galazutdinov2002} decomposed the DIB profiles they studied into a sum of overlapping Gaussian absorption features, and it was from these measurements that \citet{cami2004rotational} then established the profile variations in the $\lambda$6614 DIB. We first adopted this method, using Voigt profiles rather than Gaussians to better reproduce the red wings in some of the bands. The number of components to use in each DIB profile was determined through trial and error until a good overall fit to the DIB profile was obtained for all sightlines. This procedure resulted in good overall fits to the profile for all DIBs in all sightlines, using a Levenberg-Marquardt minimisation method, which also provided uncertainties on component positions. The central wavelengths of the individual components could then be used as a measurement for the peak location. 

\begin{figure}
\includegraphics[width=\columnwidth]{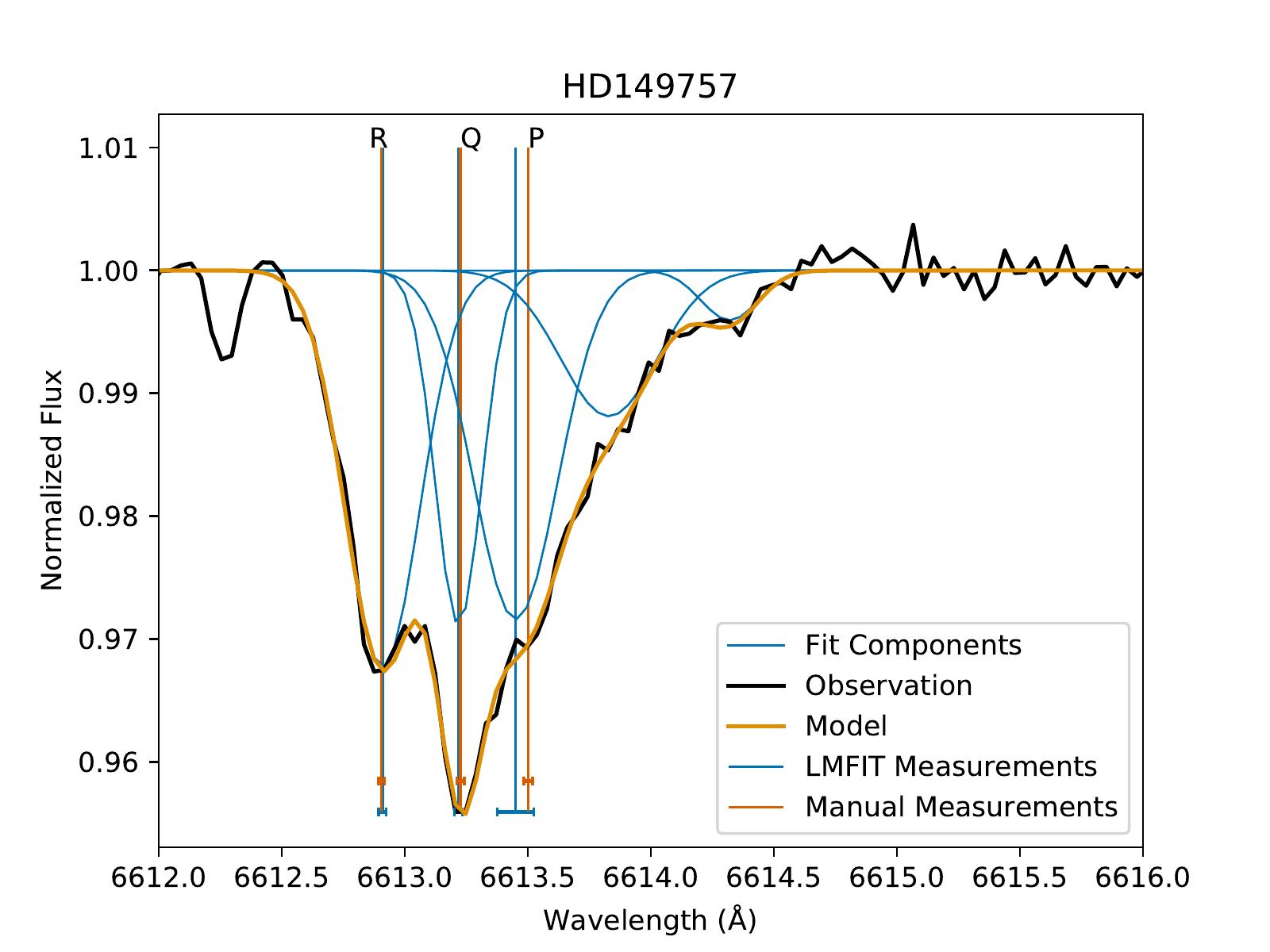}
\caption{The observed $\lambda$6614 DIB profile toward HD~149757 (black) and the best fitting model profile (orange) composed of 5 individual Voigt profiles (blue). The blue vertical lines indicate the central wavelengths of the three main components from the model fit. Dark orange vertical lines indicate the peak positions determined with the alternative manual measurement process. Especially for the third peak, both positions differ from each other, and the peak position from the model fit does not correspond to the peak absorption. Uncertainties on these measurements are indicated by the error bars at the bottom of each line. The peak positions are labelled as $P$, $Q$ or $R$ branches, consistent with our rotational contour terminology.} 
\label{fig:val_comp}
\end{figure}

However, when inspecting our results, we found that there was often a noticeable offset between the peak positions measured with this method and the peak locations apparent by eye. An example is shown in Fig.~\ref{fig:val_comp}, where the fitted components represent the first two peaks well, but not the third one. The differences between the components' central wavelength and the deepest absorption are small but are of the same magnitude as the effect we are trying to quantify. This discrepancy stems from the fitting being optimised to accurately reproduce the overall profile rather than accurately measuring the peak position. A contributing factor is likely that the profiles themselves are intrinsically asymmetric while the fitting used symmetric individual components. The data used by \citet{Galazutdinov2002} was at a much higher resolving power ($R\sim220,000)$ than our EDIBLES data, providing more detail in the structure of the profiles. Although we tried several modifications to the automated fitting process, we were unable to measure the peak positions adequately with multiple symmetric components. If the geometry of the DIB carriers was known, detailed rotational band profile modelling would address this issue, but it remains unknown.

Given the difficulties of using this automated fitting method, we decided to perform manual measurements of the peak positions by marking the points of deepest absorption in each branch, taking into account the profile shape and possible noise contributions. To exclude that this approach yields a biased result, we repeated each measurement a total of five times, each time adding random noise to the data. This allowed us to obtain an independent estimate of the involved uncertainties. The noise was drawn from a normal distribution with a standard deviation equal to the root-mean-square value of the continuum. From this set of manual measurements, we determined mean peak positions and their standard deviations. We find that the standard deviation (with values typically on the order of a few times 10$^{-2}$ \AA) is consistent with the uncertainties on the peak positions determined from the automated fitting described above, so we are sufficiently confident that the uncertainty estimate is reasonable. We adopt these manual measurements for the remainder of our analysis. The results of the manual peak measurements are listed in Tables~\ref{table:6614_manual}, \ref{table:5797_manual}, and \ref{table:6379_manual}. The peak separations derived from these measurements are summarised in Table~\ref{table:results}.

\begin{table*}
    \caption{Peak separations for each DIB in each sightline. $\nu_{QP}$ is the separation between the peak values of the Q- and P-branch frequencies, $\nu_{RQ}$ is the separation between the peak values of the R- and Q-branches, and $\nu_{PR}$ is the separation between the peak values of the P- and R-branches as defined in Sect.~\ref{Sec:Analysis}. Values were calculated from the measurements presented in Tables~\ref{table:6614_manual}, \ref{table:5797_manual} and \ref{table:6379_manual}.}
    \label{table:results}
    \centering
    \begin{tabular}{c|ccc|c|c}
\hline
\hline
\multicolumn{1}{c|}{} & \multicolumn{3}{c|}{6614} & \multicolumn{1}{c|}{5797} &  \multicolumn{1}{c}{6379}\\
\cline{1-6}
\multicolumn{1}{c|}{Target} & $\nu_{QP}$ & $\nu_{RQ}$ & $\nu_{PR}$ & $\nu_{PR}$ & $\nu_{PR}$ \\
\multicolumn{1}{c|}{} & [cm$^{-1}$] & [cm$^{-1}$] & [cm$^{-1}$] & [cm$^{-1}$] & [cm$^{-1}$] \\
\hline
HD23180 & 0.57 $\pm$ 0.06 & 0.71 $\pm$ 0.08 & 1.27 $\pm$ 0.09 & 0.6 $\pm$ 0.1 & 0.38 $\pm$ 0.04  \\  
HD24398 & 0.67 $\pm$ 0.05 & 0.68 $\pm$ 0.02 & 1.34 $\pm$ 0.05 & 0.68 $\pm$ 0.05 & 0.44 $\pm$ 0.06   \\
HD144470  & 0.70 $\pm$0.04 & 0.69 $\pm$0.05 & 1.39 $\pm$0.06 & 0.7  $\pm$0.1  & 0.46 $\pm$0.05 \\
HD147165  & 0.67 $\pm$0.05 & 0.71 $\pm$0.04 & 1.38 $\pm$0.06 & 0.77 $\pm$0.07 & 0.68 $\pm$0.05 \\
HD147683  & 0.65 $\pm$0.08 & 0.66 $\pm$0.05 & 1.30 $\pm$0.09 & 0.62 $\pm$0.08 & 0.54 $\pm$0.06 \\
HD149757  & 0.63 $\pm$0.05 & 0.73 $\pm$0.04 & 1.36 $\pm$0.05 & 0.80 $\pm$0.05 & 0.50 $\pm$0.08 \\
HD166937  & 0.68 $\pm$0.05 & 0.78 $\pm$0.06 & 1.46 $\pm$0.07 & 0.76 $\pm$0.05 & 0.58 $\pm$0.02 \\
HD170740  & 0.66 $\pm$0.03 & 0.67 $\pm$0.01 & 1.33 $\pm$0.03 & 0.69 $\pm$0.03 & 0.51 $\pm$0.04 \\
HD184915  & 0.64 $\pm$0.06 & 0.72 $\pm$0.04 & 1.37 $\pm$0.05 & 0.64 $\pm$0.05 & 0.49 $\pm$0.04 \\
HD185418  & 0.58 $\pm$0.04 & 0.69 $\pm$0.04 & 1.27 $\pm$0.03 & 0.58 $\pm$0.06 & 0.44 $\pm$0.05 \\
HD185859  & 0.54 $\pm$0.05 & 0.73 $\pm$0.05 & 1.27 $\pm$0.05 & 0.67 $\pm$0.05 & 0.44 $\pm$0.03 \\
HD203532  & 0.65 $\pm$0.04 & 0.64 $\pm$0.07 & 1.29 $\pm$0.07 & 0.7  $\pm$0.1  & 0.46 $\pm$0.04 \\
\hline
\end{tabular}
\end{table*}

\section{Rotational contour variations}
\label{Sec:Analysis}

We interpret our line profile variations by comparing them to profile changes as expected for rotational contours of typically large molecules. 

\subsection{Rotational contours}

We assume that the three DIB profiles arise from the rotational contours of a large molecule, with each substructure component corresponding to the $P$-, $Q$-, and $R$-branches (these branches are indicated in Fig.~\ref{fig:val_comp}; each branch consists of numerous unresolved rotational lines). As a first step, we restrict the analysis to linear or spherical top molecules, as in that case, their rotational energy levels depend on only a single rotational constant $B$; we also assume that the rotational constant in the excited state does not differ too much from that in the ground state (i.e. we assume that $\Delta B/B << 1$). These assumptions allow a relatively straightforward analysis, in line with the fact that many identified species in dark clouds fulfil this criterion, but obviously, this puts a very strong constraint on the data interpretation. We further assume that each carrier molecule has a Boltzmann rotational energy distribution with a single rotational excitation temperature $T_\text{rot}$. In Appendix~\ref{Sect:App:formalism}, we use standard molecular spectroscopy formalism to derive expressions for the allowed transitions under these assumptions, as a function of $B$ and $T_\text{rot}$. The measured wavelength of each sub-peak then corresponds to absorption originating from the rotational level with the highest population, which has rotational quantum number $J_\text{max}$. For a given rotational constant $B$, a higher $T_\text{rot}$ leads to a higher $J_\text{max}$, so variations in the peak separation are due to changes in the rotational temperature (Eqs.~\ref{Eq:nu_P_T}--\ref{Eq:nu_R_T}). With our approximations, the total intensities in the $P$ and $R$ branches relative to the $Q$ branch should be similar, and vary little with rotational temperature. Hence, the integrated intensity of the individual $P$ and $R$ branches should be similar as well. Given that the bands overlap and show some asymmetries, we are unable to test this in the observations. 

\subsection{Variations in the rotational temperatures}

For each of the three DIBs, we measure the separation between the components we ascribe to the $P$- and $R$-branches. Under our assumptions, this provides the value of the product $B\cdot T_\text{rot}$ from: 
\begin{equation}
    B\cdot T_{\rm rot} \approx \frac{hc(\nu_R-\nu_P)^2}{8ak} = \frac{0.180}{a}(\Delta\nu_{RP})^2\hspace{0.2cm}{\rm [cm^{-1}~\cdot~K]},
\label{Eq:BT}
\end{equation}
where $a=1$ for linear molecules and $a=2$ for spherical tops, $h$ is the Planck constant, $c$ is the speed of light, $k$ is the Boltzmann constant, and $\nu_{R}$ and $\nu_{P}$ are the frequencies (expressed in cm$^{-1}$) corresponding to the R- and P-branch peaks respectively, and $\Delta\nu_{RP} \equiv \nu_R - \nu_P$ (see Appendix~\ref{Sect:App:formalism} for the derivation). Because each DIB must have the same carrier along all lines of sight, each DIB also has the same $B$ for all sightlines. Thus, the changing peak separation implies that the rotational temperature varies between lines of sight.

\begin{figure}
    \resizebox{\hsize}{!}{\includegraphics{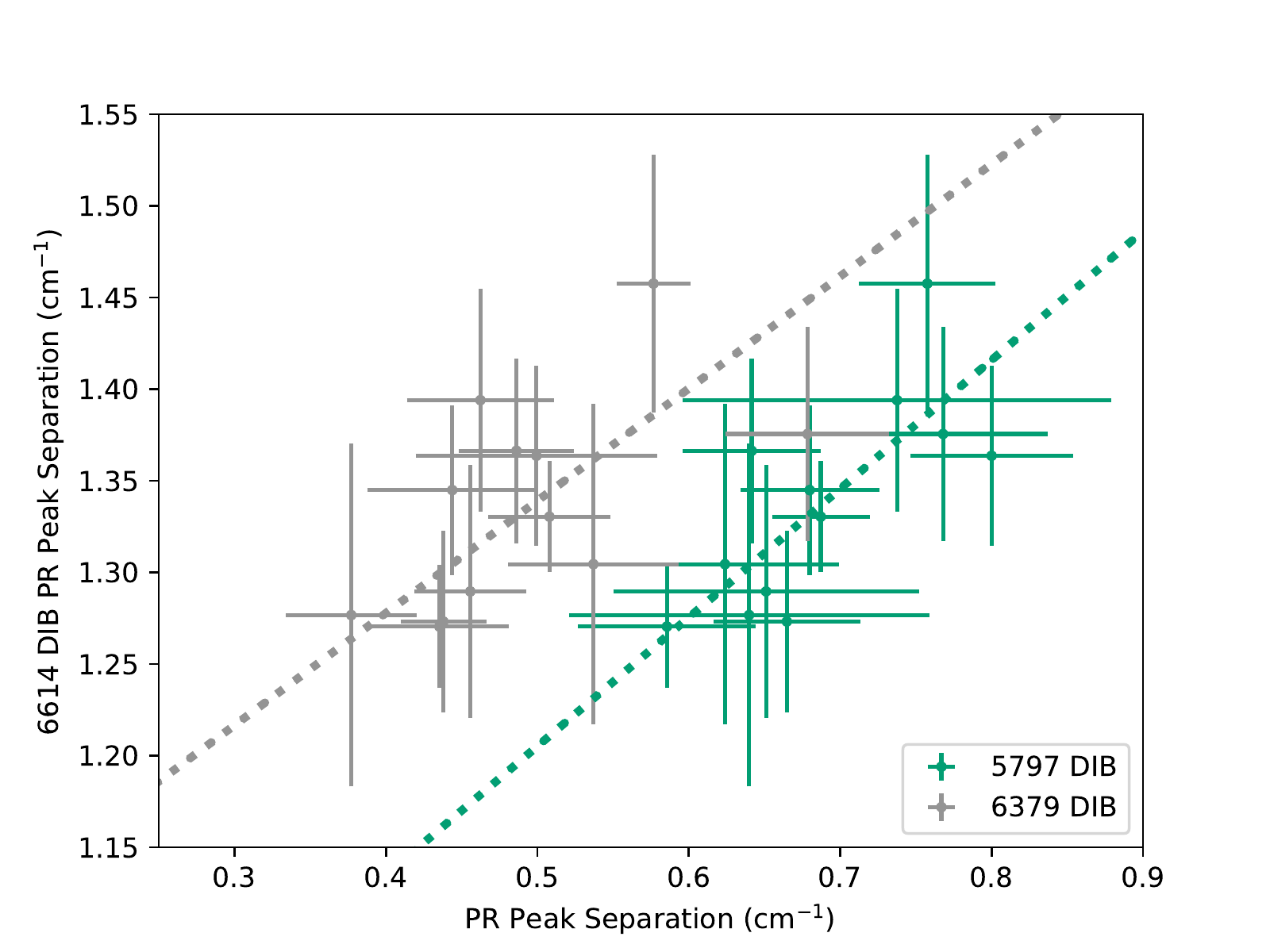}}
    \caption{$P$- to $R$-branch separation in all sightlines of the $\lambda$6614 DIB (vertical axis) as a function of the separation of the $\lambda\lambda$5797 (green) and 6379 (grey) DIBs on the horizontal axis (data values are listed in Table~\ref{table:results}). The dotted lines are linear models of the data using orthogonal-distance fitting of the 6614 measurements as a function of the 5797 values (green) and 6379 values (grey). 
}    
    \label{Fig:PR_separation_allDIBs}
\end{figure}

Figure~\ref{Fig:PR_separation_allDIBs} shows the $P$- to $R$-branch separation in all sightlines of the $\lambda$6614 DIB as a function of the separation of the $\lambda\lambda$5797 and 6379 DIBs. We find that the $P$- to $R$-branch separations do vary between sightlines, but the range in separations for each DIB is typically not much larger than the measurement uncertainties. However, as can be seen in Fig.~\ref{Fig:PR_separation_allDIBs}, the variations in the peak separation are correlated. The linear Pearson correlation coefficients are 0.74 for the correlation between the 6614 and 5797 DIBs and 0.60 for the correlation between the 6614 and 6379 DIBs. 
The peak separations of the different DIBs thus tend to increase in the same sightlines indicating that the same physical processes that cause an increase in the rotational temperature in one DIB cause increases in that of the other DIBs as well. 

We can use Eq.~(\ref{Eq:BT}) to quantify those temperature increases. The peak separation of the $\lambda$6614 DIB varies from $\sim$1.25 to $\sim$1.45 (Fig.~\ref{Fig:PR_separation_allDIBs} and Table~\ref{table:results}) and thus 
\begin{equation}
    T_{6614}^{\rm max} = 1.35 \times T_{6614}^{\rm min},
\label{Eq:6614_Tminmax}
\end{equation}
where the subscript refers to the DIB and the superscript to sightlines with the maximum and minimum peak separations respectively. Similarly, the peak separation of the $\lambda$5797 DIB ranges from $\sim$0.6 to $\sim$0.85 while that of the $\lambda$6379 ranges from $\sim$0.4 to $\sim$0.7 and thus
\begin{equation}
    T_{5797}^{\rm max} = 2.0 \times T_{5797}^{\rm min}
\label{Eq:5797_Tminmax}
\end{equation}
\begin{equation}
    T_{6379}^{\rm max} = 3.1 \times T_{6379}^{\rm min}.
\label{Eq:6379_Tminmax}
\end{equation}
We note that the value for $T^{\rm max}_{6379}/T^{\rm min}_{6379}$ is largely determined by the peak separation of 0.68 in HD~147165; if we discard this point, we find a ratio similar to that for the $\lambda$5797 DIB. 
Such large changes in the rotational temperatures were unexpected and are difficult to explain if the rotational temperatures are high. These variations would be more reasonable for very low rotational temperatures (for example, a temperature in the range 3~K to 9~K). Alternatively, our explicit assumption of linear or spherical top molecules might not hold -- in which case, other factors play a role in the peak separation as well.

\subsection{Rotational constant and rotational temperature of the $\lambda$6614 DIB carrier}

\begin{figure}
\includegraphics[width=\columnwidth]{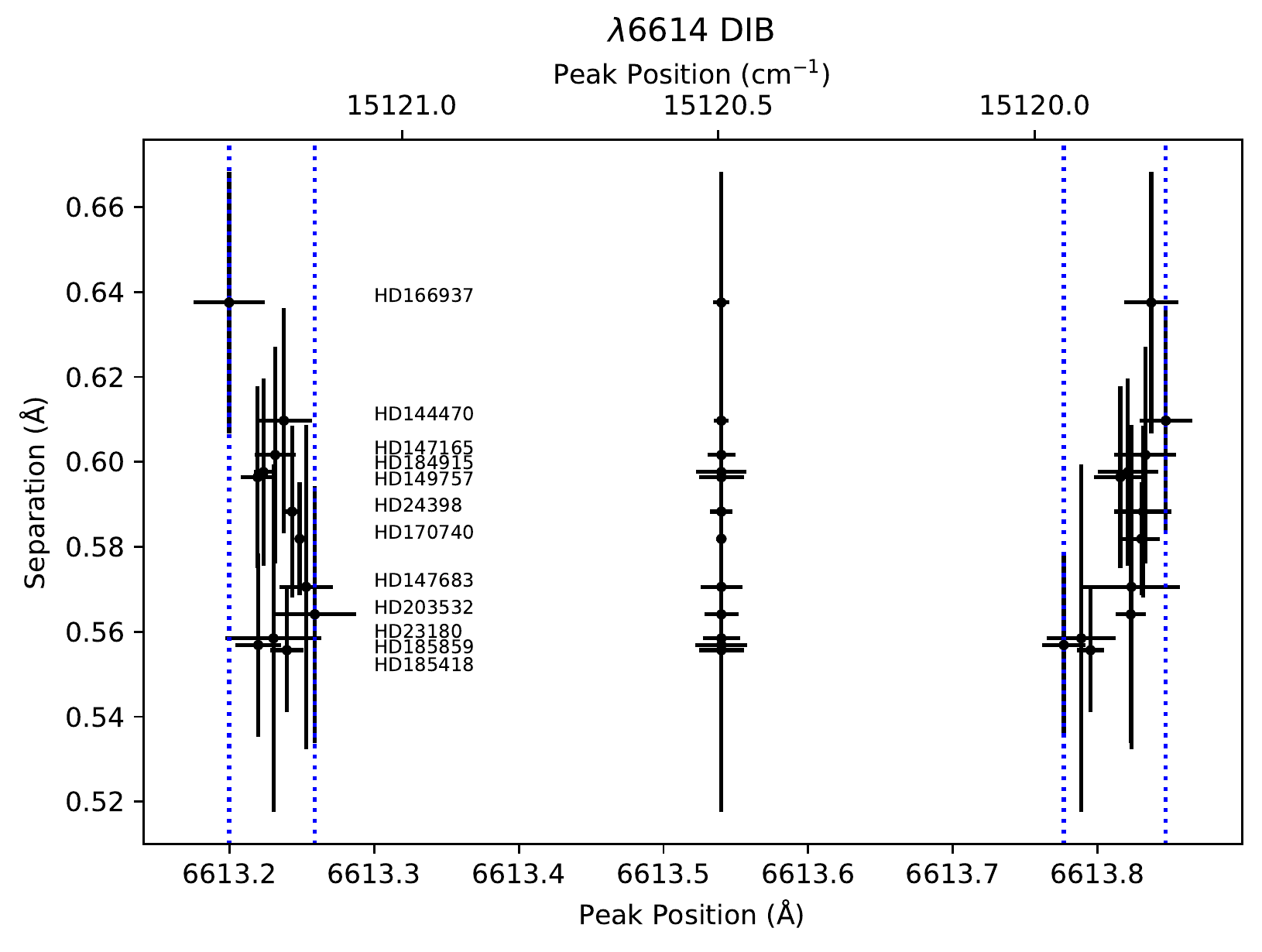}
\caption{Measured peak positions of the $\lambda$6614 DIB in our target sightlines. Positions have been shifted such that the Q-branch lies at 6613.54~{\AA} in all lines of sight.} 
\label{Fig:6614peakpos}
\end{figure}

Unlike the other two DIBs, $\lambda$6614 DIB has a triple-peak substructure. In the framework of rotational contours, the central peak corresponds to the unresolved $Q$-branch, the position of which is expected to be almost insensitive to $T_\text{rot}$ (see Appendix~\ref{Sect:App:formalism}). Our measurements confirm that this is indeed the case. After correcting for the radial velocities of the clouds from the Na lines, the measured peak positions of the central peak differ by less than 0.05{\AA} and show no correlation with the peak separation. The differences might arise from errors in the wavelength calibration, which are uncertain by a similar amount \citep{Cox2017}. We, therefore, use the wavelength of the $Q$-branch peak as a reference point. 

Figure~\ref{Fig:6614peakpos} shows the peak positions for the three sub-peaks in the $\lambda$6614 DIB after aligning their central peaks; this presentation of the data is similar to that used in \citet{cami2004rotational} but with larger uncertainties due to the lower spectral resolution of the EDIBLES data set. This indicates that the separations of both the blue ($R$) and red ($P$) peaks from the central peak change systematically, both moving away from the central peak. 

Expressing the peak separations relative to this central ($Q$-branch) peak allows us to directly determine the rotational constant $B$ independently of the rotational temperature (see Eqs.~(\ref{Eq:nuR_nuQ_J}) and (\ref{Eq:nuQ_nuP_J})): 
\begin{equation}
    (\nu_R - \nu_Q) - (\nu_Q-\nu_P) = 2B(1+\frac{\Delta B}{B}) \approx 2B.
\end{equation}

This difference is too small to measure reliably for each sightline individually, but by taking the weighted mean of the 12 sightlines (assuming they are independent measurements), we find that the rotational constant of the $\lambda$6614 DIB carrier is  
\begin{eqnarray}
B_{6614} & = & (22.2\pm8.9)\times10^{-3}~{\rm cm}^{-1}.
\label{Eq:B6614}
\end{eqnarray}
This is the value that we use for the remainder of this paper. 
We note that this value is compatible with the value of $B_{6614} = (16.4\pm3.1)\times10^{-3}$~cm${-1}$derived by \citet{cami2004rotational} using a slightly different method and a more accurate data set. With the value of $B$ known, we can then use Eq.~(\ref{Eq:BT}) to determine the excitation temperatures for each sightline, albeit with large uncertainties. Assuming a linear geometry for the $\lambda$6614 DIB carrier, the range in peak separation of $\sim$1.25--1.45~cm$^{-1}$ across our sightlines then corresponds to rotational temperatures in the range 12.7--17.1~K, slightly lower than the temperatures in \citet{cami2004rotational} due to the slightly larger rotational constant. Within the 1$\sigma$ uncertainties on $B_{6614}$, this range could be as low as 9.0--12.2~K and as high as 21.1--28.4~K. For a spherical geometry, the resulting temperatures are a factor of 2 smaller, that is, the nominal temperature range is 6.3--8.5~K, but within the uncertainties on $B_{6614}$, this range could be as low as 4.5--6.1~K or as high as 10.6--14.2~K.

\subsection{Relationships between the rotational constants}

The other two DIBs do not exhibit the triple-peak substructure, and thus do not allow us to determine $B$ and $T_{\rm rot}$ independently. However, we can gain some insight by comparing them to the $\lambda$6614 DIB using Eq.~\ref{Eq:BT}: 
\begin{equation}
    B_{5797} = B_{6614} \frac{T_{6614}}{T_{5797}}\frac{a_{6614}}{a_{5797}}\frac{(\Delta\nu_{RP}^{5797})^2}{(\Delta\nu_{RP}^{6614})^2} = \frac{B_{6614}}{a_{5797}}\frac{2.93~{\rm K}}{T_{5797}^{\rm min}}
\end{equation}

\begin{equation}
    B_{6379} = B_{6614} \frac{T_{6614}}{T_{6379}}\frac{a_{6614}}{a_{6379}}\frac{(\Delta\nu_{RP}^{6379})^2}{(\Delta\nu_{RP}^{6614})^2} = \frac{B_{6614}}{a_{6379}}\frac{1.30~{\rm K}}{T_{6379}^{\rm min}},
\end{equation}
where we have used the minimum peak separations and correspondingly that  \mbox{$T_{6614}^{\rm min}a_{6614}=12.7$~K}. Using the values for the maximum peak separations together with Eqs.~(\ref{Eq:6614_Tminmax})--(\ref{Eq:6379_Tminmax}) yields the same result. However we do not know $T_{5797}^{\rm min}$ or $T_{6379}^{\rm min}$, which are required to determine the rotational constants. As a firm lower limit to these temperatures, we can use the temperature of the cosmic microwave background (CMB), as the rotational excitation temperature of highly polar molecules are expected to be close to the CMB value \citep{cn_cmb}. Thus, using $T^{\rm min}=2.725$~K, we can determine upper limits to the rotational constants. The values, however, depend on the geometry of the carrier (through $a_{5797}$ and $a_{6379}$); we thus find
\begin{eqnarray}
    B_{5797}^{\rm linear} & \le & (23.8\pm9.6)\times10^{-3}~{\rm cm}^{-1}\label{Eq:B5797_limit_lin}\\
    B_{6379}^{\rm linear} & \le & (10.6\pm4.2)\times10^{-3}~{\rm cm}^{-1}\label{Eq:6379_limit_lin},
\end{eqnarray}
as firm upper limits to the rotational constants of these DIB carriers if they are linear molecules and
\begin{eqnarray}
        B_{5797}^{\rm spherical} & \le & (11.9\pm4.8)\times10^{-3}~{\rm cm}^{-1}\label{Eq:B5797_limit_spher}\\
    B_{6379}^{\rm spherical} & \le & (5.3\pm2.1)\times10^{-3}~{\rm cm}^{-1},
    \label{Eq:6379_limit_spher}
\end{eqnarray}
if they are spherical tops.

We have no prior information about these rotational temperatures, so $T^\text{min}$ could be substantially higher than the CMB value. However, the minimum temperatures must be sufficiently low to allow the rotational temperature to vary by a factor of 2 to 3 between sightlines. If we assume that the minimum temperatures for all DIBs are the same, that is $T_{5797}^{\rm min} = T_{6379}^{\rm min} = T_{6614}^{\rm min} = 12.7$~K, a value 4.6 times higher than the cosmic microwave background, the rotational constants are consequently lower by the same factor. In the absence of well-defined information, this choice is arbitrary, but we adopt it for the rest of our analysis, keeping in mind that the only strong constraints we have are the upper limits in Eqs.~(\ref{Eq:B5797_limit_lin})--(\ref{Eq:6379_limit_spher}). We discuss this issue further in Sect.~\ref{sect:discuss}. Under this assumption, we then find: 
\begin{eqnarray}
    B_{5797}^{\rm linear} & = & (5.1\pm2.0)\times10^{-3}~{\rm cm}^{-1}\label{Eq:B5797_linear_final}\\
    B_{6379}^{\rm linear} & = & (2.3\pm0.9)\times10^{-3}~{\rm cm}^{-1}\label{Eq:B6379_linear_final},
\end{eqnarray}
and 
\begin{eqnarray}
    B_{5797}^{\rm spherical} & = & (2.6\pm1.0)\times10^{-3}~{\rm cm}^{-1}\label{Eq:B5797_spherical_final}\\
    B_{6379}^{\rm spherical} & = & (1.1\pm0.4)\times10^{-3}~{\rm cm}^{-1}\label{Eq:B6379_spherical_final},
\end{eqnarray}
where uncertainties are derived from the statistical uncertainties on $B_{6614}$; their systematic uncertainties are much larger than this because of the unknown temperatures. 

With these values for the rotational constants, the rotational excitation temperature for the $\lambda$5797 DIB then changes from 12.7~K to 25.4~K across the sightlines considered here while that of the 6379 DIB changes from 12.7 to 38.1~K. Recall, however, that all three DIB carriers may not have the same $T_{\rm min}$ -- this is an assumption.

\section{DIB carrier size estimates}
\label{sect:estimates}

\begin{figure*}
\resizebox{\hsize}{!}{\includegraphics{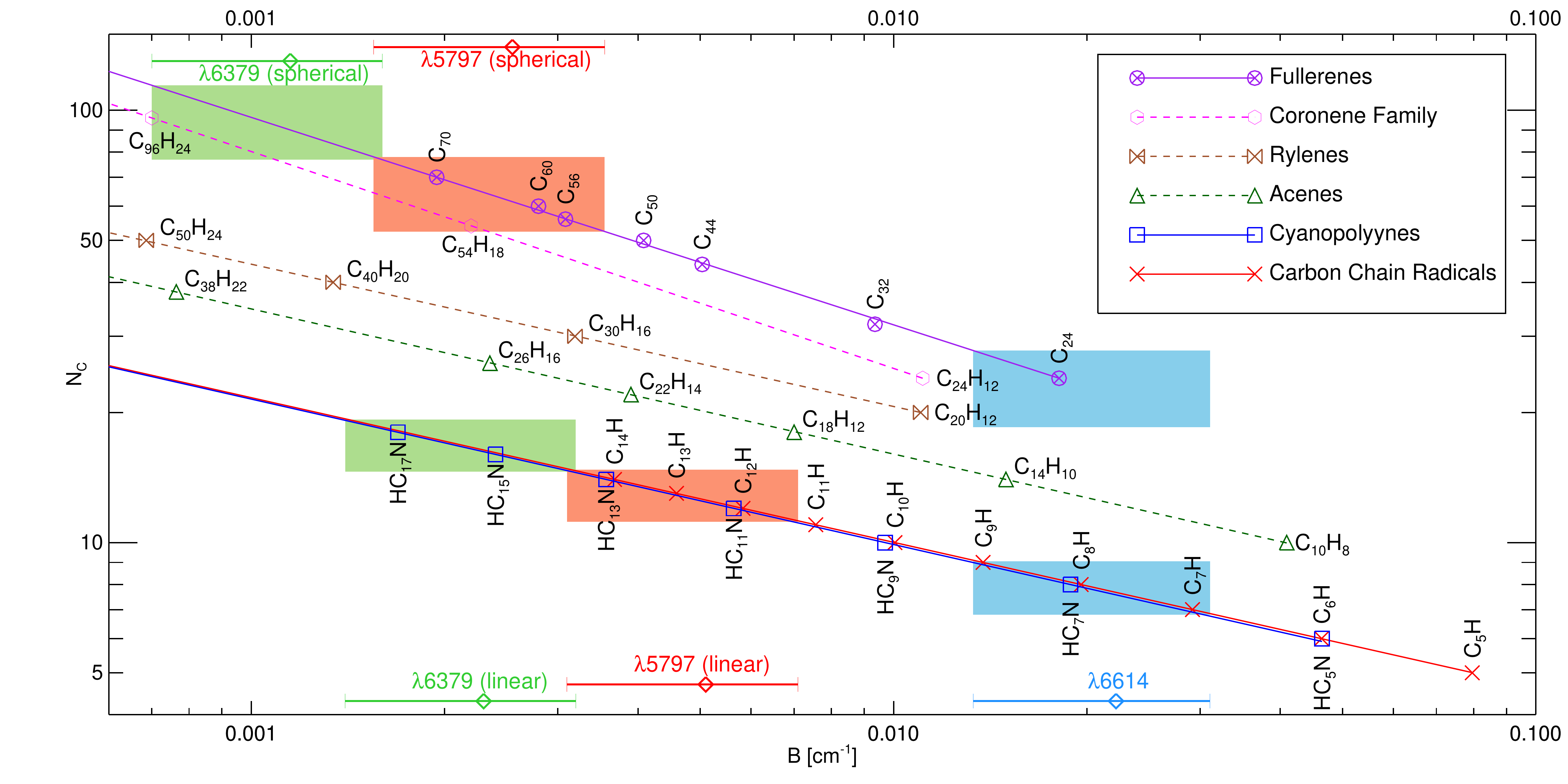}}
\caption{\label{Fig:NC_vs_B}%
Sizes of various molecules (expressed as equivalent number of carbon atoms $N_C$) as a function of their rotational constants $B$ for experimentally measured or theoretically calculated values from the literature (see main text for details). The lines are linear models fitted to these values for individual molecular families and extrapolated to larger sizes. We used the solid lines to determine DIB carrier sizes for linear and spherical species; the dashed lines for PAHs are for illustrative purposes only.
The shaded areas represent our results for the $\lambda\lambda$6379 (green), 5797 (red) and 6614 (blue) DIBs assuming either a spherical geometry (top) or linear species (bottom). The width of these shaded areas indicate our nominal 1-$\sigma$ confidence intervals for the derived $B$ values (also indicated at the top and bottom of the figure, with the nominal value indicated by the $\diamond$ symbols) while the height of these shaded areas represents the corresponding confidence interval for the derived $N_C$ values. We note that the $B$-value for the $\lambda$6614 DIB is independent of the assumed geometry.
%
}
\end{figure*}

The difference in $P$- to $R$-branch peak separation between the DIBs indicates that the three DIB carriers may be molecules of very different sizes -- unless their rotational excitation temperatures are very different. We compared the derived rotational constants to literature values (both experimental and theoretical) for several molecular families (Fig.~\ref{Fig:NC_vs_B}). Because we assumed a linear or spherical top geometry (where the profile can be described by a single rotational constant $B$), we consider such geometries first. 

We start by comparing the measured rotational constants to those of linear species. Rotational constants are available for several acetylenic free radicals (C$_n$H) up to C$_{14}$H \citep{1986A&A...164L...5G, 1988A&A...189L..13P, C7H_Radical, 1997ApJS..113..105M, doi:10.1063/1.477161}. The linear model fitted to these data in Fig.~\ref{Fig:NC_vs_B} has a slope of $-2.99$, close to the value of $-3$ expected for uniform solid rods rotating around their midpoints ($B \propto N_C^{-3}$). We expect this to be a good approximation for linear molecules, and thus we can extrapolate to larger species if needed. Within our uncertainties, the derived $B$ value for the $\lambda$6614 DIB carrier (indicated by the lower blue box in Fig.~\ref{Fig:NC_vs_B}) is compatible with those of C$_7$H, C$_8$H or C$_9$H , which are $B = 29.2 \times 10^{-3}$~cm$^{-1}$, 19.6$ \times 10^{-3}$~cm$^{-1}$ and 13.8$ \times 10^{-3}$~cm$^{-1}$, respectively \citep{C7H_Radical, 1997ApJS..113..105M}. 
For the $\lambda$5797 DIB carrier, the upper limit to the rotational constant in Eq.~(\ref{Eq:B5797_limit_lin}) implies lower limits to the size of the DIB carrier similar to the values for the $\lambda$6614 DIB: the smallest size possible would be C$_7$H or similar sized chains. If instead we use the $B$ value from Eq.~(\ref{Eq:B5797_linear_final}), we find this value is consistent with chains in the C$_{12}$H--C$_{14}$H range (lower red box in Fig.~\ref{Fig:NC_vs_B}). 
Similarly, we find an absolute lower limit to the size of the $\lambda$6379 DIB carrier corresponding to C$_{10}$H ($B=10.0\times10^{-3}$~cm$^{-1}$). The $B$ value from Eq.~(\ref{Eq:B6379_linear_final}) is slightly larger than C$_{14}$H; using our extrapolation yields chains in the range C$_{15}$H -- C$_{19}$H with a nominal closest match to C$_{16}$H. We stress here that the rotational constants of the species we use here are only taken as a proto-typical example to derive an approximate $B$-value; the electronic spectra of C$_6$H, C$_8$H, C$_{10}$H and C$_{12}$H have been recorded, and these do not match any DIBs. 

Performing the same comparison for cyanopolyynes leads to identical size estimates but swapping one C atom for an N atom. The $\lambda$6614 rotational constant is close to the $B$ value reported for HC$_{7}$N ($18.9\times10^{-3}$~cm$^{-1}$, \citealt{Arnau_cyclopolyynes}); the $\lambda$5797 to that of HC$_{11}$N (5.6$\times10^{-3}$~cm$^{-1}$, \citealt{1997ApJS..113..105M}) and $\lambda$6379 to HC$_{15}$N (2.64$\times10^{-3}$~cm$^{-1}$, \citealt{2000ApJS..129..611M}). Using other species \citep[for instance methylpolyynes or methylcyanopolyynes][]{2000ApJS..129..611M} yields very similar results. We have summarised the derived sizes in Table~\ref{table:results}. 

Spherical top species are similarly well described by a single rotational constant $B$, and we thus next considered fullerene species with a cage geometry. This is in line with the recent identification of C$_{60}^+$ as a DIB carrier (see Sect.~\ref{Sec:introduction}), and a further motivation may be that large PAHs may fragment along a chain of smaller fullerenes \citep{2014ApJ...797L..30Z}. 
Our derived rotational constant for the $\lambda$6614 DIB is compatible with the calculated constant of $B = 18.1\times10^{-3}~{\rm cm}^{-1}$ for a C$_{24}$ cage by \citet{Bernstein_2017}. We can also use the rotational constant of C$_{60}$ ($B = 2.8\times10^{-3}~{\rm cm}^{-1}$; see, for example, \citealt{Changala49}), C$_{70}$ ($B = 1.9\times10^{-3}~{\rm cm}^{-1}$; \citealt{NEMES199725}) and those of several other fullerenes \citep[][rotational constants from private communication]{2019MNRAS.485.1137C} for comparison. For uniform spherical shells, $B\propto N_C^{-2}$; this corresponds well to the linear fit in Fig.~\ref{Fig:NC_vs_B} that we used to extrapolate to other size fullerenes. Note though that some fullerenes (for example, C$_{70}$) are not perfect spherical tops but slightly oblate or prolate. 
With these additional data, we find that the uncertainties on the $B$ value of the $\lambda$6614 DIB allow cages in the range 19--27 C atoms. The upper limit to the rotational constant of the $\lambda$5797 DIB carrier (Eq.~\ref{Eq:B5797_limit_spher}) implies a size larger than C$_{32}$ for this DIB. The $B$-value from Eq.~(\ref{Eq:B5797_spherical_final}) is closest to that of C$_{60}$, and the uncertainties allow a range of 52--78 C atoms. The $\lambda$6379 DIB carrier must be even larger. From Eq.~(\ref{Eq:6379_limit_spher}) we find a size at least comparable to C$_{44}$; using the values from Eq.~(\ref{Eq:B6379_spherical_final}) we find an extrapolated size of 90 C atoms, and a nominal range between 77 and 114 C atoms. These derived sizes for each DIB are listed in Table~\ref{table:results}. 

For comparison, we also considered (planar) PAH species, but with the caveat that for such species, the line profile is no longer primarily determined by only one rotational constant $B$, but one also needs to consider the other rotational constants $A$ and $C$ and the quantum number $K$. This complicates the contour interpretation and would require more detailed profile modelling in most cases. Using our measurements and the derived $B$ values to estimate the size of this type of carrier is thus no longer well justified and we, therefore, refrain from doing so. We present the discussion below for illustrative purposes only. 

Members of the coronene family are thought to be some of the most stable PAHs and have been compared to the DIBs as well \citep[see for example][]{Tan:benzo[ghi]perylene}. Figure~\ref{Fig:NC_vs_B} shows the rotational constants for coronene (C$_{24}$H$_{12}$; $B = 11.1\times10^{-3}$~cm$^{-1}$; \citealt{2007CP....332..353M}), circumcoronene (C$_{54}$H$_{18}$; $B = 2.2\times10^{-3}$~cm$^{-1}$; \citealt{2007CP....332..353M}), and N-circumcircumcoronene (C$_{96}$H$_{24}$; $B = 0.7\times10^{-3}$~cm$^{-1}$; \citealt{2005ApJ...632..316H}). The slope of the linear model fitted in Fig.~\ref{Fig:NC_vs_B} is consistent with the $B\propto N_C^{-2}$ relation expected for uniform disks. For a given $B$ value, the size of a coronene-family PAH molecule is only slightly lower than the corresponding sizes of fullerenes.

Another family of PAH molecules that have been proposed as DIB carrier candidates are the polyacenes \citep{2011ApJ...728..154S, 2019A&A...625A..41O}. The smallest two members of this class are naphthalene \citep[C$_{10}$H$_8$; $B = 40.9\times10^{-3}~{\rm cm}^{-1}$;][]{2007CP....332..353M} and anthracene \citep[C$_{14}$H$_{10}$; $B=15.0\times10^{-3}~{\rm cm}^{-1}$;][]{2006A&A...456..161M}. 

Due to the elongated geometry of polyacenes, their rotational constants $B$ do not scale as uniform disks, but as chains, such that $B\propto N_C^{-3}$. For such a species, the profile of a DIB would primarily be determined by the rotational constant $A$ and quantum number $K$, rather than by $B$ and $J$ as in our approximation \citep[see][for details]{2019A&A...625A..41O}. The acene sizes corresponding to a given $B$-value are larger than the carbon chains but smaller than the members of the coronene family. Rylenes have slightly larger sizes than the acenes; Fig.~\ref{Fig:NC_vs_B} shows rotational constants for perylene, terrylene, quaterrylene, and pentarylene \citep{2007CP....332..353M}.

\begin{table}
\caption{DIB carrier sizes expressed as the number of equivalent carbon atoms inferred for different geometries. The ranges are derived from the 1$\sigma$ confidence intervals on $B$. }
\label{size_approx} 
\centering     
\begin{tabular}{llccc}
\hline \hline
 & &  $\lambda$6614 & $\lambda$5797 & $\lambda$6379 \\
\hline
\multicolumn{5}{l}{Linear}\\
& C$_n$H radicals &  7--9          & 12--14         & 15--19 \\
& cyanopolyynes   &  7--9          & 12--14         & 15--19 \\
\multicolumn{5}{l}{Cages}\\
& fullerenes      &  19--27        & 52--78         & 77--114 \\
\hline
\end{tabular}
\end{table}

\section{Discussion}
\label{sect:discuss}

We have found that the $\lambda\lambda$6614, 5797, and 6379 DIBs show a systematic increase in peak separations in the same sightlines (Fig.~\ref{Fig:PR_separation_allDIBs}). This is not simply a broadening of the DIBs but a systematic shift in the absorption peak wavelengths, which move away from each other for each of the three DIBs in a correlated way. All three carriers must therefore respond in similar ways to changes in their environment, but with a different magnitude. This observation supports the interpretation of DIB line profiles as rotational contours, as changes in the rotational excitation temperature would naturally produce such shifts. It is difficult to explain this effect using proposed alternative (non-rotational contour) explanations of the profile shapes, such as isotopic substitution \citep{1996MNRAS.282.1372W} or blends of multiple unrelated DIBs \citep[][]{2015ApJ...801....6B}.

If the peak separations are due to changes in rotational temperature, we expect a relationship with environmental parameters that cause rotational excitation. Depending on the species, the rotational temperature can depend on the kinetic temperature, gas density, and UV radiative pumping followed by internal conversion. These parameters (particularly for the latter process) are not directly observable. The molecular hydrogen fraction $f({\rm H}_2)$ provides a rough indication of the UV radiation field, but only indirectly, and we find no significant correlation between the measured peak separations and $f({\rm H}_2)$. \citet{2009A&A...498..785K} reported that the width and substructure of $\lambda$6196 correlates with the rotational excitation temperature of C$_2$. The C$_2$ excitation temperature is generally different from the kinetic temperature due to radiative pumping \citep{1982ApJ...258..533V, 1989ApJ...340..273V}. For the sightlines in our sample for which C$_2$ temperatures are available, there may be a weak correlation between the C$_2$ excitation temperature and the peak separations; however, due to the large uncertainties, this trend is not significant at this point. 

As we noted in Sect.~\ref{Sect:DIBselection}, all three DIBs studied here show extended tails to the red (ETRs) toward Herschel~36 \citep{2013ApJ...773...41D,2013ApJ...773...42O} but these ETRs are much more pronounced for the $\lambda\lambda$5797 and 6614 DIBs. Assuming linear molecules, these authors concluded that the carriers of $\lambda\lambda$5797 and 6614 are polar molecules, while that of the $\lambda$6379 must be a non-polar species. Because polar molecules have much larger dipole moments, they should cool much faster by spontaneous emission and therefore reach a lower steady-state excitation temperature than their non-polar counterparts. For the diffuse ISM, the rotational excitation temperatures of highly polar molecules are close to the temperature of the cosmic microwave background \citep{cn_cmb}. This should then lead to much less variation in the peak separations for the polar ($\lambda\lambda$5797 and 6614) versus the non-polar ($\lambda6379$) DIB carriers. However, we find that all three DIBs exhibit pronounced variations.

For planar species such as PAHs, additional processes affect the rotational distribution, for example, the rocket effect and cooling cascades that slightly favour the $\Delta J=+1$ transitions. \citet{1997A&A...324..661R} considered these processes for the $\lambda$5797 carrier. They found that under diffuse ISM conditions, $T_{\rm rot}$ is 18--35~K for carriers 15--30 carbon atoms in size, close to the values we inferred for this DIB. Similar effects should apply to the rotational excitation of fullerenes. This then further supports that the carriers of the three DIBs we study could be due to fairly small PAH-like or fullerene-like species.

The variations in peak separation are purely observational and do not depend on our derived $B$ values. We assumed that the rotational contours could be described by $B$ and $J$ alone (that is, by linear or spherical top species) and that the rotational population has a Boltzmann distribution with a single excitation temperature $T_{\rm rot}$. As mentioned above, for other species, including polyacenes, the profile is determined by $K$ and the rotational constant $A$ \citep{2019A&A...625A..41O}. However, our assumptions are valid for linear species and fullerenes, and planar species fall between those two families (see Fig.~\ref{Fig:NC_vs_B}). 

With those caveats, our size estimates are summarised in Table~\ref{size_approx}. The rotational constants of the different DIB carriers scale with their $P$- to $R$-branch separation and depend on their assumed geometry. We set their absolute values by scaling to the $B$ value for the $\lambda$6614 DIB carrier,  which was determined independently of the rotational temperature because it had three substructure peaks. Our value of $B=(22.2\pm8.9)\times 10^{-3}$~cm$^{-1}$ is compatible with the value of $B=(16.4\pm3.1)\times 10^{-3}$~cm$^{-1}$ derived by \citet{cami2004rotational} for $\lambda6614$ even though we are using lower resolution observations and different sightlines. The derived value has a large uncertainty, and there is some bias: a larger value of $B$ would be easier to measure; hence it is more likely that the true value of $B$ is smaller than our derived value. If our measured $B$ value for $\lambda$6614 DIB is incorrect, all shaded areas in Fig.~\ref{Fig:NC_vs_B} would shift by the same amounts (to the left if the true value is smaller) and thus imply larger carrier sizes than listed in Table~\ref{size_approx}. Our size estimates are smaller than some previous estimates \citep{Ehrenfreund_rotation,Kerr96}, which often assumed a much higher rotational excitation temperature, but still generally point to fairly large molecular species. Small PAHs exposed to UV radiation photochemically dehydrogenate and fragment and are thus not expected to survive in the ISM. The size below which this process dominates is uncertain but is likely to be in the range of 25--50 C atoms, with some authors viewing 35 C atoms as a reasonable value \citep{2003ApJ...584..316L, 2005pcim.book.....T,Tielens:PAHreview,2013A&A...552A..15M}. Our derived sizes for the $\lambda\lambda$6614 and 5797 carriers are close to (or smaller than) this limit. 
%

\section{Conclusions}

The EDIBLES survey allows us to select a large number of single cloud lines of sight for which clear variations can be seen in the band profiles of the $\lambda\lambda$6614, 5797 and 6379 DIBs. The variations are found to be directly correlated with environmental conditions, specifically changes in the rotational excitation temperature. These changes offer an additional tool to constrain the sizes of possible carrier molecules. 

We independently determined the rotational constant $B$ and rotational temperature $T_{\rm rot}$ of the $\lambda$6614 carrier, assuming that the carriers are linear or spherical molecules. For the other two DIBs, we estimated rotational constants for linear and spherical geometries by assuming that all three DIBs have the same minimum temperature. Carrier sizes were estimated by comparison with literature data for plausible molecular families. The range in excitation temperatures is in good agreement with theoretical calculations for the $\lambda$5797 DIB. However, our carrier size estimates are smaller than many previous determinations. 

The EDIBLES data allow one to look into the line profile variations of DIBs and to link this to differing environmental conditions that, in turn, allow one to derive structural information. In the present study, for obvious reasons, we have made the explicit choice to interpret the rotational contours as originating from linear or spherical species. This puts clear constraints on the size of possible DIB carriers. Similar work, starting from other molecular geometries, will come with the challenge that the unresolved band contours may not provide sufficient information for clear interpretations. In this case, higher resolution data will be necessary. The present work helps in addressing the right lines of sight.  

\begin{acknowledgements}
We thank the anonymous referee for providing us with insightful comments that allowed us to greatly improve this paper. HM, JC, AF and HF acknowledge support from an NSERC Discovery Grant. JC, AF and HF were also supported through a SERB Accelerator Award from Western University. 
This work is based on observations collected at the European Organisation for Astronomical Research in the Southern Hemisphere under ESO programme  194.C-0833.
This research has made use of NASA's Astrophysics Data System Bibliographic Services. This research made use of Astropy,\footnote{http://www.astropy.org} a community-developed core Python package for Astronomy \citep{astropy:2013, astropy:2018}. 
\end{acknowledgements}

\bibliographystyle{aa}
\bibliography{ref}

\clearpage 

\begin{appendix}

\section{Rotational contour formalism}
\label{Sect:App:formalism}

\subsection{Basic assumptions and equations}
We consider a linear or spherical top molecule in some (lower) electronic and vibrational state. Under interstellar conditions, this will most often be the ground state, but that is not required. Within this lower electronic and vibrational state, there are numerous rotational states with energies $E(J)$ determined (to first order) by the rotational constant $B$ and the rotational quantum number $J$: 
\begin{equation}
    E(J) = BJ(J+1).
\end{equation}
In the following, we express both $B$ and $E$ in wavenumbers (cm$^{-1}$). 

From this lower electronic and vibrational state, transitions occur to some higher (excited) electronic and/or vibrational state. The rotational energy levels within the upper state are given by
\begin{equation}
    E(\jp) = \nu_0 + \bp\jp(\jp+1),
\end{equation}
where $\nu_0$ is the energy of the electronic/vibrational transition. 

The selection rules of molecular spectroscopy allow transitions between upper and lower rotational levels when $\Delta J=\pm1$, and in some cases also $\Delta J=0$. This leads to three possible sets of transitions: $P$-branch lines for which $\jp = J-1$; $Q$-branch lines for which $\jp = J$; and $R$-branch lines for which $\jp = J+1$. The frequencies $\nu$ of these transitions (again in cm$^{-1}$) are then given by the difference between the upper state energy and the lower state energy: 
\begin{eqnarray}
\nu_{P} & = & \nu_0 - J(\bp+B) + J^2(\bp-B)\\
   & = & \nu_0 - J(2B+\Delta B) + J^2\Delta B\label{Eq:nu_P}\\
\nu_{Q} & = & \nu_0 + J(J+1)\Delta B\\
\label{Eq:nu_Q}
\nu_{R} & = & \nu_0 + 2\bp + J(3\bp-B) + J^2(\bp-B)\\
   & = & \nu_0 + 2(B+\Delta B) + J(2B+3\Delta B) + J^2\Delta B
\label{Eq:nu_R}
\end{eqnarray}
where $\Delta B = \bp - B$ and we have expressed the frequencies in terms of the lower state $J$ levels.\\ 

The frequency differences between the $P$, $R$ and $Q$ branch lines originating from the same lower state $J$ are then:
\begin{eqnarray}
\nu_R - \nu_Q & = & 2(J+1)(B + \Delta B)\label{Eq:nuR_nuQ_J}\\
\nu_Q - \nu_P & = & 2J(B + \Delta B)\label{Eq:nuQ_nuP_J}\\
\nu_R - \nu_P & = & 2(2J+1)(B + \Delta B)
\label{Eq:nuR_nuP_J}
\end{eqnarray}

The strength of each of the transitions is determined by the combination of the intrinsic oscillator strength for each line and the population distribution of the lower state rotational levels. We assume that the rotational population follows a Boltzmann distribution characterised by the rotational temperature $T_{\rm rot}$: 
\begin{equation}
    \frac{n_J}{N} = \frac{g_J}{P(T_{\rm rot})}e^{-hcE_J/kT_{\rm rot}},
\label{Eq:Boltzmann}
\end{equation}
where $g_J$ is the statistical weight of rotational level $J$ and $P(T_{\rm rot})$ is the partition function at temperature $T_{\rm rot}$. The statistical weight is given by 
\begin{equation}
    g_J = (2J+1)^a,
\label{Eq:g_J}
\end{equation}
where $a=1$ for linear species and $a=2$ for a spherical geometry. From Eq.~(\ref{Eq:Boltzmann}), we find that the highest population occurs for rotational level $J_{\rm max}$ given by
\begin{equation}
    J_{\rm max} = \sqrt{\frac{akT_{\rm rot}}{2hcB}} - \frac{1}{2}.
\label{Eq:Jmax}
\end{equation}

We interpret each of the three DIB profiles as a rotational contour, assuming that each substructure peak corresponds to the $P$-, $Q$-, or $R$-branch transition originating from $J_{\rm max}$. This is equivalent to assuming that the oscillator strengths for the individual $J$ lines are either constant or do not vary strongly between adjacent values of $J$. The frequencies of the substructure peaks can then be expressed in terms of the rotational temperature by substituting Eq.~(\ref{Eq:Jmax}) into Eqs.~\ref{Eq:nu_P}--\ref{Eq:nu_R}: 
\begin{eqnarray}
\nu_P & = & \nu_0 + B + \frac{3\Delta B}{4} + \frac{\Delta B}{2B}\frac{akT_{\rm rot}}{hc} - (B + \Delta B)\sqrt{\frac{2akT_{\rm rot}}{hcB}}
\label{Eq:nu_P_T}\\
\nu_Q & = & \nu_0 - \frac{\Delta B}{4} + \frac{\Delta B}{2B}\frac{akT_{\rm rot}}{hc}
\label{Eq:nu_Q_T}\\
\nu_R & = & \nu_0 + B + \frac{3\Delta B}{4} + \frac{\Delta B}{2B}\frac{akT_{\rm rot}}{hc} + (B + \Delta B)\sqrt{\frac{2akT_{\rm rot}}{hcB}}.
\label{Eq:nu_R_T}
\end{eqnarray}

\noindent Taking the difference between these expressions provides the peak separations (in cm$^{-1}$): 
\begin{eqnarray}
\nu_R - \nu_Q & = & (B + \Delta B)\left(\sqrt{\frac{2akT_{\rm rot}}{hcB}} + 1\right)\label{Eq:nuR_nuQ}\\
\nu_Q - \nu_P & = & (B + \Delta B)\left(\sqrt{\frac{2akT_{\rm rot}}{hcB}} - 1\right)\label{Eq:nuQ_nuP}\\
\nu_R - \nu_P & = & 2(B + \Delta B)\sqrt{\frac{2akT_{\rm rot}}{hcB}}.
\label{Eq:nuR_nuP}
\end{eqnarray}

\subsection{Interpreting DIB profiles}

As the $P$-branch consists of the lower-energy transitions, it appears on the red side of the observed DIB profiles. For each of the three DIBs, we measure the separation between the $P$- and $R$-branch peaks. Equation~(\ref{Eq:nuR_nuP}) shows that this separation depends on $B$, $\Delta B$ and $T_{\rm rot}$. Assuming that $\Delta B/B \ll 1$, we rearrange and simplify Eq.~(\ref{Eq:nuR_nuP}):
\begin{equation}
    B\cdot T_{\rm rot} \approx \frac{hc(\nu_R-\nu_P)^2}{8ak}.
\label{Eq:BT_app}
\end{equation}

Because the rotational constant is specific to each DIB carrier, but the same in different lines of sight, any significant variation in the peak separation $(\nu_R - \nu_P)$ can only be due to changes in the rotational temperature $T_{\rm rot}$. This provides the relative changes in rotational temperature between lines of sight; determining the absolute value requires knowledge of the rotational constant $B$. 

The $\lambda$6614 DIB has a three-peak profile, so we can also measure its $Q$ branch. Equation~(\ref{Eq:nu_Q_T}) relates the (central) $Q$-branch peak position to the rotational excitation temperature. If the difference in rotational temperatures between sightlines are small or zero, we expect the central peak to appear at the same wavelength in all sightlines. Measurable changes in the peak position would only occur for large rotational temperature variations: a typical value of $\Delta B / B\sim 1\%$ leads to a peak shift of no more than $\sim$0.3 cm$^{-1}$ for rotational temperatures changing from 20~K to 100~K. For the range in rotational temperatures determined in Sect.~\ref{Sec:Analysis}, we find a peak shift that is at least an order of magnitude smaller. After shifting the spectra to the interstellar rest frame (using the velocities in Table~\ref{target_data}), the central peak positions of the $\lambda$6614 DIB scatter around a mean value, but we did not find a systematic effect (for example, we did not find a correlation with the peak separations). The scatter is thus most likely the consequence of small uncertainties in the wavelength calibration or differences in velocity distribution between the DIBs and atomic species in the same interstellar cloud. 

We use the central peak position as a reference point. Measuring the peak separations of the $P$ and $R$-branch peaks relative to the $Q$-branch peak in principle determines the rotational constant $B$ from Eqs.~(\ref{Eq:nuR_nuQ_J}) and (\ref{Eq:nuQ_nuP_J}): 
\begin{equation}
    (\nu_R - \nu_Q) - (\nu_Q-\nu_P) = 2B(1+\frac{\Delta B}{B}) \approx 2B.
\end{equation}
Once $B$ is determined, we can use Eq.~(\ref{Eq:BT_app}) to infer $T_{\rm rot}$.

\section{Interstellar Na lines}
\label{Sect:Na_Doublet_plots}

\begin{figure*}
\centering
\resizebox{0.8\hsize}{!}{
\includegraphics[width=\columnwidth]{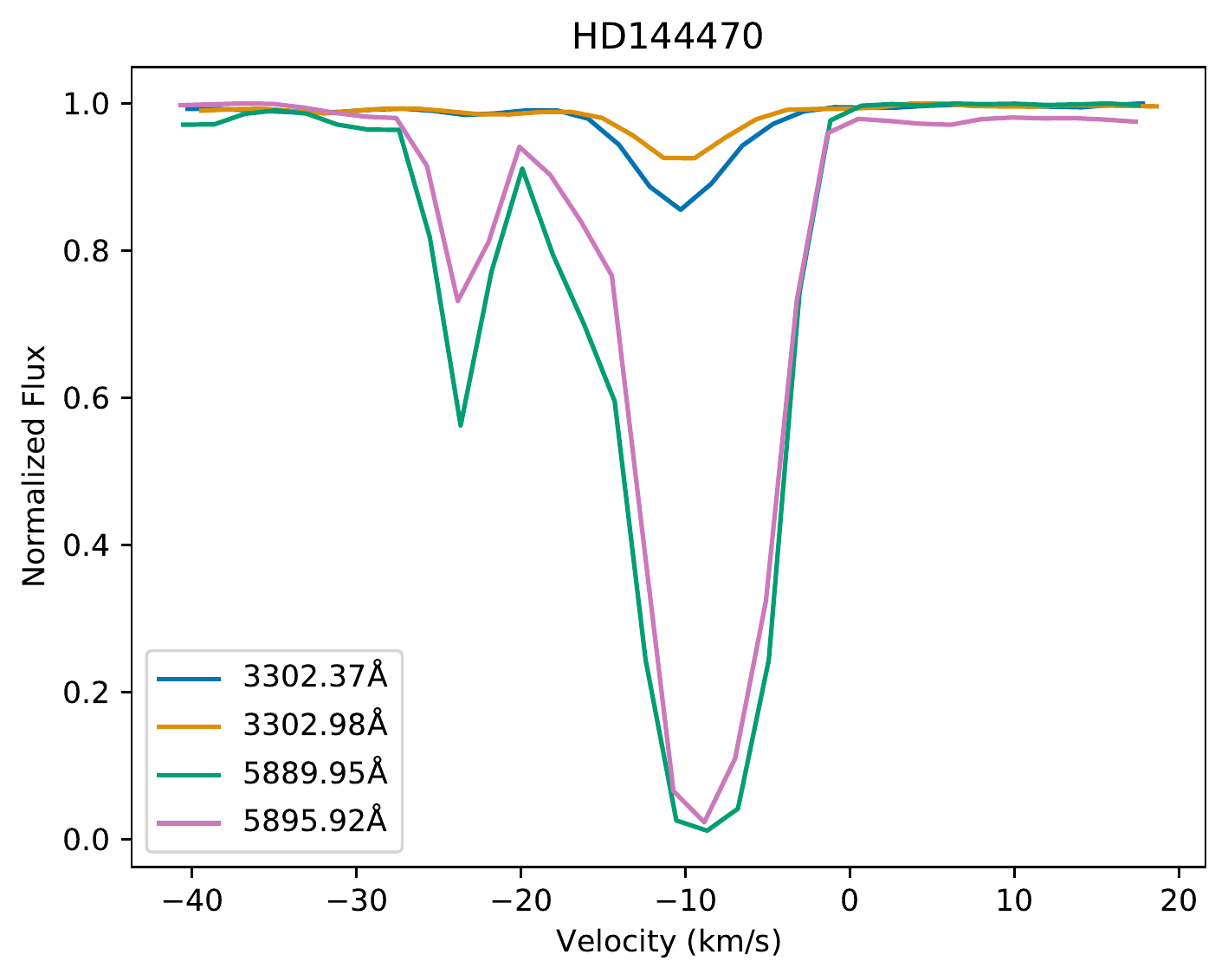}
\includegraphics[width=\columnwidth]{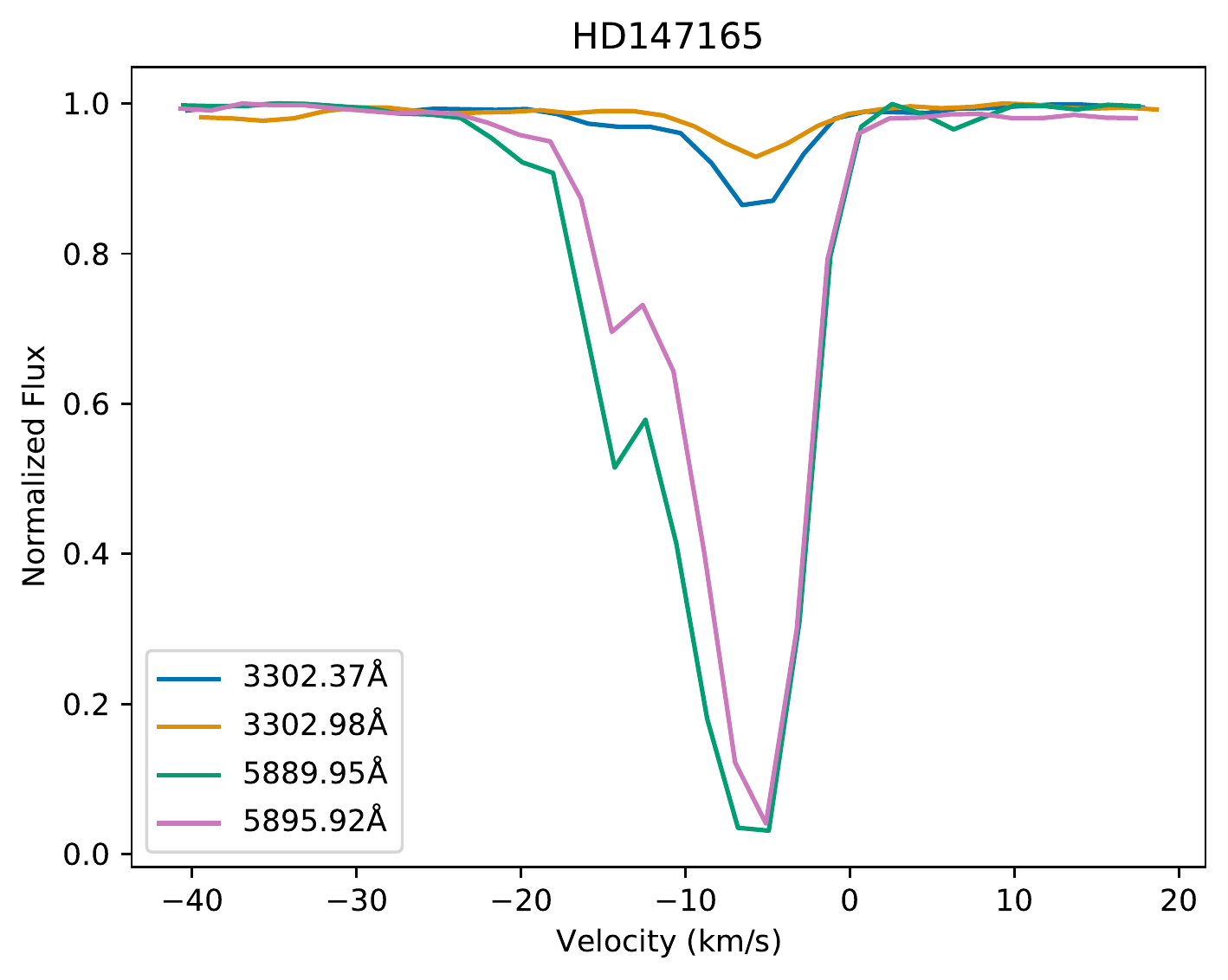}}\\
\resizebox{0.8\hsize}{!}{
\includegraphics[width=\columnwidth]{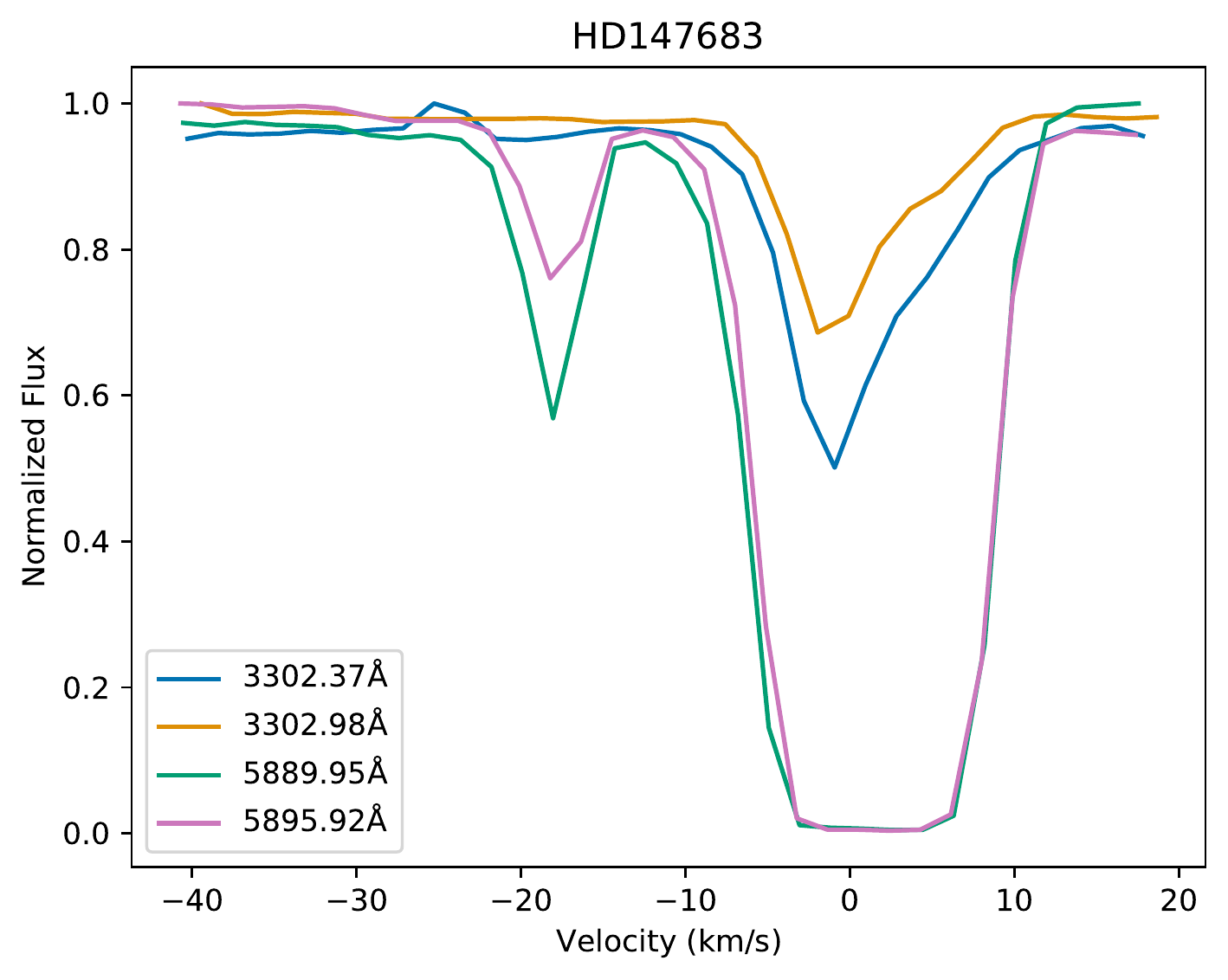}
\includegraphics[width=\columnwidth]{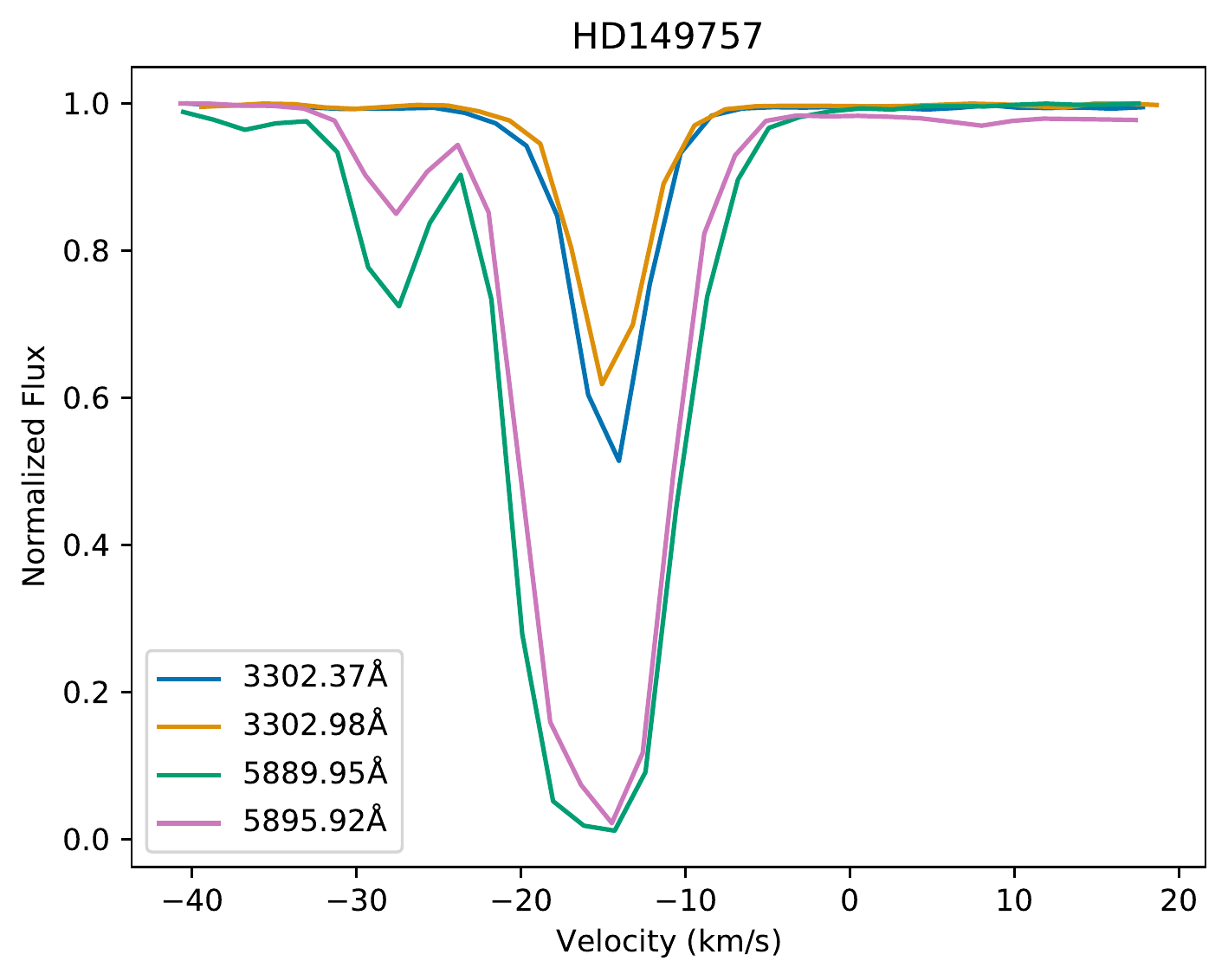}}\\
\resizebox{0.8\hsize}{!}{
\includegraphics[width=\columnwidth]{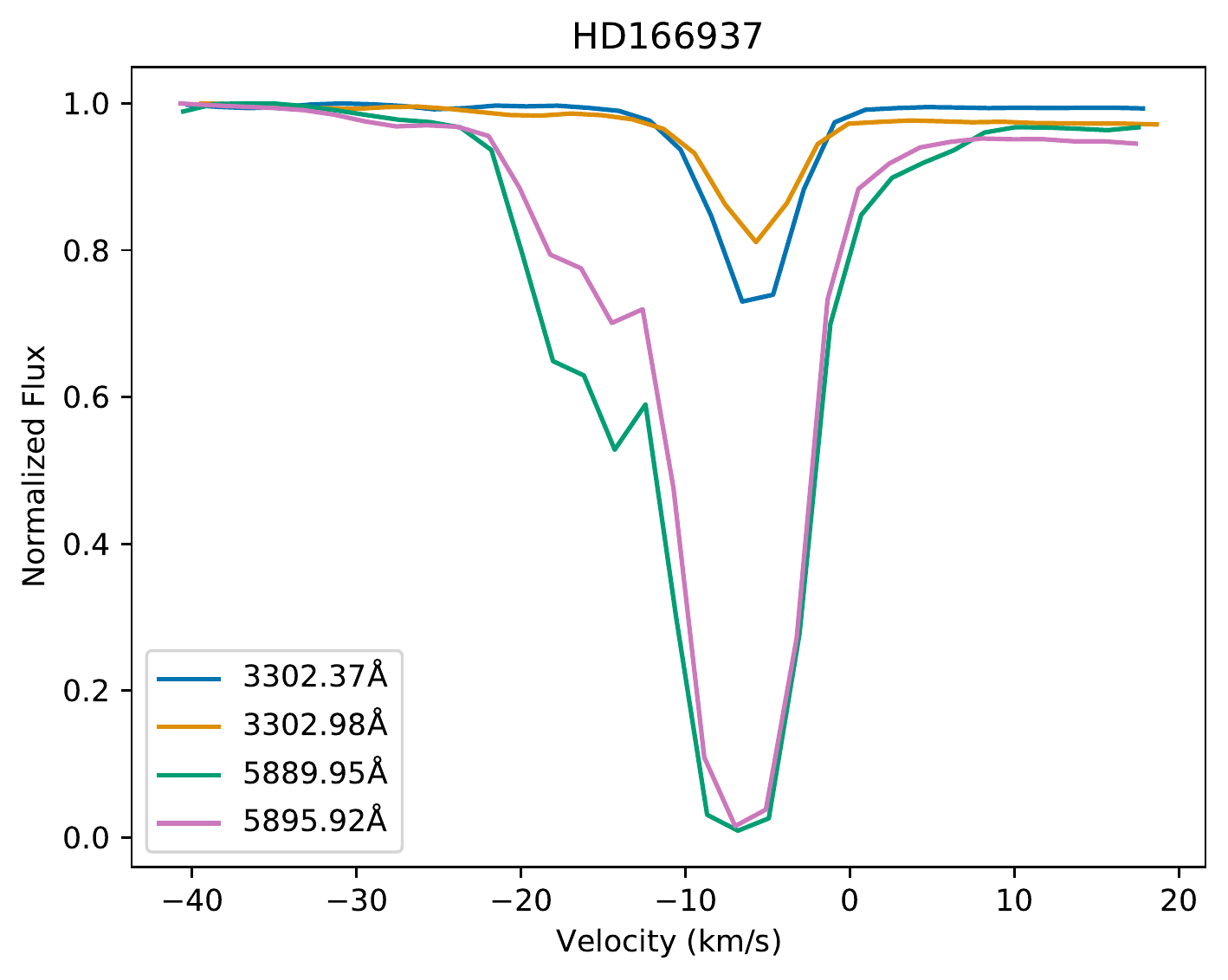}
\includegraphics[width=\columnwidth]{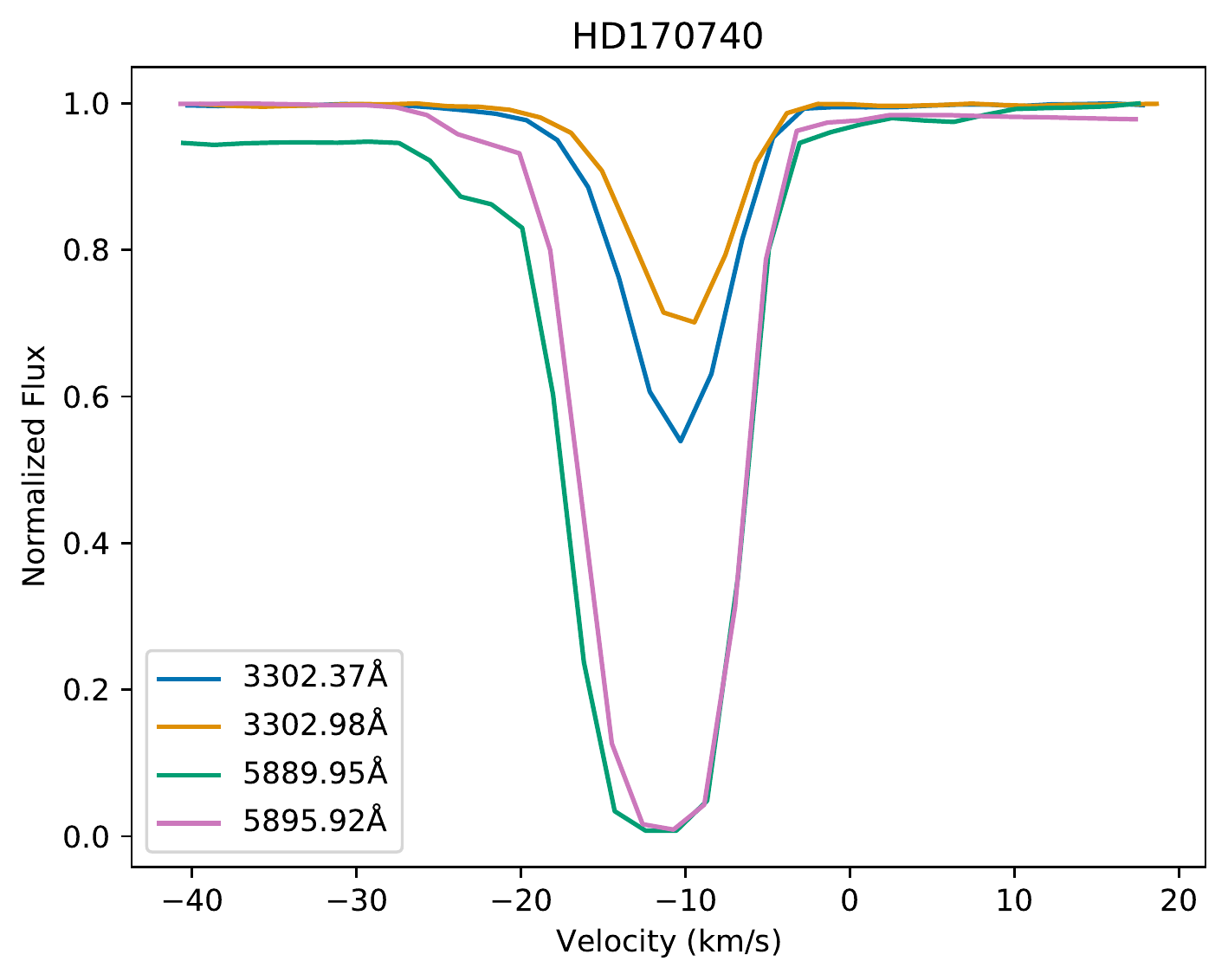}}
\caption{\label{Fig:Na1}EDIBLES spectra of the interstellar \ion{Na}{i} D lines and \ion{Na}{i} UV doublet at 3302~\AA\ for each of our targets, shown in velocity space and normalised. The lines at 3302~\AA\ show only one dominant component, our criterion for single cloud sightlines. The D lines are always saturated, producing broader profiles and making weaker components visible as well.}
\end{figure*}

\begin{figure*}
\centering
\resizebox{0.8\hsize}{!}{
\includegraphics[width=\columnwidth]{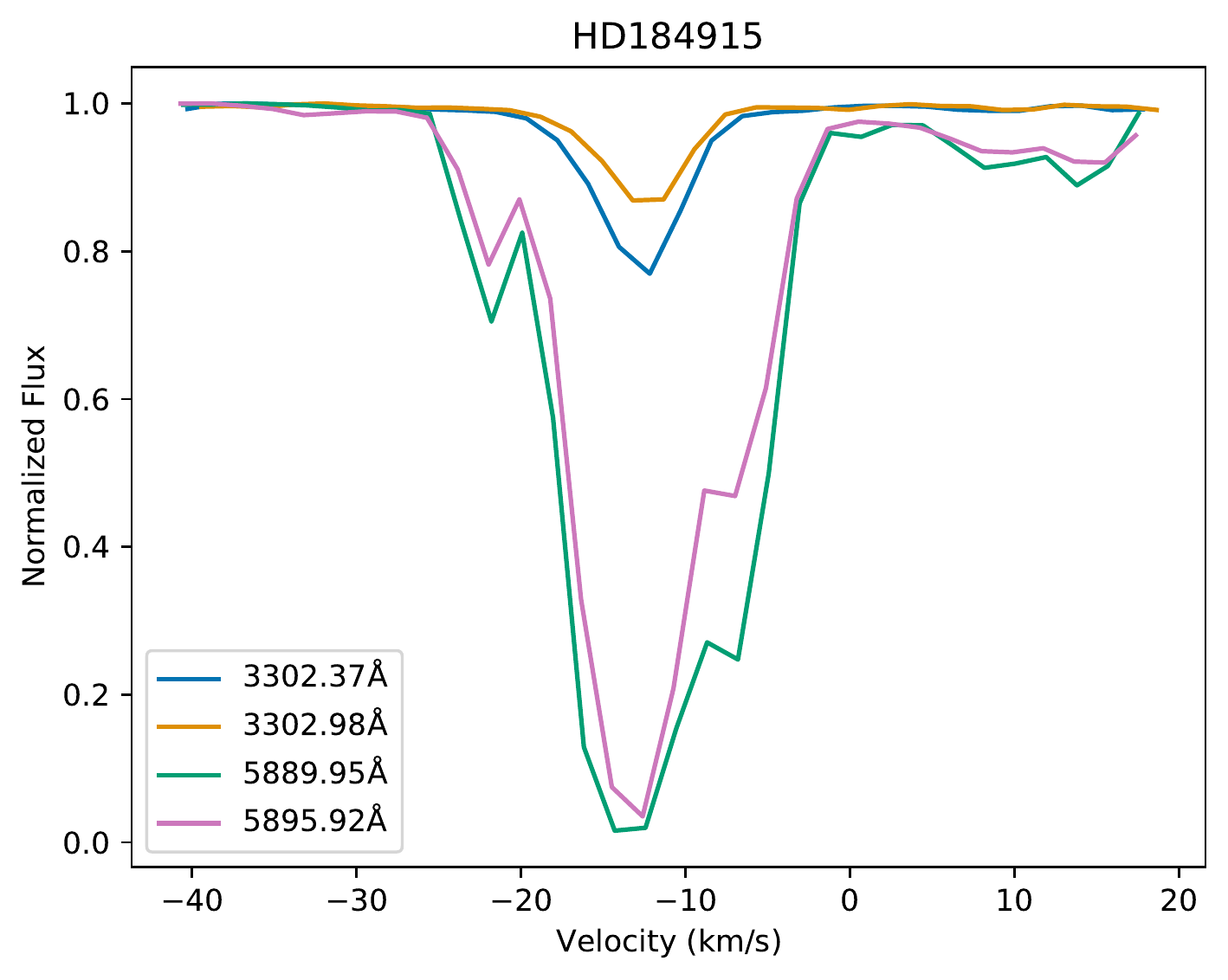}
\includegraphics[width=\columnwidth]{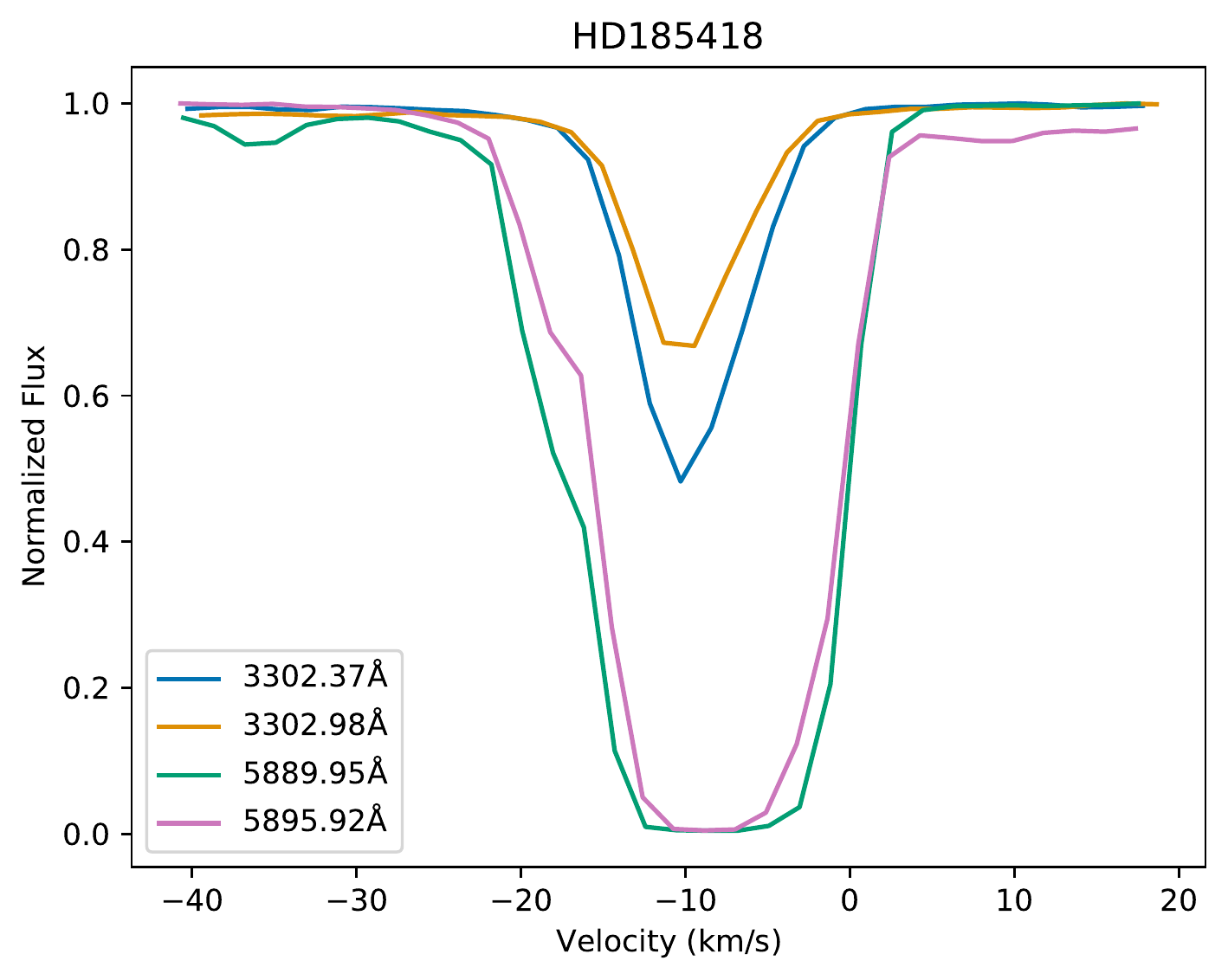}}\\
\resizebox{0.8\hsize}{!}{
\includegraphics[width=\columnwidth]{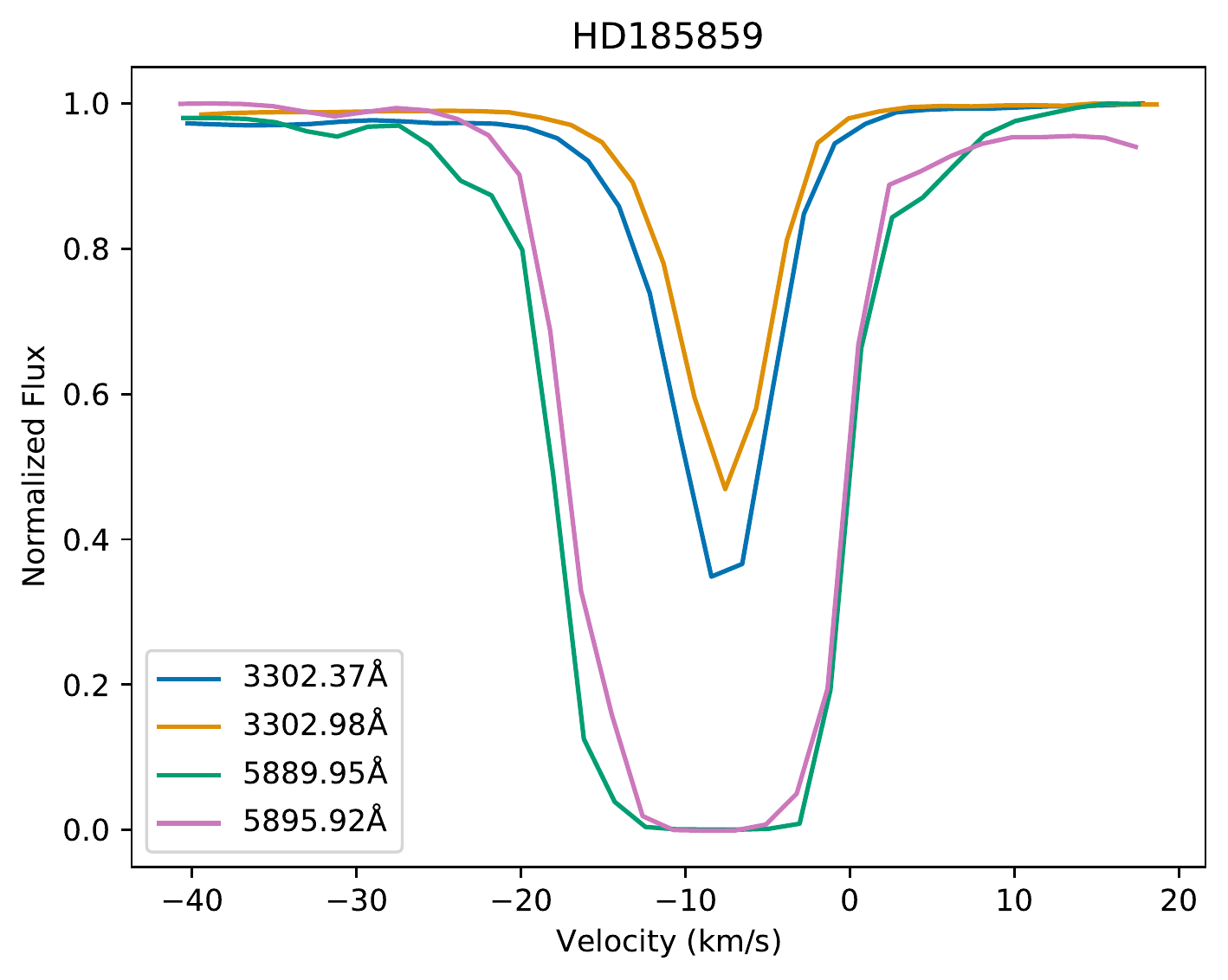}
\includegraphics[width=\columnwidth]{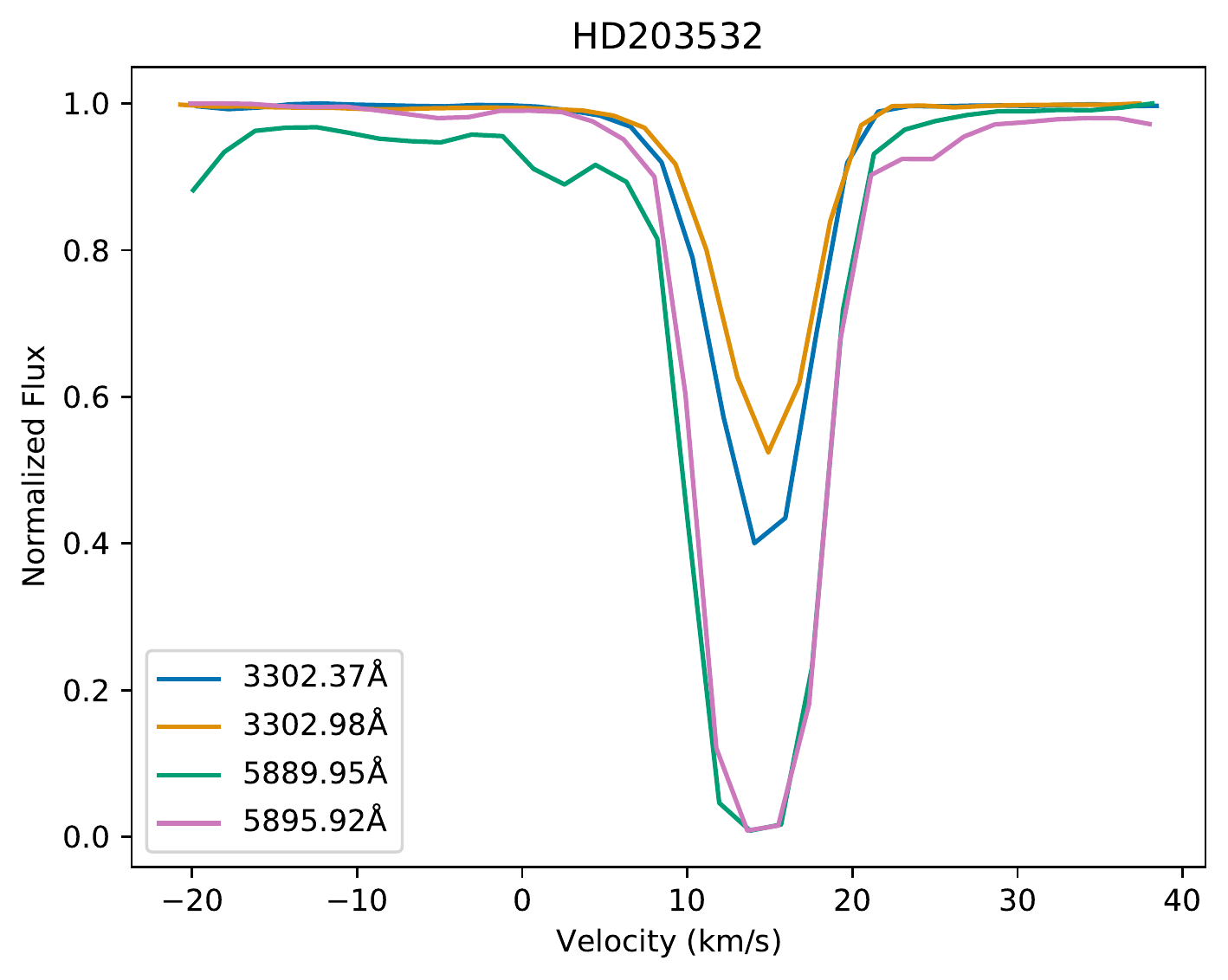}}\\
\resizebox{0.8\hsize}{!}{
\includegraphics[width=\columnwidth]{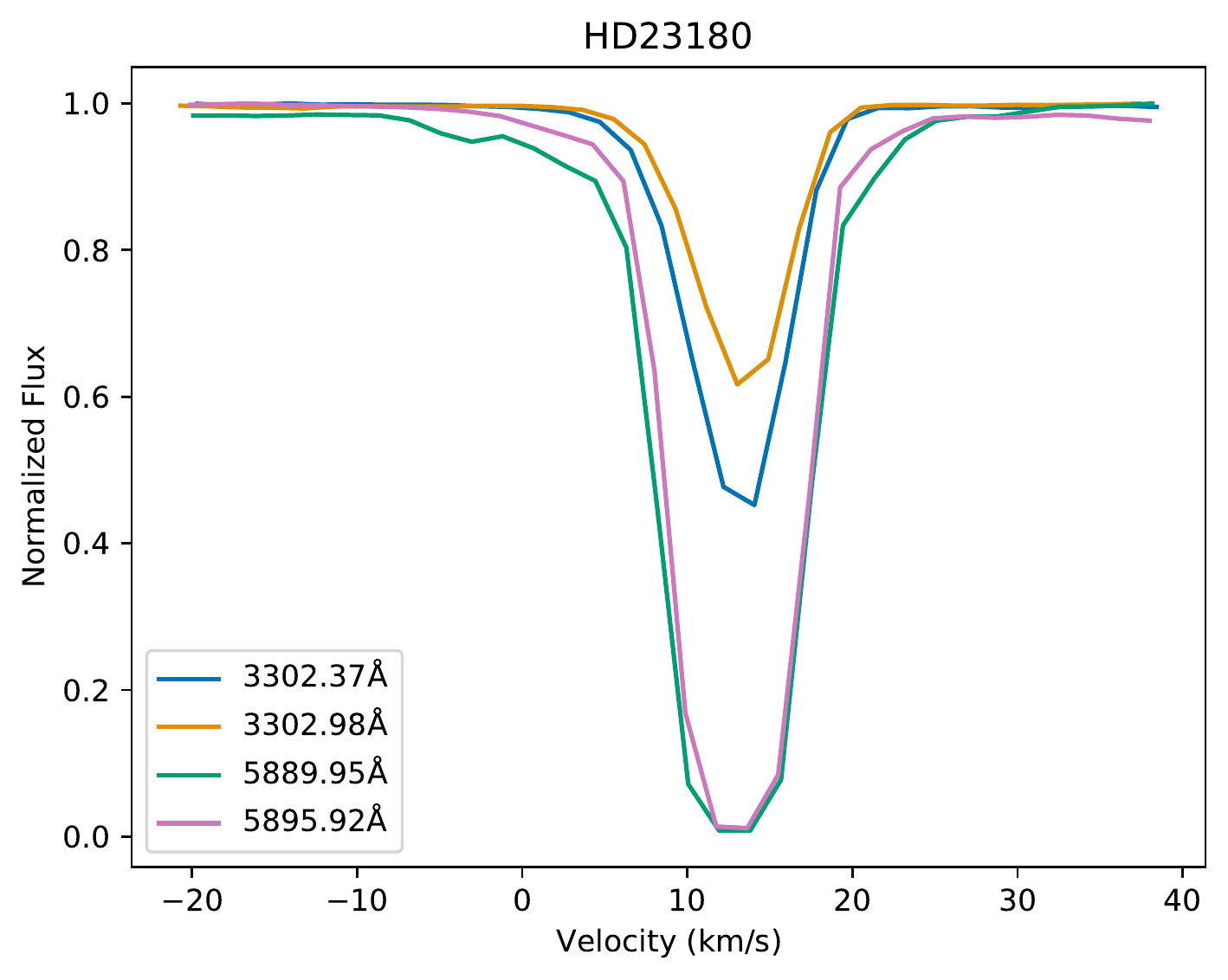}
\includegraphics[width=\columnwidth]{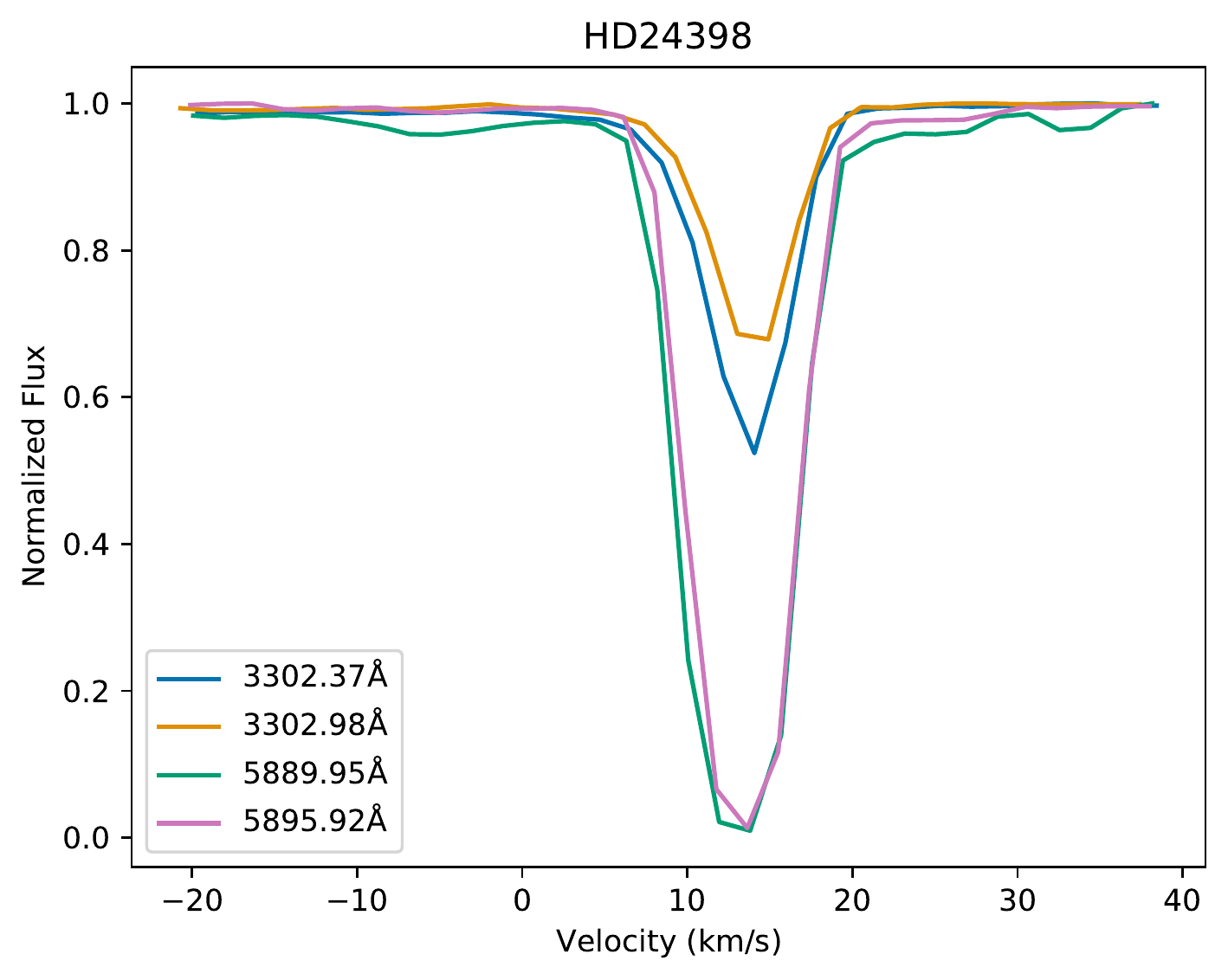}}
\caption{\label{Fig:Na2} continued.}
\end{figure*}

Figures~\ref{Fig:Na1} and \ref{Fig:Na2} show the EDIBLES observations of the \ion{Na}{i} D lines and \ion{Na}{i} UV doublet at 3302~\AA\ for each of our targets.

\section{Measurements for the $\lambda$6614 DIB.}
\label{Sect:App_6614}
The measurements of the $\lambda$6614 DIB are listed in Table~\ref{table:6614_manual} and are shown in Figs.~\ref{6614_fit_results_1}--\ref{6614_fit_results_2}.  

\begin{table*}
\caption{Manually determined peak position measurements for the $\lambda$6614 DIB.}   
\label{table:6614_manual} 
\centering     
\begin{tabular}{l c c c }   
\hline\hline 
Target    &    Peak 1 & Peak 2 & Peak 3 \\
          &  [\AA] & [\AA] & [\AA]\\
HD23180 & 6613.52 $\pm$ 0.03 & 6613.83 $\pm$ 0.01 & 6614.07 $\pm$ 0.02 \\
HD24398 & 6613.540 $\pm$ 0.004 & 6613.837 $\pm$ 0.008 & 6614.13 $\pm$ 0.02 \\
 HD144470 &  6613.03 $\pm$0.02 &  6613.33 $\pm$0.01 &  6613.64 $\pm$0.02 \\
 HD147165 &  6613.11 $\pm$0.01 &  6613.42 $\pm$0.01 &  6613.71 $\pm$0.02 \\
 HD147683 &  6613.24 $\pm$0.02 &  6613.53 $\pm$0.01 &  6613.81 $\pm$0.03 \\
 HD149757 &  6612.91 $\pm$0.01 &  6613.23 $\pm$0.02 &  6613.50 $\pm$0.02 \\
 HD166937 &  6613.07 $\pm$0.03 &  6613.41 $\pm$0.01 &  6613.71 $\pm$0.02 \\
 HD170740 &  6613.000 $\pm$0.003 &  6613.292 $\pm$0.003 &  6613.58 $\pm$0.01 \\
 HD184915 &  6612.95 $\pm$0.01 &  6613.27 $\pm$0.02 &  6613.55 $\pm$0.02 \\
 HD185418 &  6613.06 $\pm$0.01 &  6613.36 $\pm$0.02 &  6613.62 $\pm$0.01 \\
 HD185859 &  6613.07 $\pm$0.02 &  6613.39 $\pm$0.02 &  6613.63 $\pm$0.02 \\
 HD203532 &  6613.58 $\pm$0.03 &  6613.86 $\pm$0.01 &  6614.14 $\pm$0.01 \\
\hline   
\end{tabular}

\end{table*}
\begin{figure*}
\centering
\resizebox{0.8\hsize}{!}{
\includegraphics[width=\columnwidth]{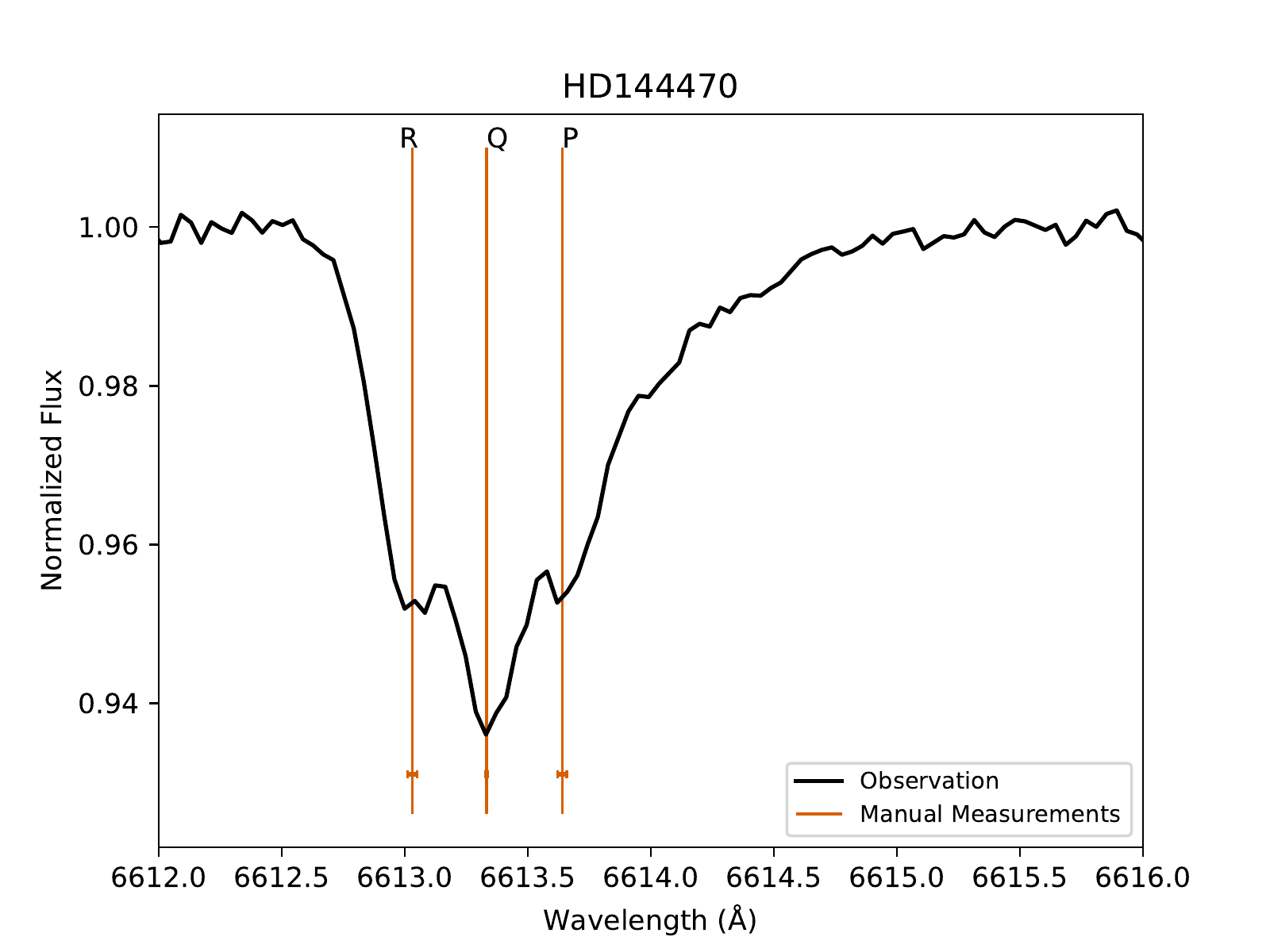}
\includegraphics[width=\columnwidth]{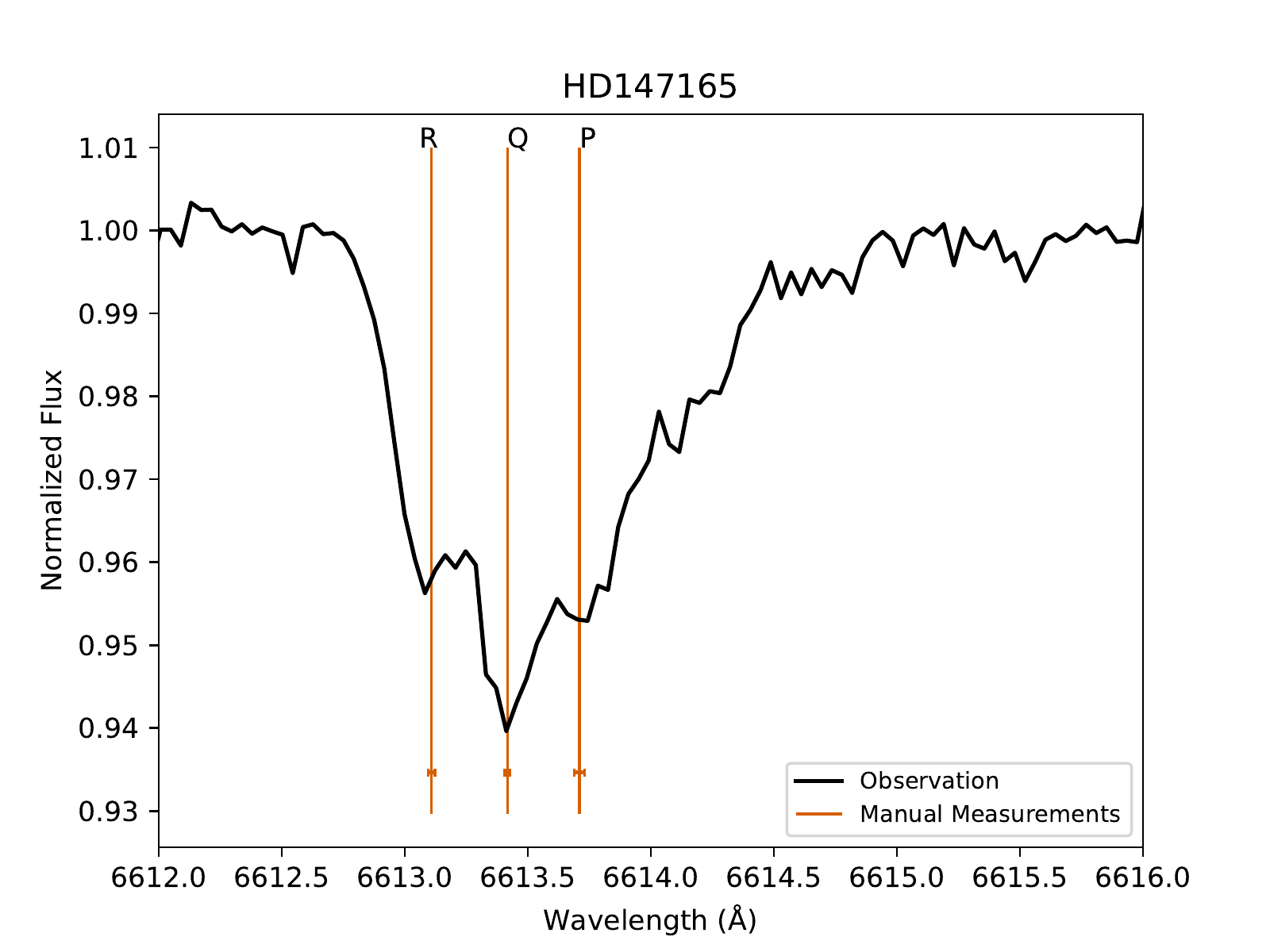}}
\resizebox{0.8\hsize}{!}{
\includegraphics[width=\columnwidth]{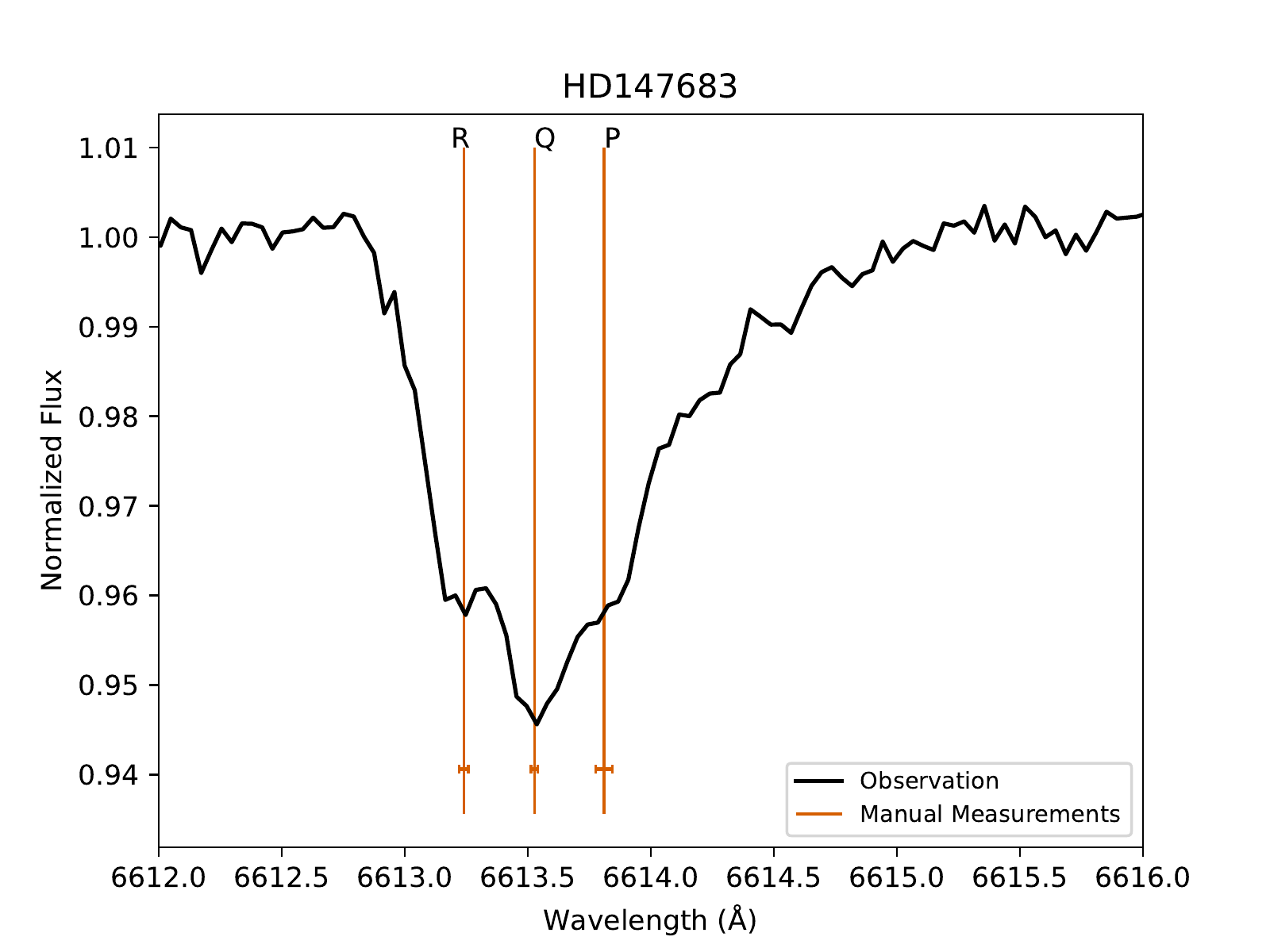}
\includegraphics[width=\columnwidth]{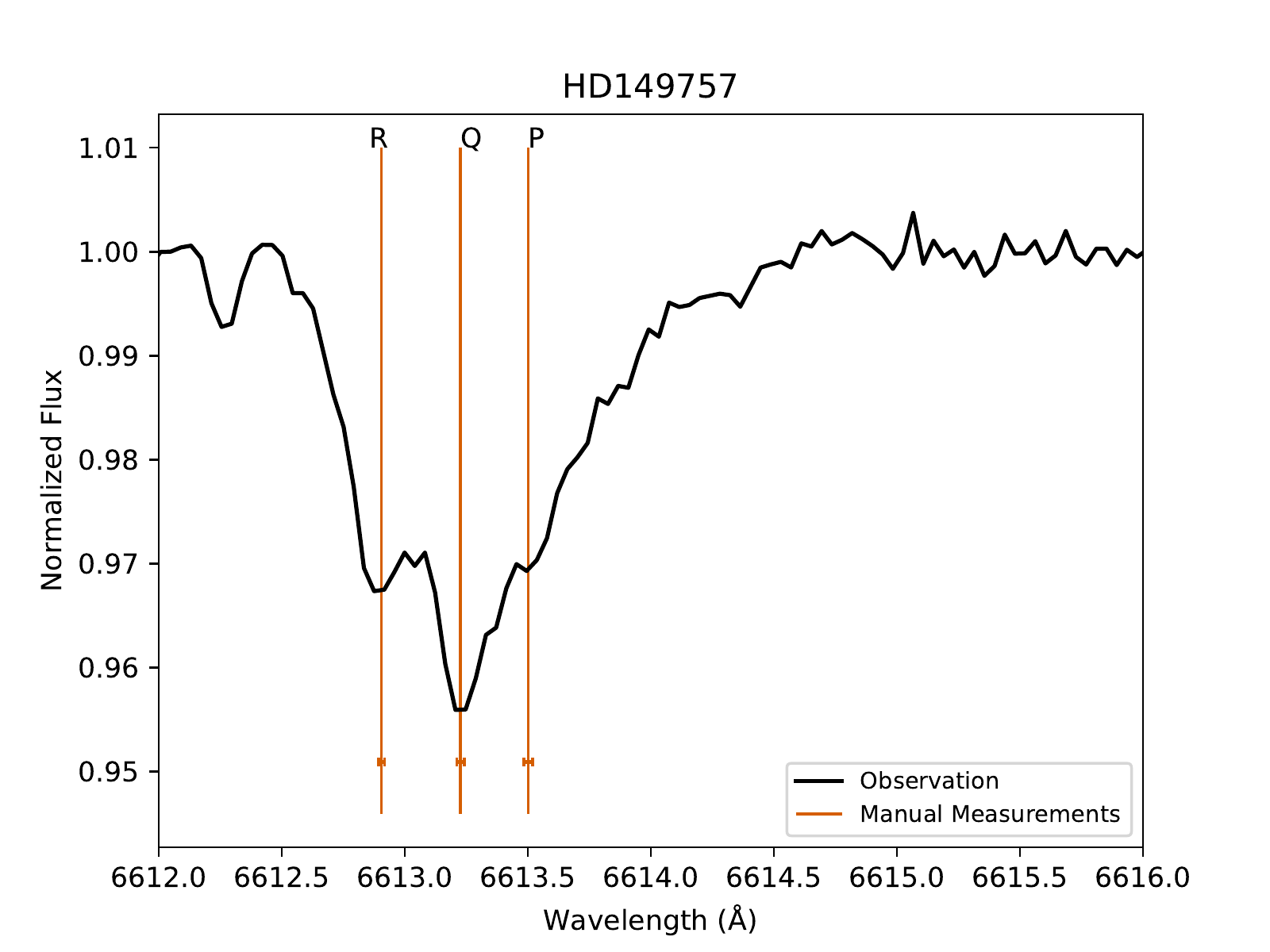}}
\resizebox{0.8\hsize}{!}{
\includegraphics[width=\columnwidth]{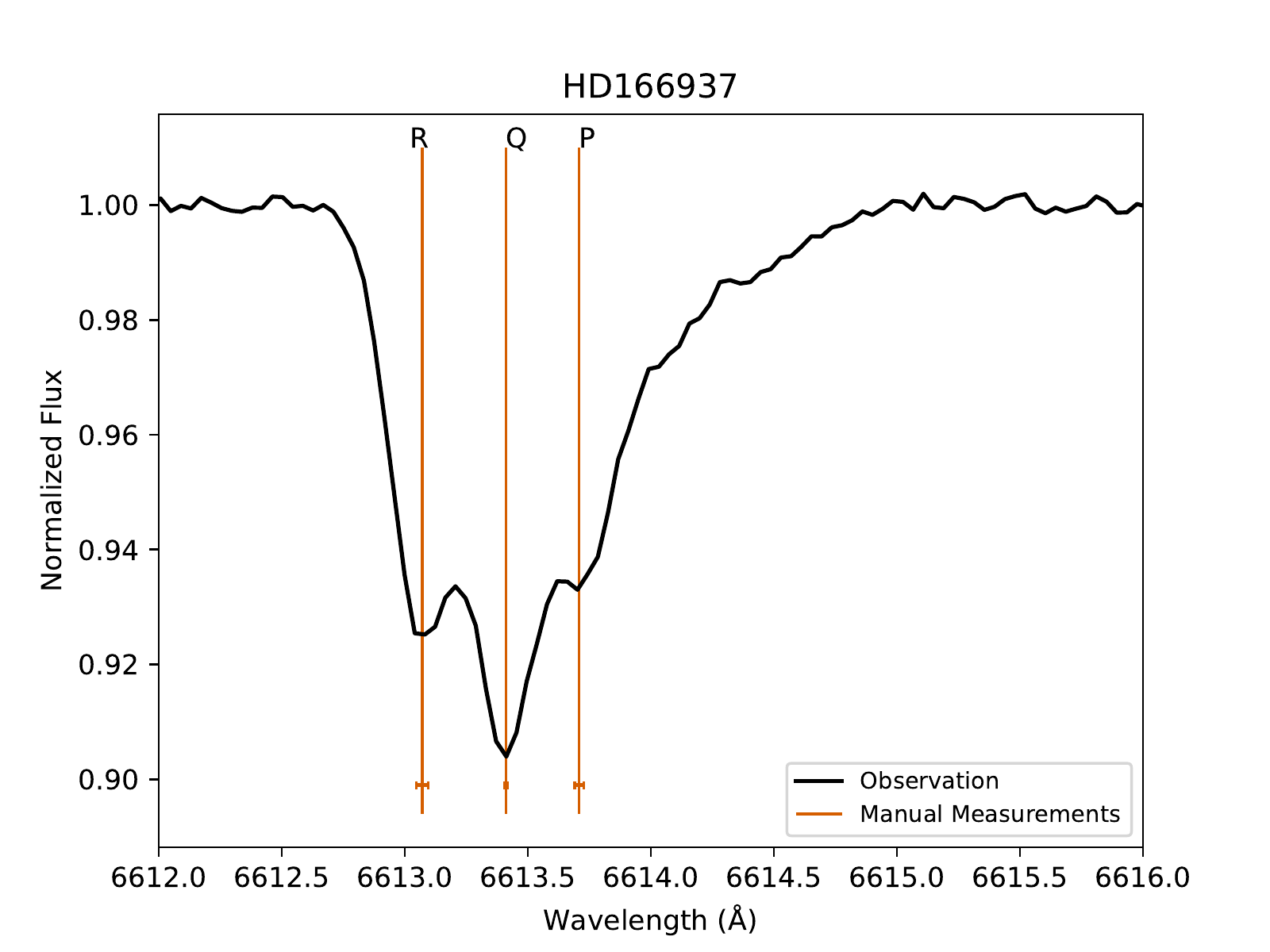}
\includegraphics[width=\columnwidth]{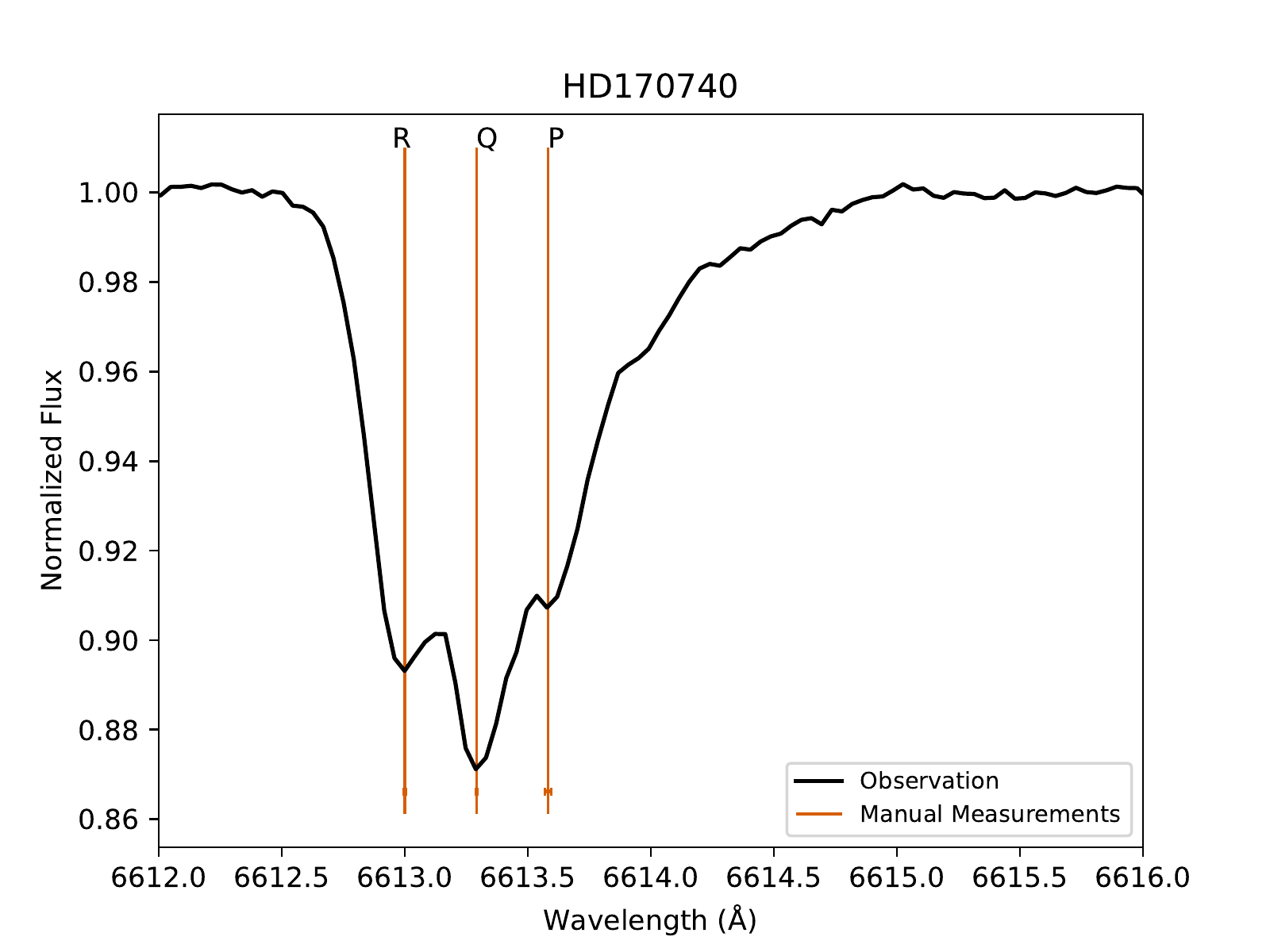}}
\caption{The observed $\lambda$6614 DIB profile  (black line) and the location of the peaks (orange lines) with indicated error for each of our targets.}
\label{6614_fit_results_1}
\end{figure*}
\begin{figure*}
\centering

\resizebox{0.8\hsize}{!}{
\includegraphics[width=\columnwidth]{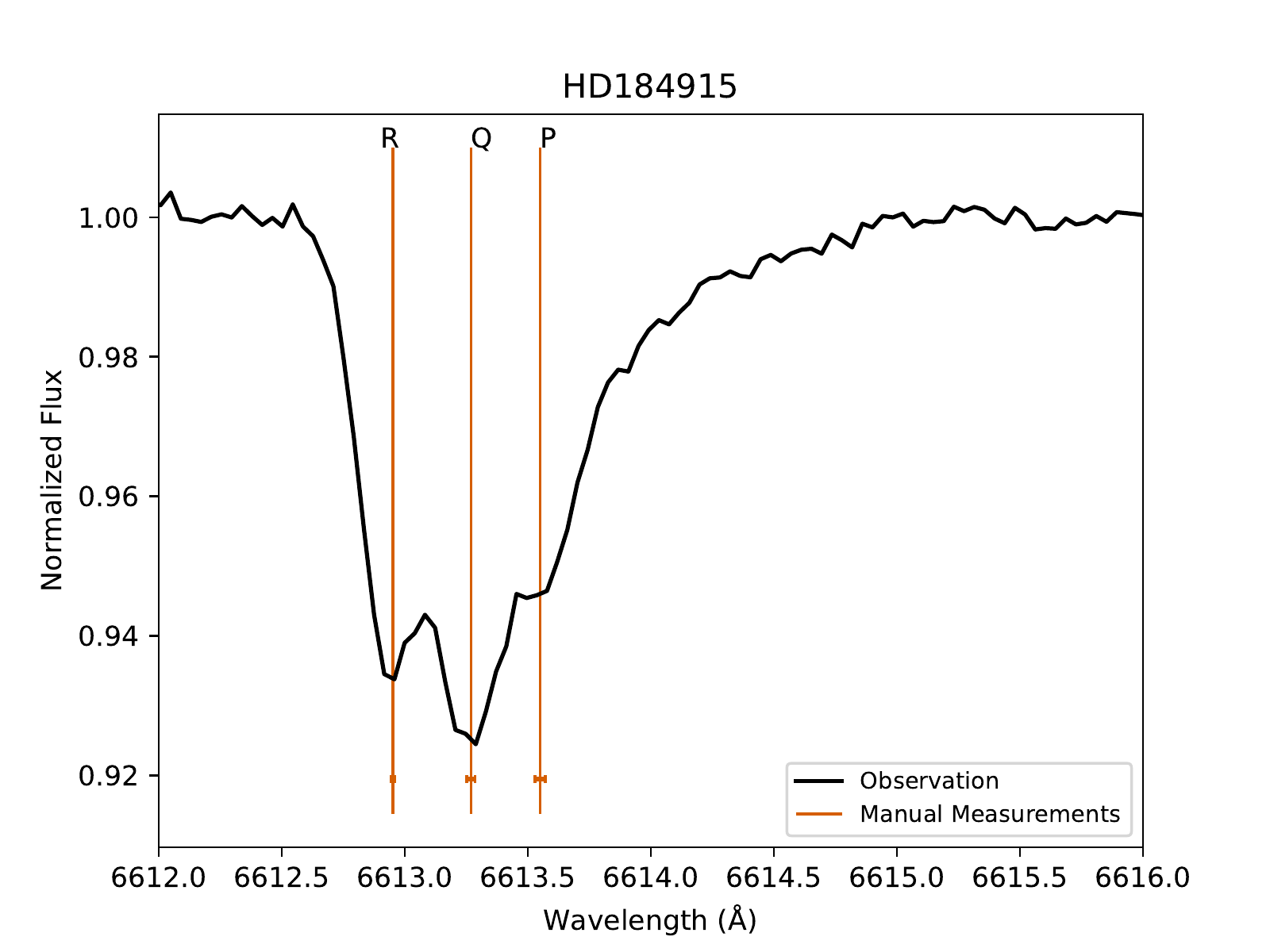}
\includegraphics[width=\columnwidth]{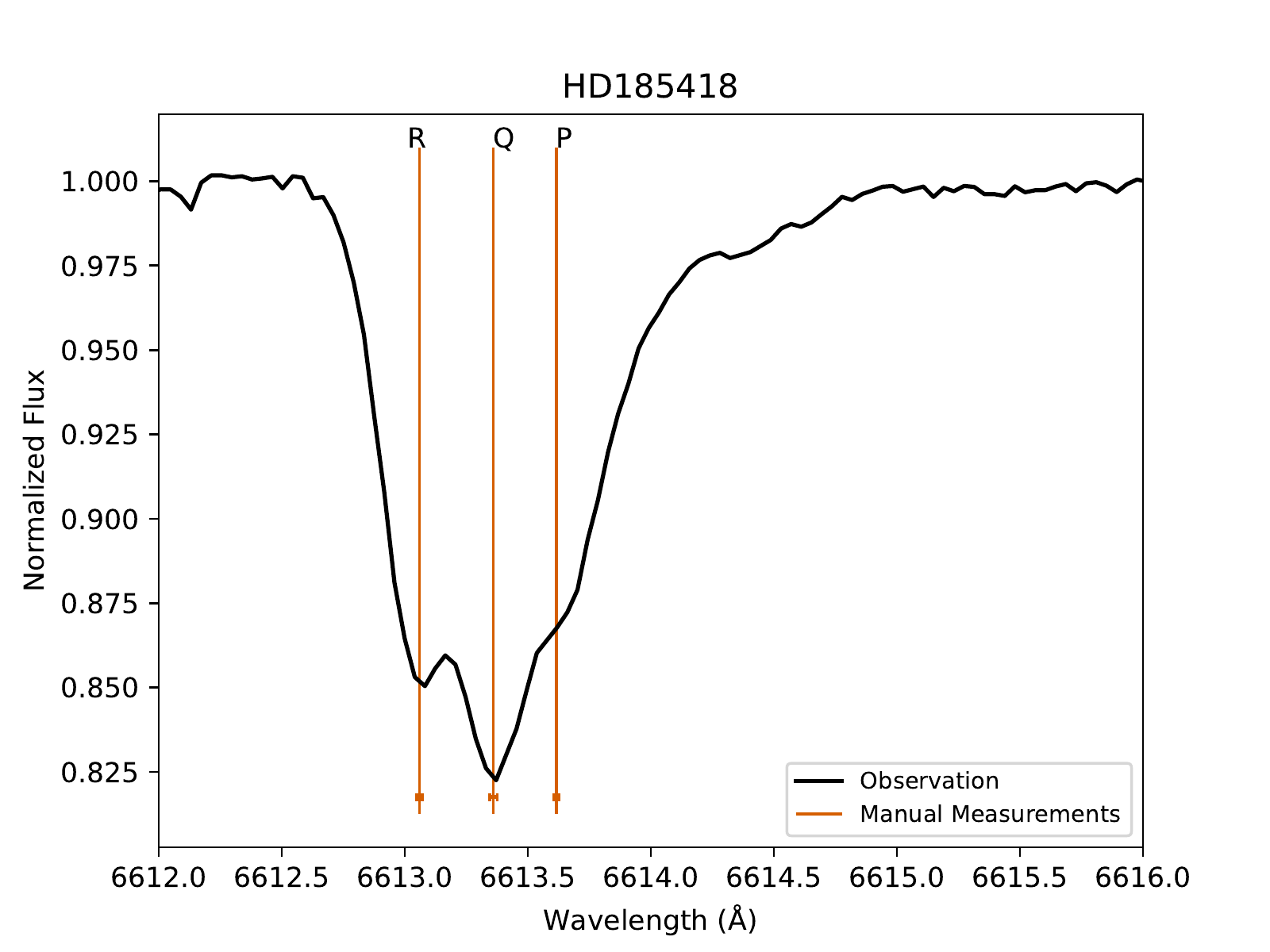}}
\resizebox{0.8\hsize}{!}{
\includegraphics[width=\columnwidth]{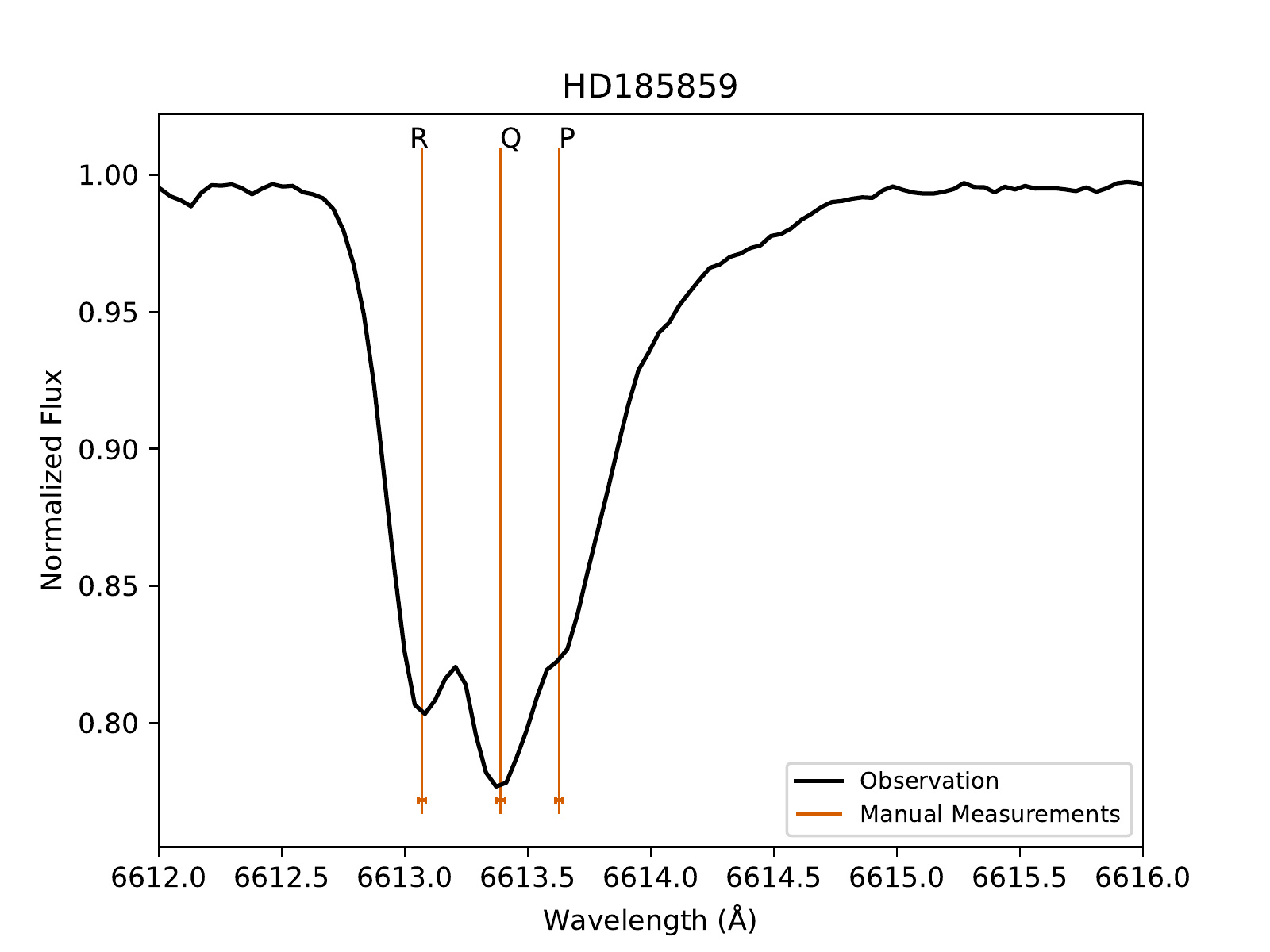}
\includegraphics[width=\columnwidth]{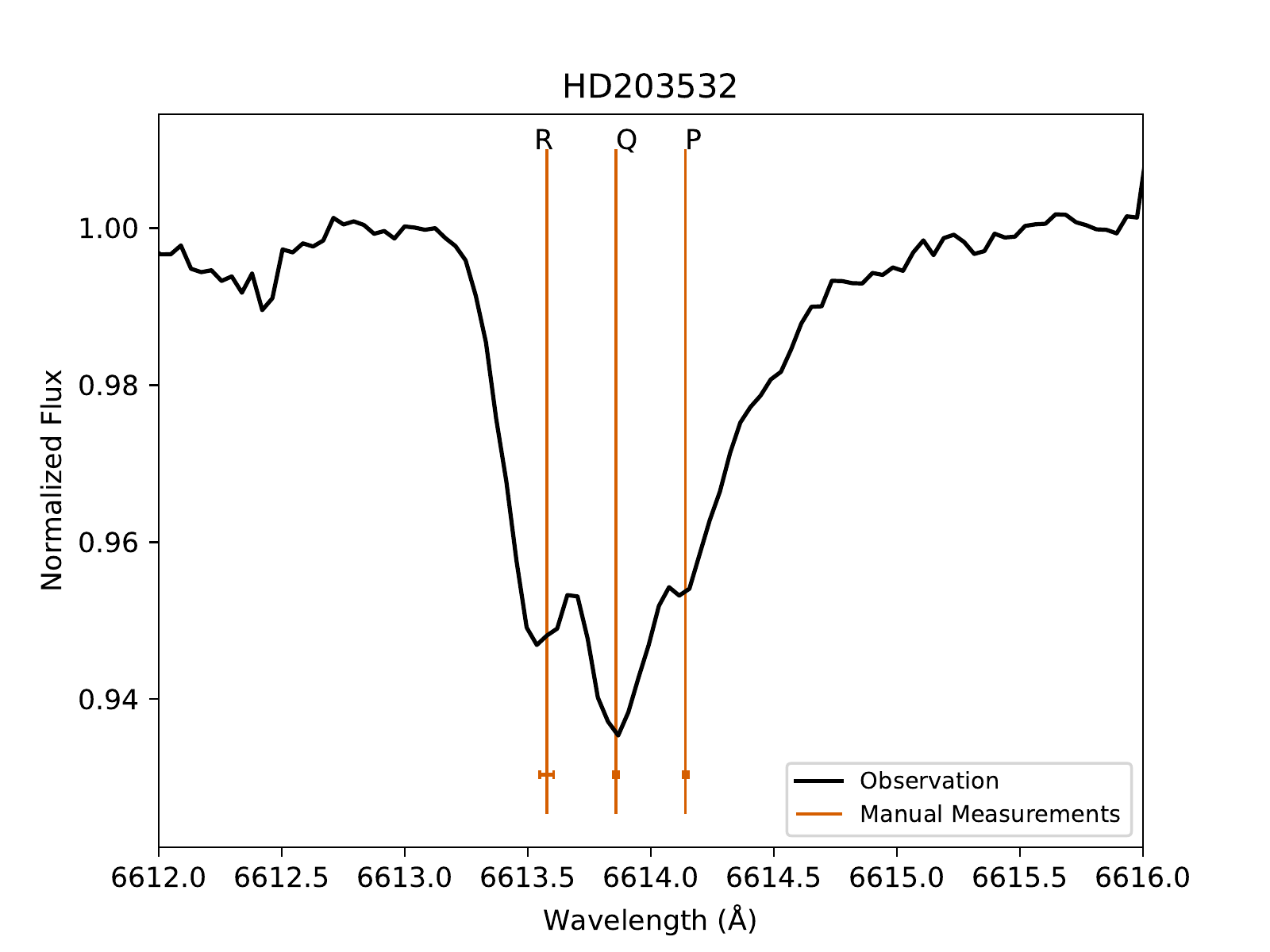}}
\resizebox{0.8\hsize}{!}{
\includegraphics[width=\columnwidth]{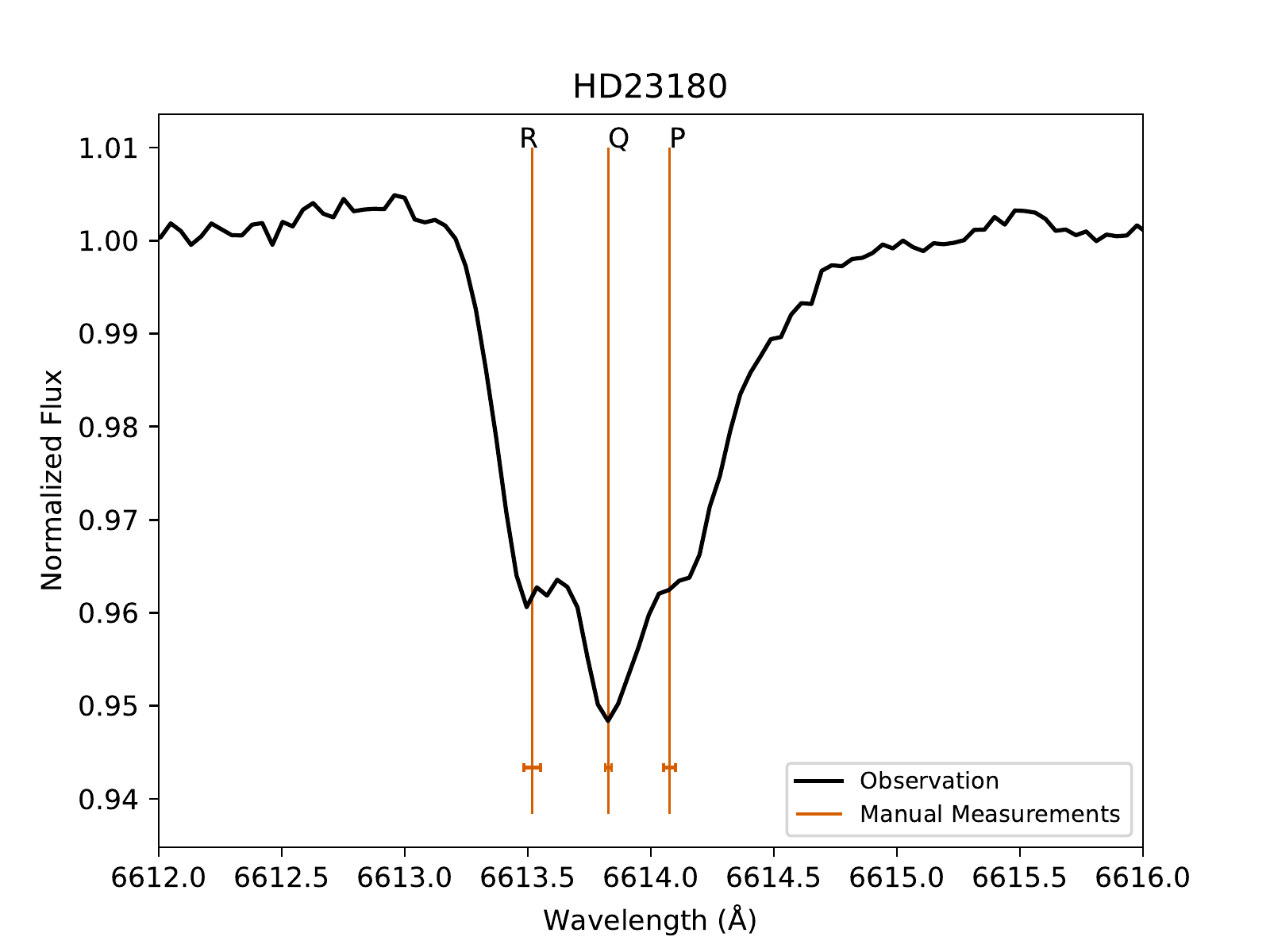}
\includegraphics[width=\columnwidth]{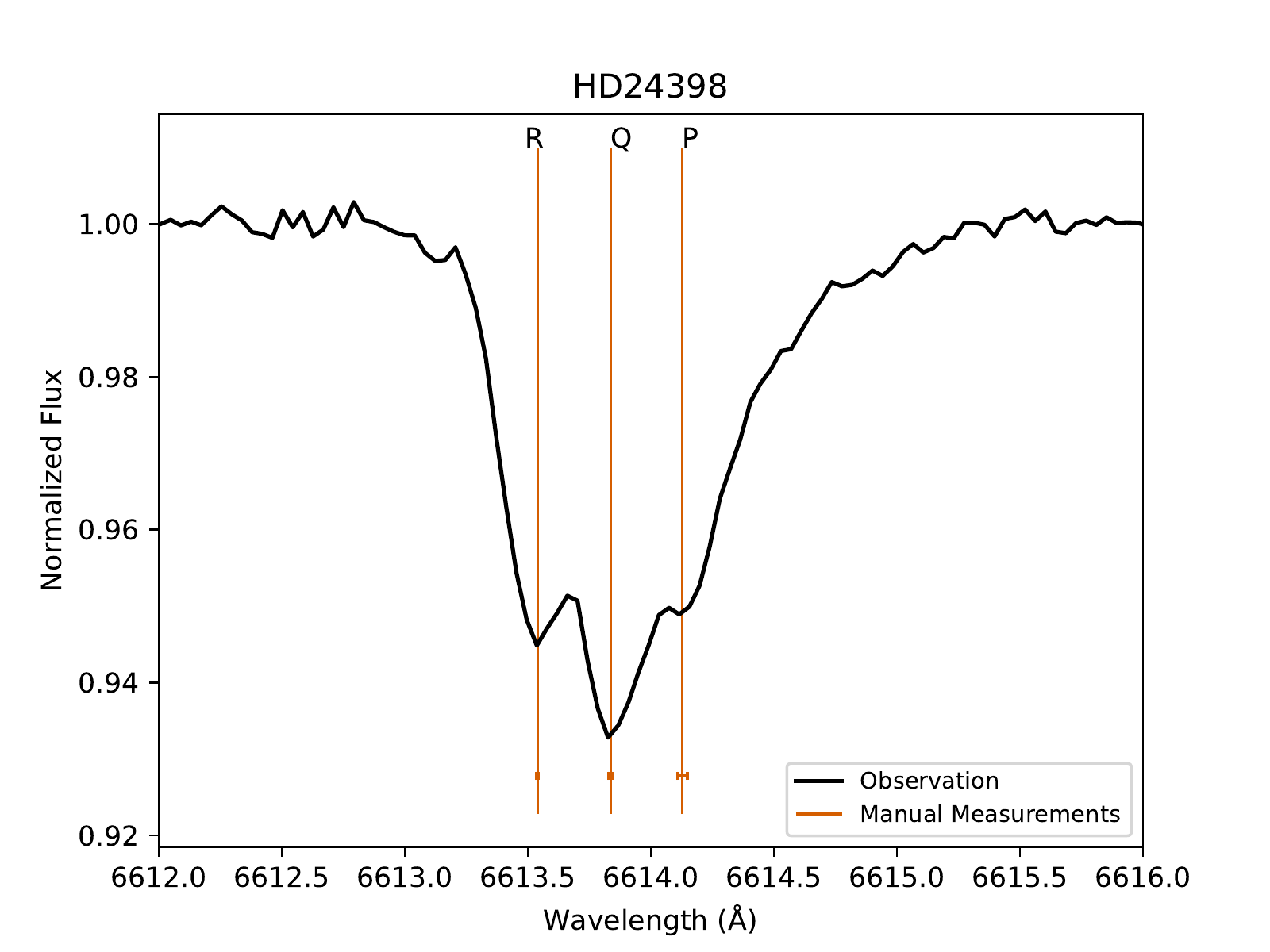}}
\caption{continued.}
\label{6614_fit_results_2}
\end{figure*}

\section{Measurements for the $\lambda$5797 DIB.}
\label{Sect:App_5797}
The measurements of the $\lambda$5797 DIB are listed in Table~\ref{table:5797_manual} and are shown in Figs.~\ref{5797_fit_results_1}--\ref{5797_fit_results_2}.

\begin{table}
\caption{Same as Table~\ref{table:6614_manual} but for the $\lambda$5797 DIB.}   
\label{table:5797_manual} 
\centering     
%
\begin{tabular}{c c c}   
\hline\hline 
Target &    Peak 1 & Peak 2  \\
 & [\AA] & [\AA]\\
 HD23180 &  5797.13 $\pm$0.02 &  5797.34 $\pm$0.04 \\
 HD24398 &  5797.14 $\pm$0.01 &  5797.369 $\pm$0.007 \\
 HD144470 &  5796.72 $\pm$0.02 &  5796.96 $\pm$0.05 \\
 HD147165 &  5796.760 $\pm$0.008 &  5797.02 $\pm$0.02 \\
 HD147683 &  5796.88 $\pm$0.02 &  5797.09 $\pm$0.02 \\
 HD149757 &  5796.55 $\pm$0.01 &  5796.82 $\pm$0.01 \\
 HD166937 &  5796.73 $\pm$0.01 &  5796.98 $\pm$0.01 \\
 HD170740 &  5796.642 $\pm$0.008 &  5796.87 $\pm$0.007 \\
 HD184915 &  5796.65 $\pm$0.01 &  5796.86 $\pm$0.007 \\
 HD185418 &  5796.71 $\pm$0.02 &  5796.91 $\pm$0.01 \\
 HD185859 &  5796.72 $\pm$0.01 &  5796.94 $\pm$0.01 \\
 HD203532 &  5797.19 $\pm$0.03 &  5797.40 $\pm$0.01 \\

\hline   
\end{tabular}
\end{table}
\begin{figure*}
\centering
\resizebox{0.8\hsize}{!}{
\includegraphics[width=\columnwidth]{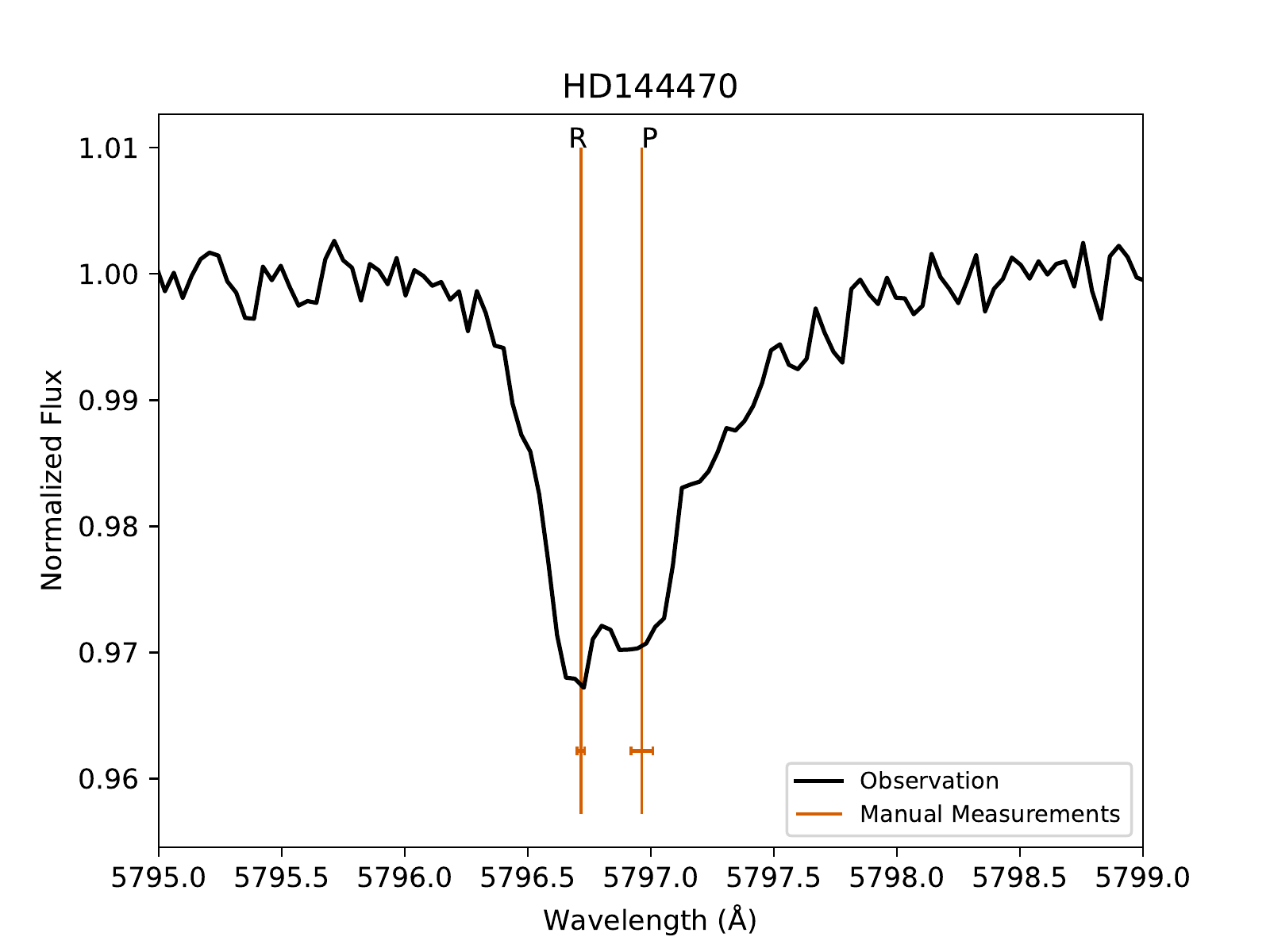}
\includegraphics[width=\columnwidth]{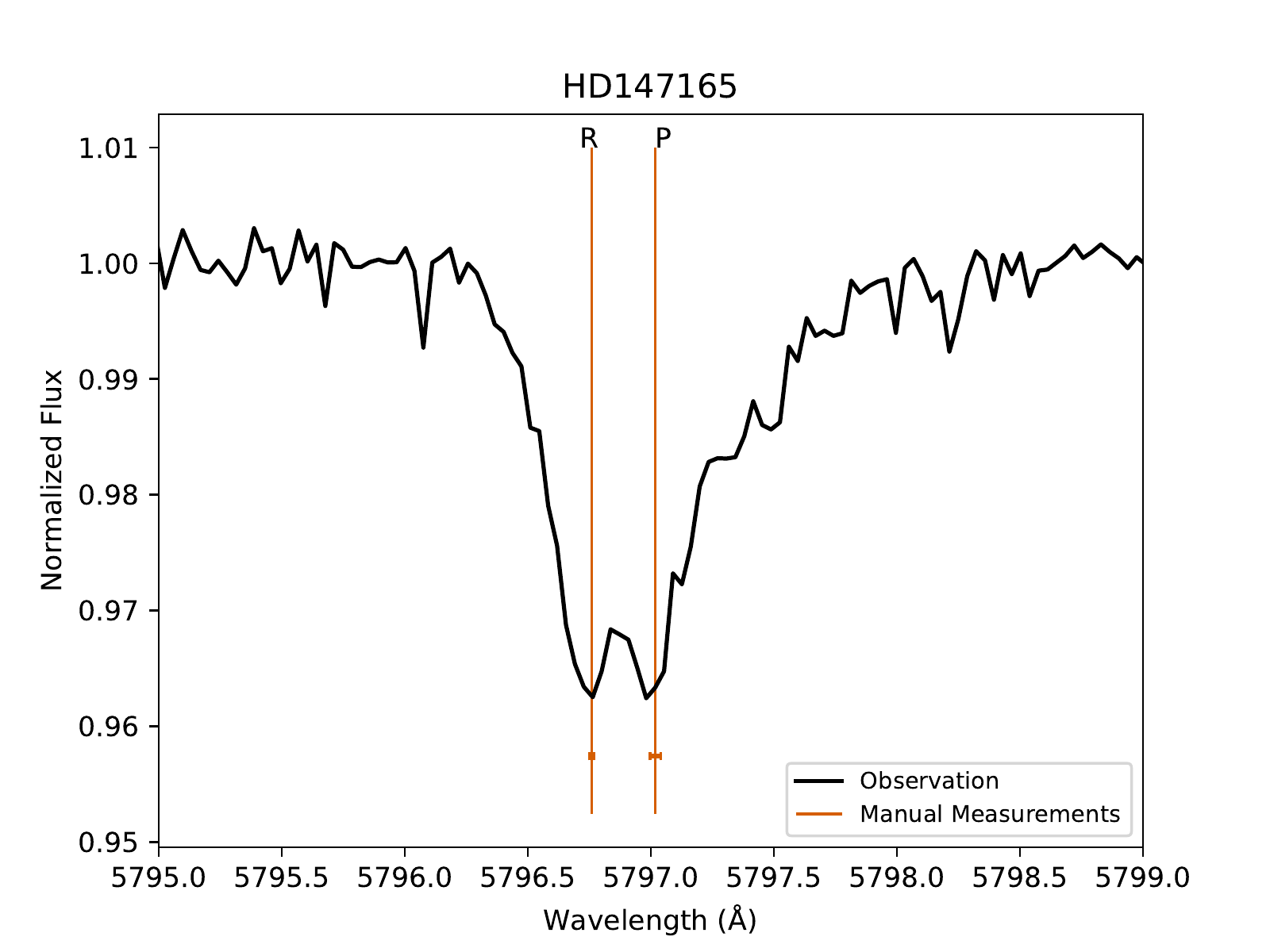}}
\resizebox{0.8\hsize}{!}{
\includegraphics[width=\columnwidth]{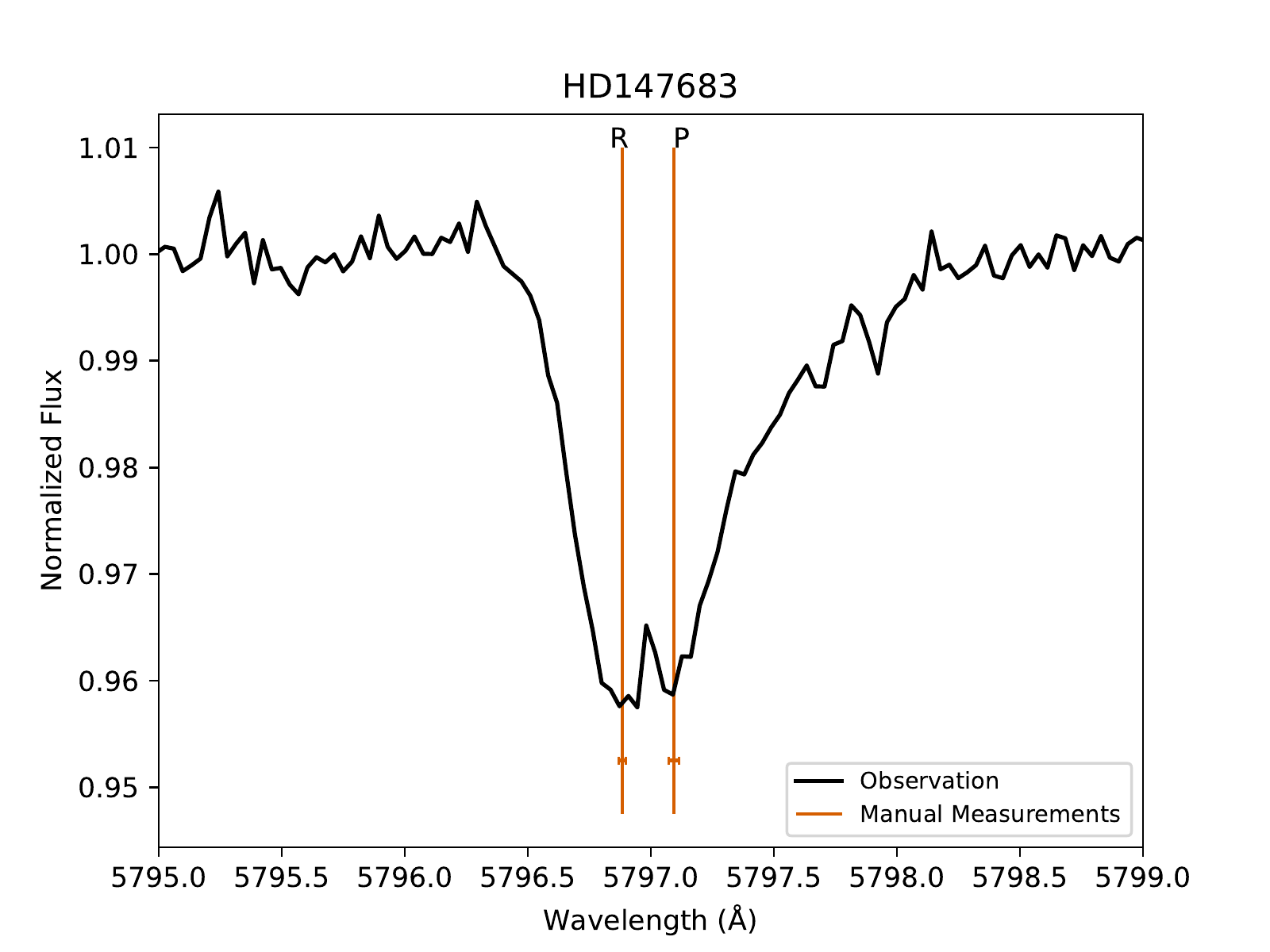}
\includegraphics[width=\columnwidth]{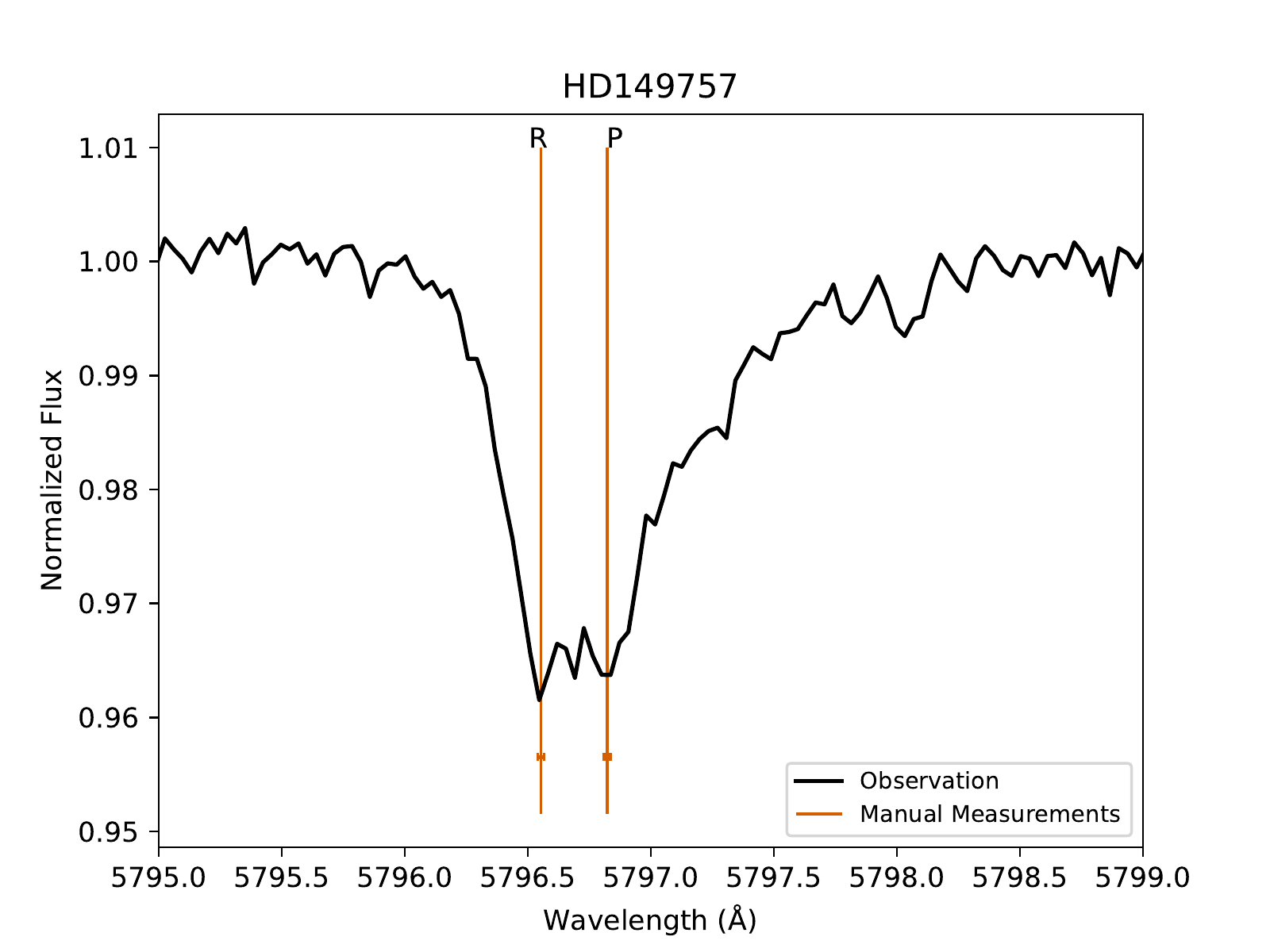}}
\resizebox{0.8\hsize}{!}{
\includegraphics[width=\columnwidth]{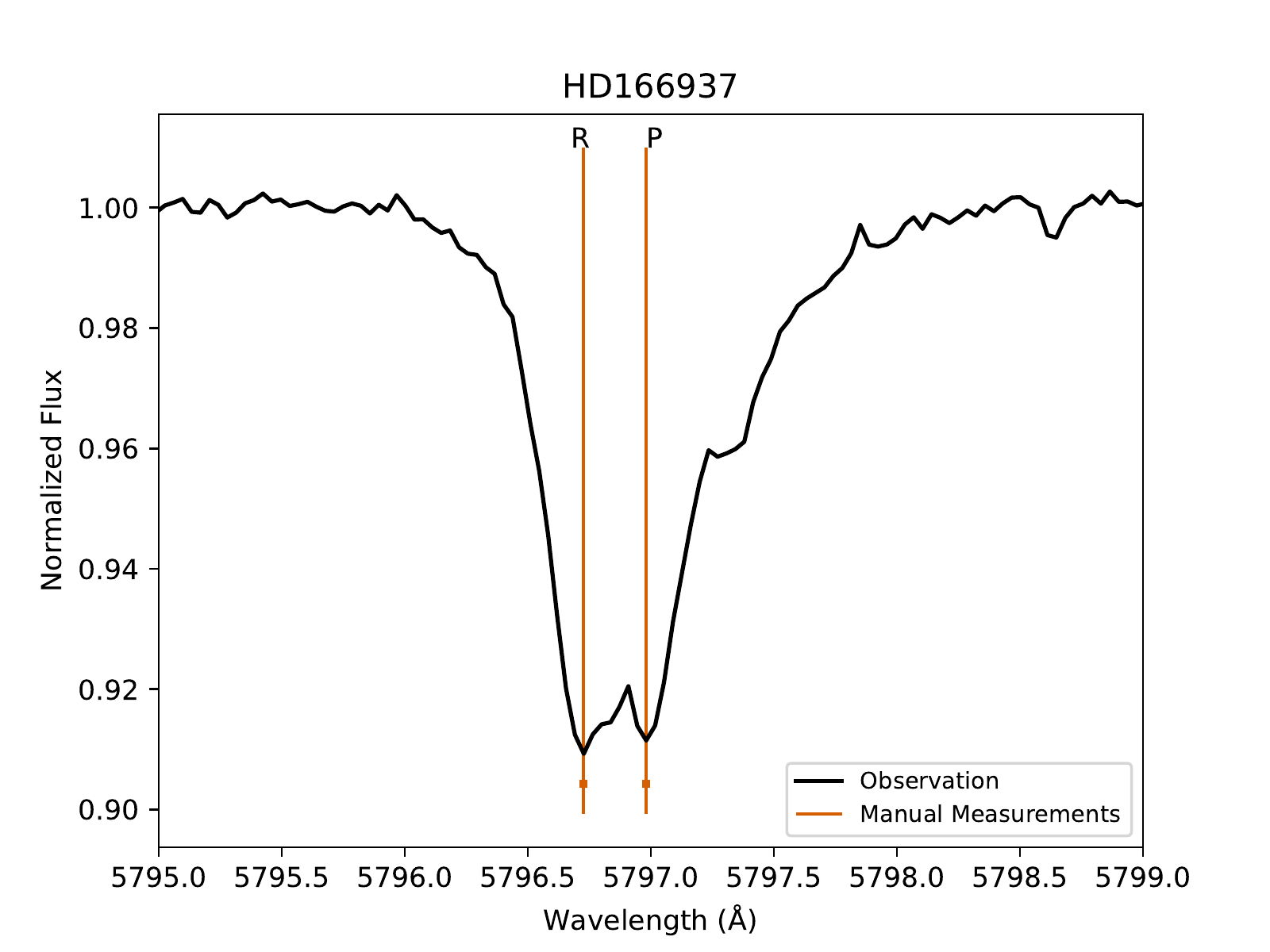}
\includegraphics[width=\columnwidth]{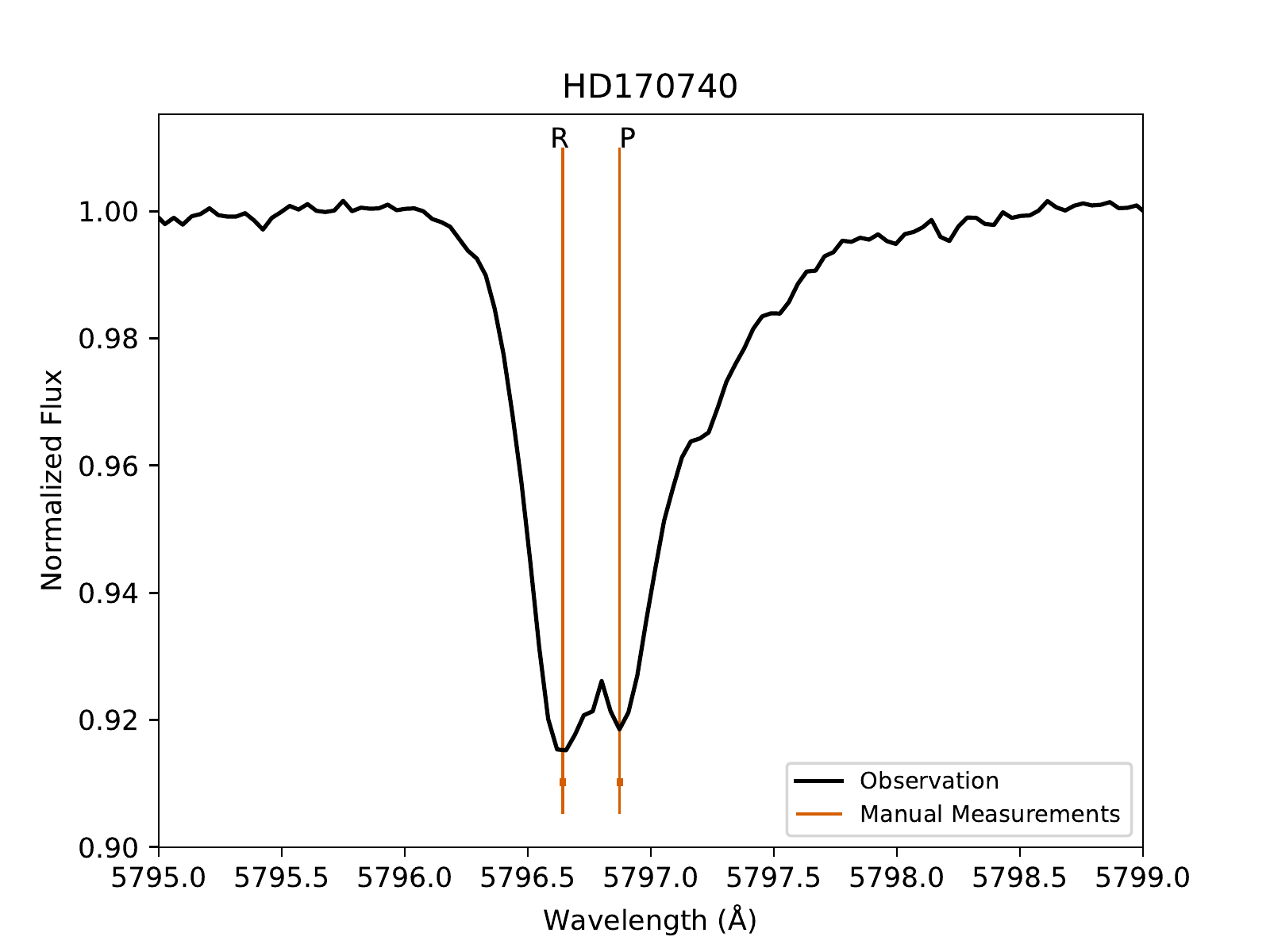}}
\caption{Same as Fig.~\ref{6614_fit_results_1} but for the 5797{\AA} DIB.}
\label{5797_fit_results_1}
\end{figure*}
\begin{figure*}
\centering

\resizebox{0.8\hsize}{!}{
\includegraphics[width=\columnwidth]{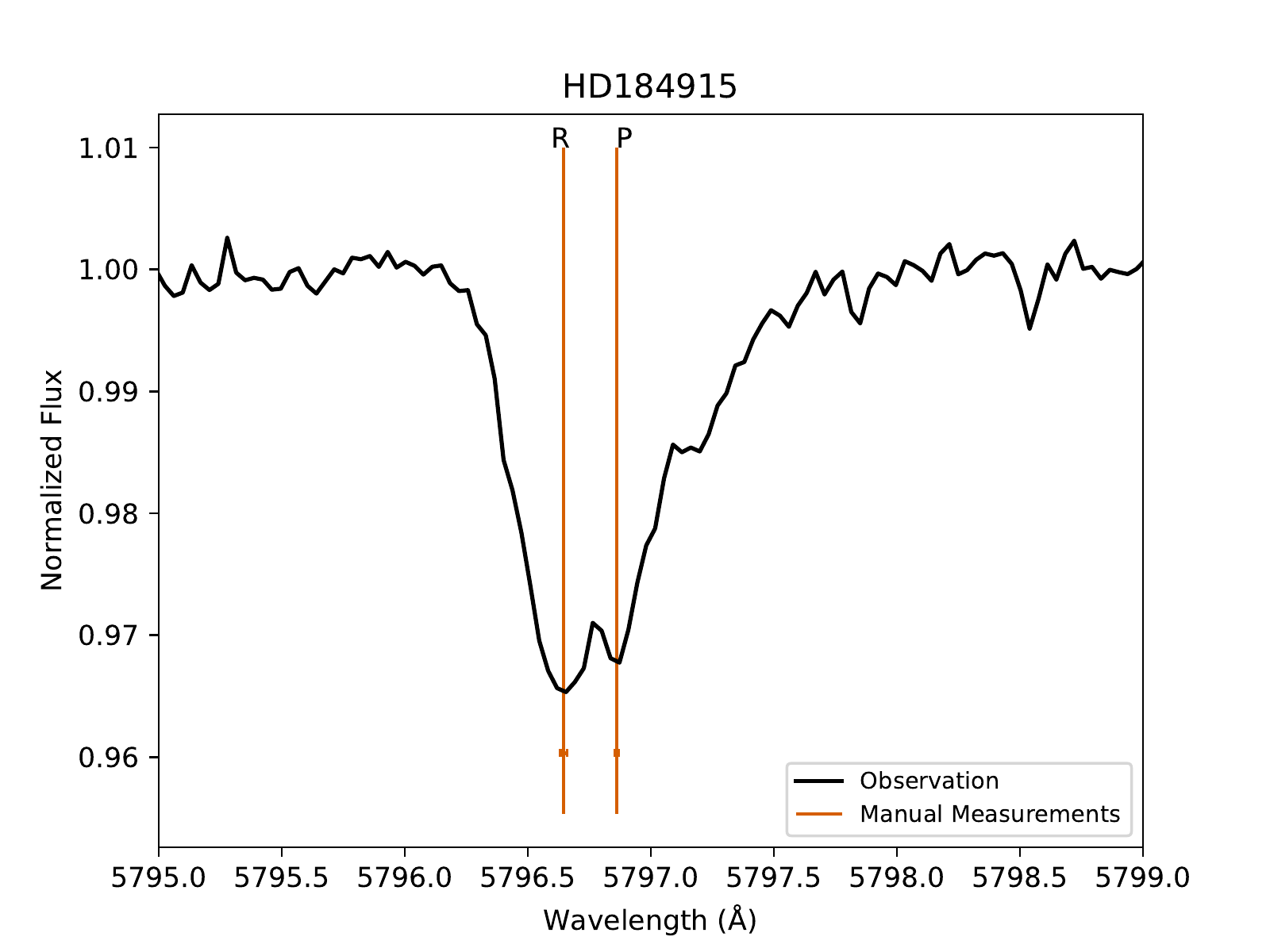}
\includegraphics[width=\columnwidth]{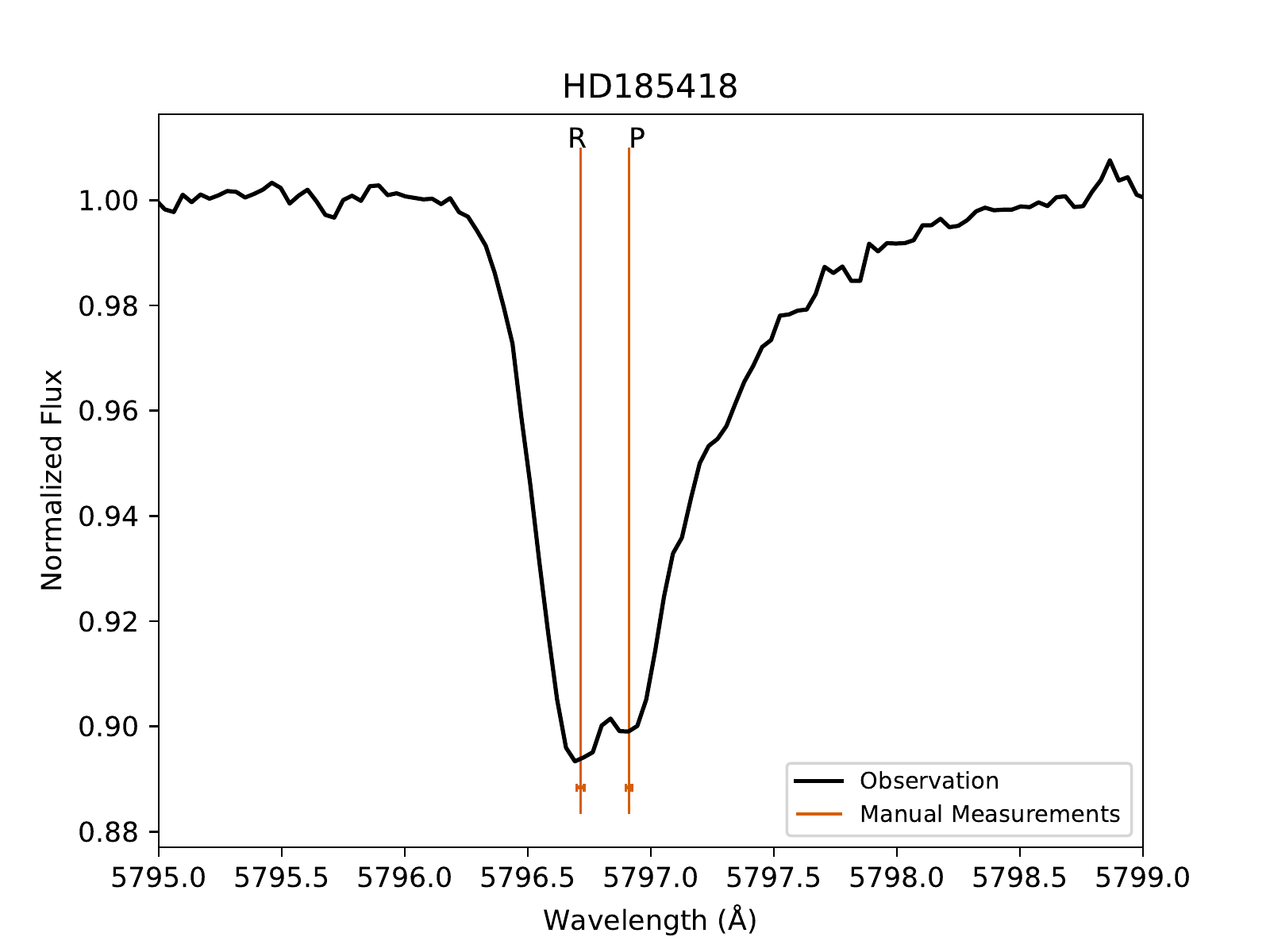}}
\resizebox{0.8\hsize}{!}{
\includegraphics[width=\columnwidth]{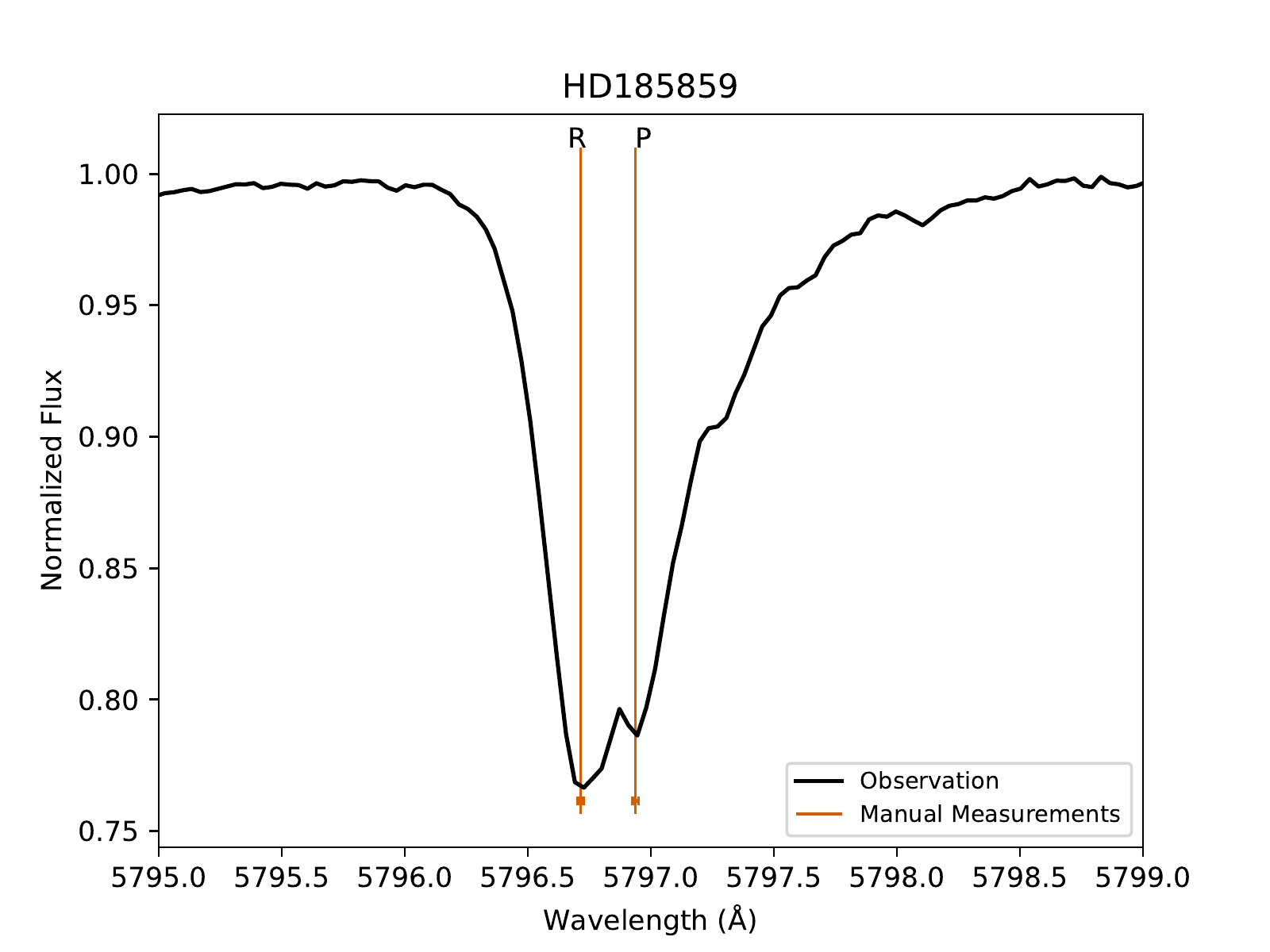}
\includegraphics[width=\columnwidth]{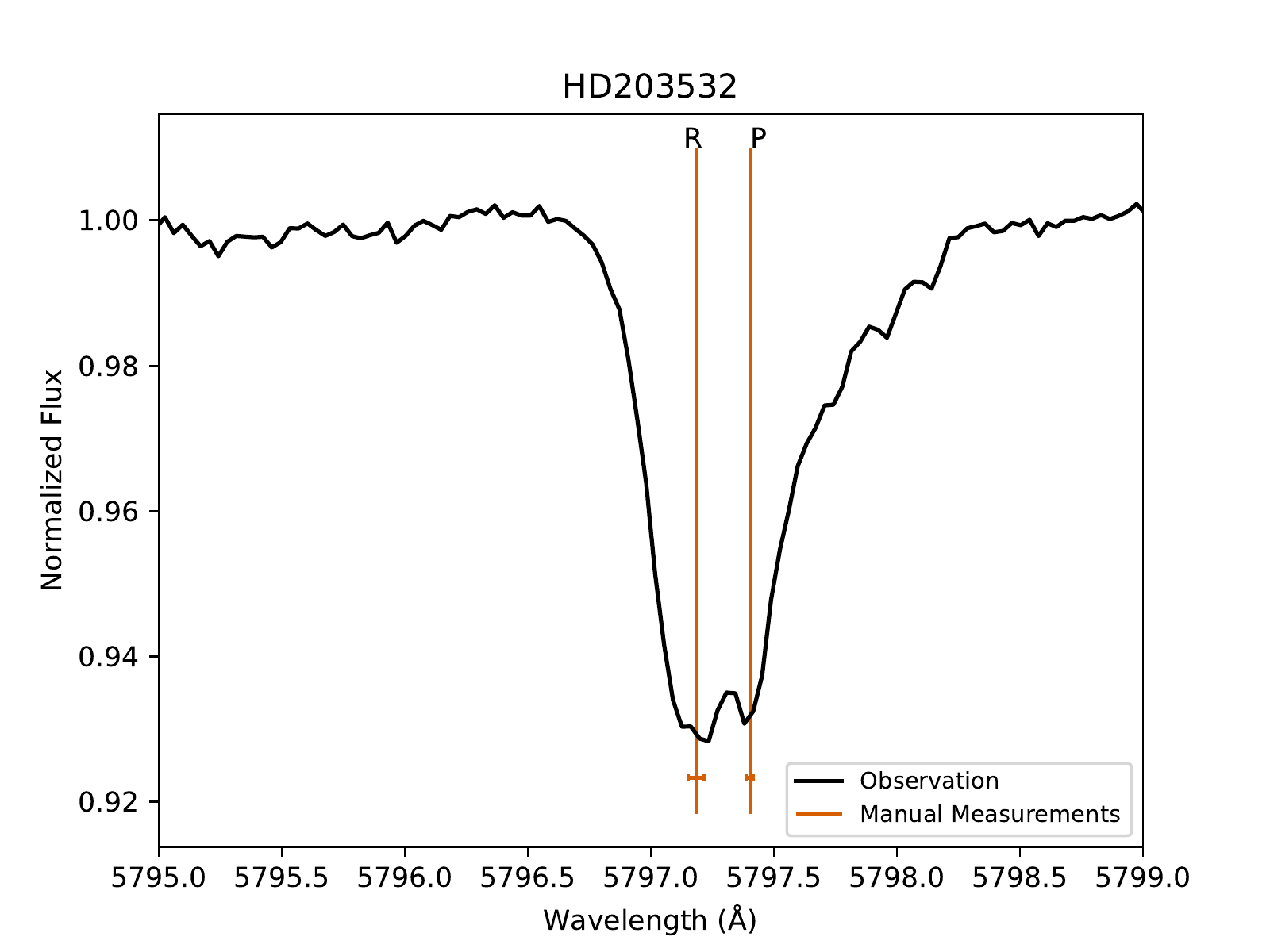}}
\resizebox{0.8\hsize}{!}{
\includegraphics[width=\columnwidth]{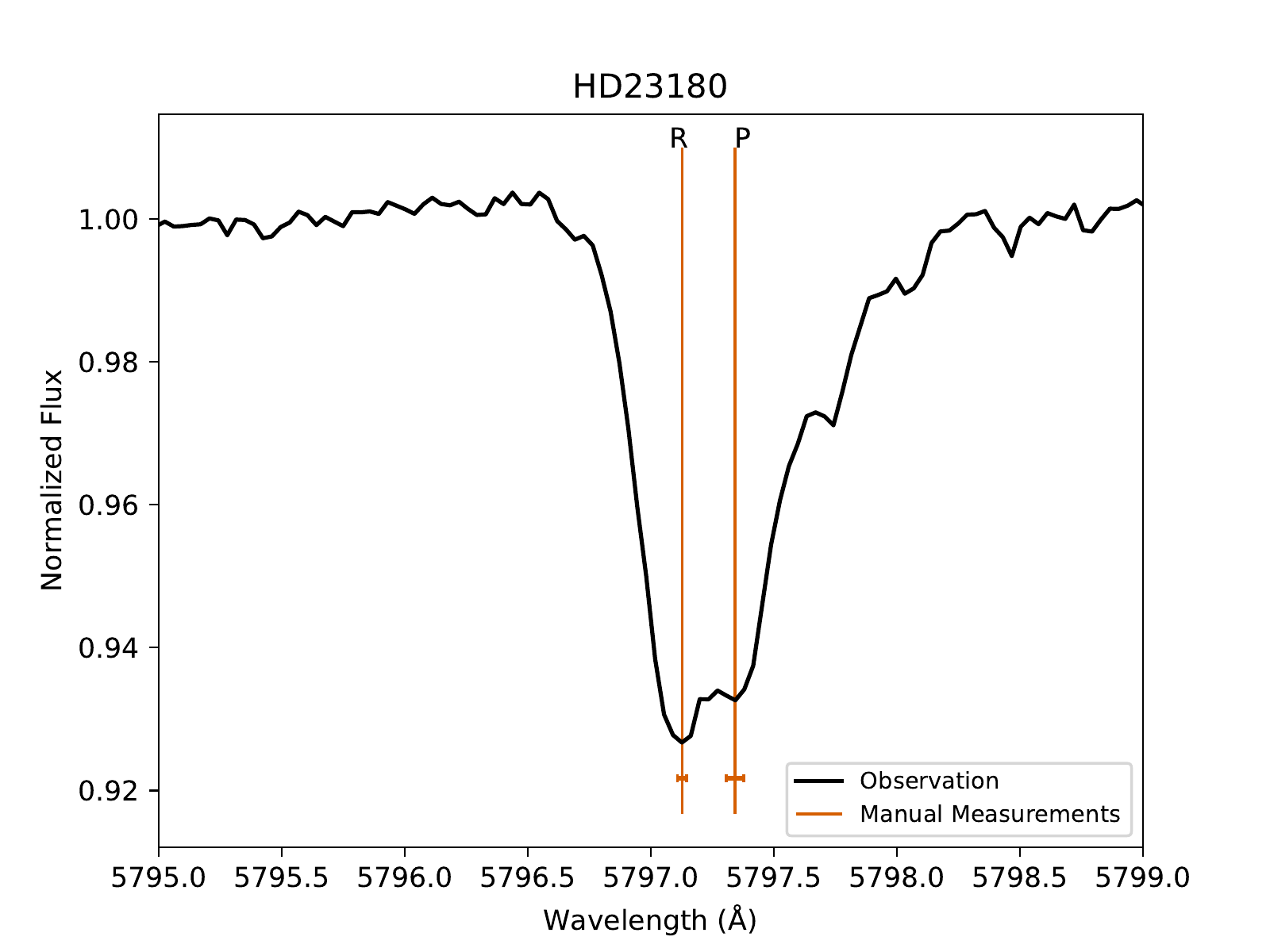}
\includegraphics[width=\columnwidth]{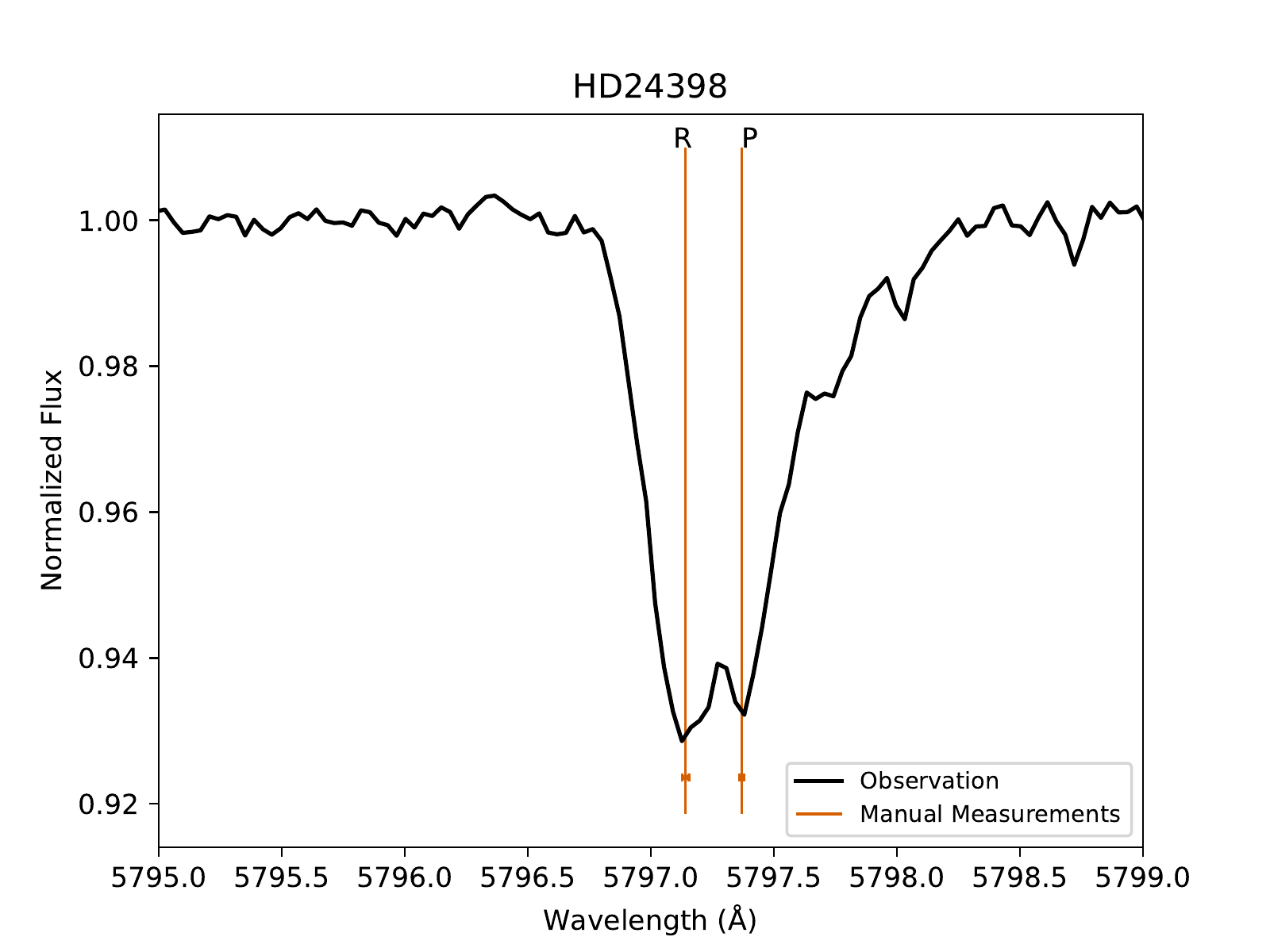}}
\caption{continued.}
\label{5797_fit_results_2}
\end{figure*}

\section{Measurements for the $\lambda$6379 DIB.}
The measurements of the $\lambda$6379 DIB are listed in Table~\ref{table:6379_manual} and shown in Figs.~\ref{6379_fit_results_1}--\ref{6379_fit_results_2}. 
\label{Sect:App_6379}

\begin{table}
\caption{Same as Table~\ref{table:6614_manual} but for the $\lambda$6379 DIB.}    
\label{table:6379_manual} 
\centering     
\begin{tabular}{c c c}   
\hline\hline 
Target &    Peak 1 & Peak 2  \\
 & [\AA] & [\AA]\\
 HD23180 &  6379.45 $\pm$0.02 &  6379.606 $\pm$0.008 \\
 HD24398 &  6379.48 $\pm$0.02 &  6379.659 $\pm$0.004 \\
 HD144470 &  6378.97  $\pm$0.02  &  6379.16 $\pm$0.01 \\
 HD147165 &  6379.005 $\pm$0.002 &  6379.281 $\pm$0.009 \\
 HD147683 &  6379.12  $\pm$0.02  &  6379.341 $\pm$0.008 \\
 HD149757 &  6378.86  $\pm$0.03  &  6379.063 $\pm$0.009 \\
 HD166937 &  6379.002 $\pm$0.008 &  6379.237 $\pm$0.006 \\
 HD170740 &  6378.92  $\pm$0.02  &  6379.124 $\pm$0.005 \\
 HD184915 &  6378.88  $\pm$0.01  &  6379.08 $\pm$0.01 \\
 HD185418 &  6378.98  $\pm$0.01  &  6379.15 $\pm$0.01 \\
 HD185859 &  6379.010 $\pm$0.007 &  6379.188 $\pm$0.009 \\
 HD203532 &  6379.48  $\pm$0.01  &  6379.666 $\pm$0.006 \\
\hline   
\end{tabular}

\end{table}

\begin{figure*}
\centering
\resizebox{0.8\hsize}{!}{
\includegraphics[width=\columnwidth]{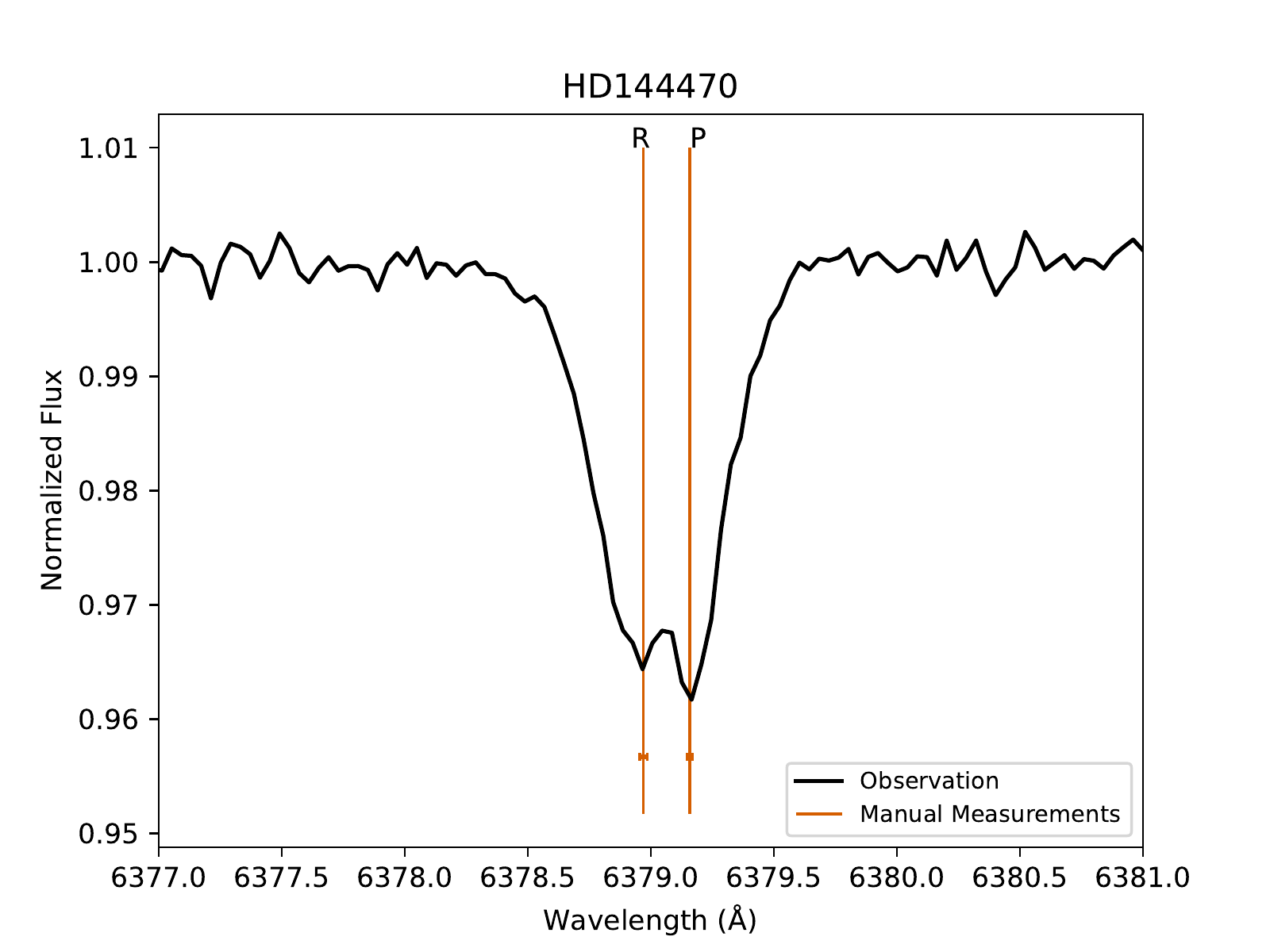}
\includegraphics[width=\columnwidth]{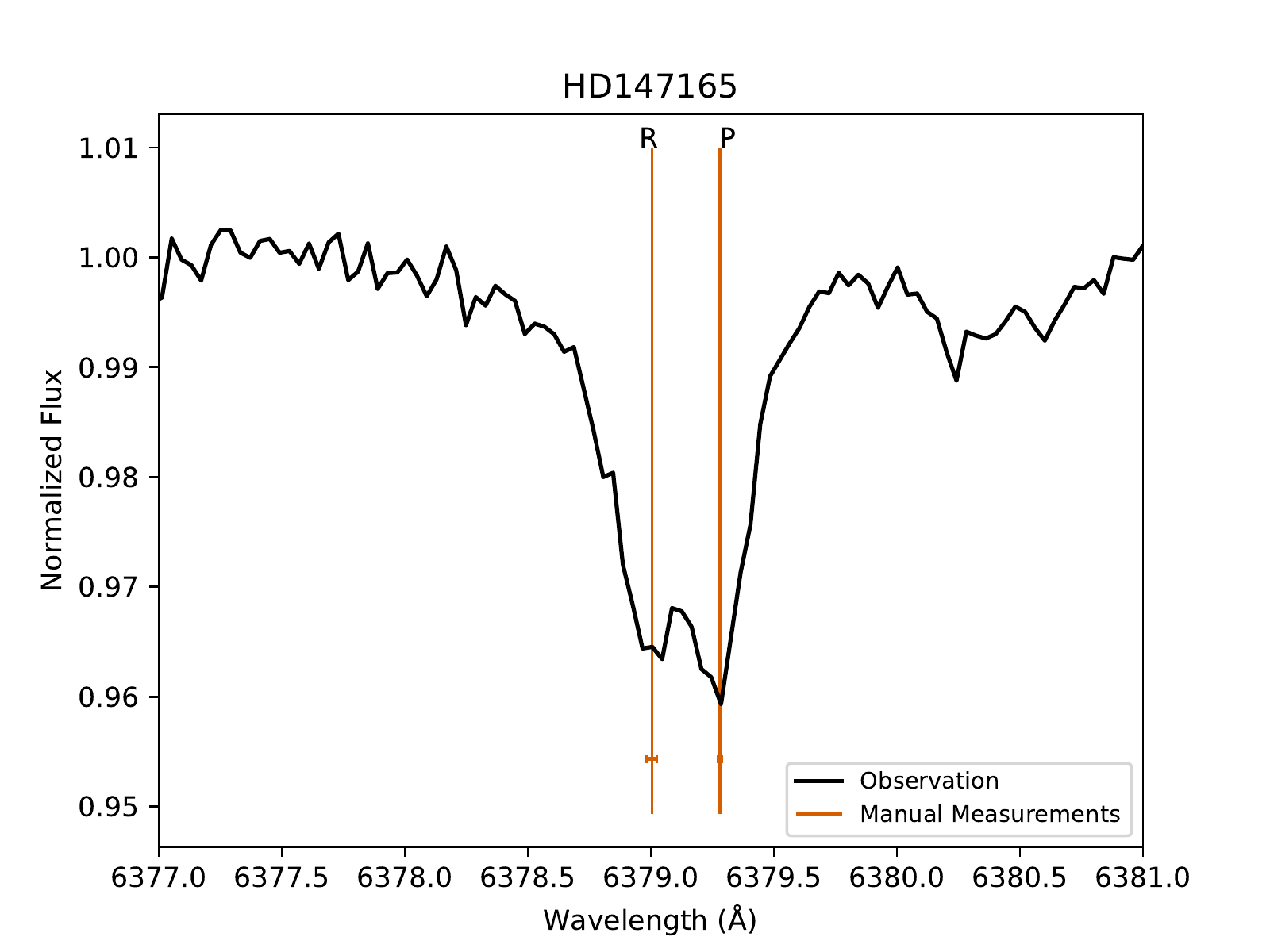}}
\resizebox{0.8\hsize}{!}{
\includegraphics[width=\columnwidth]{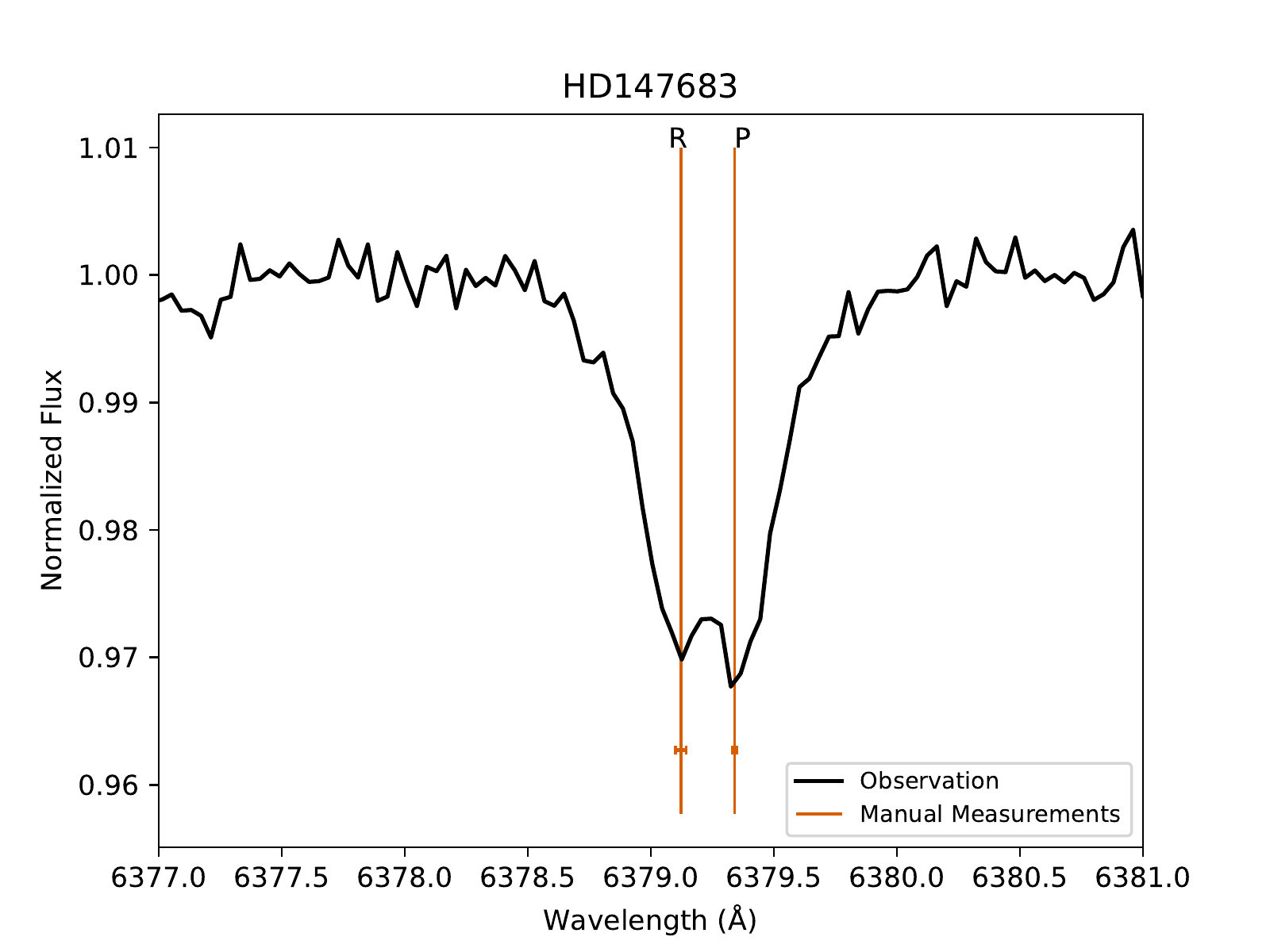}
\includegraphics[width=\columnwidth]{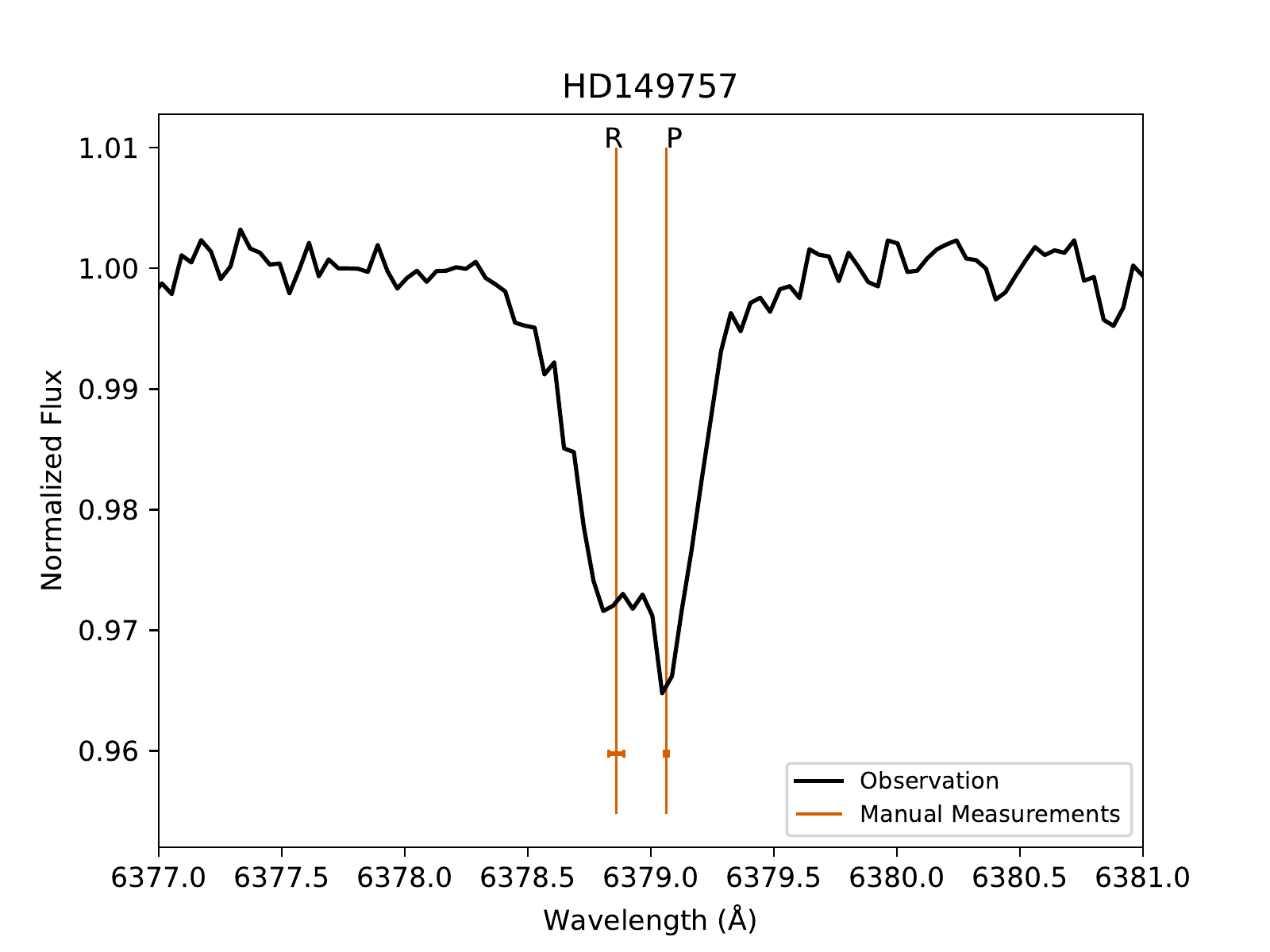}}
\resizebox{0.8\hsize}{!}{
\includegraphics[width=\columnwidth]{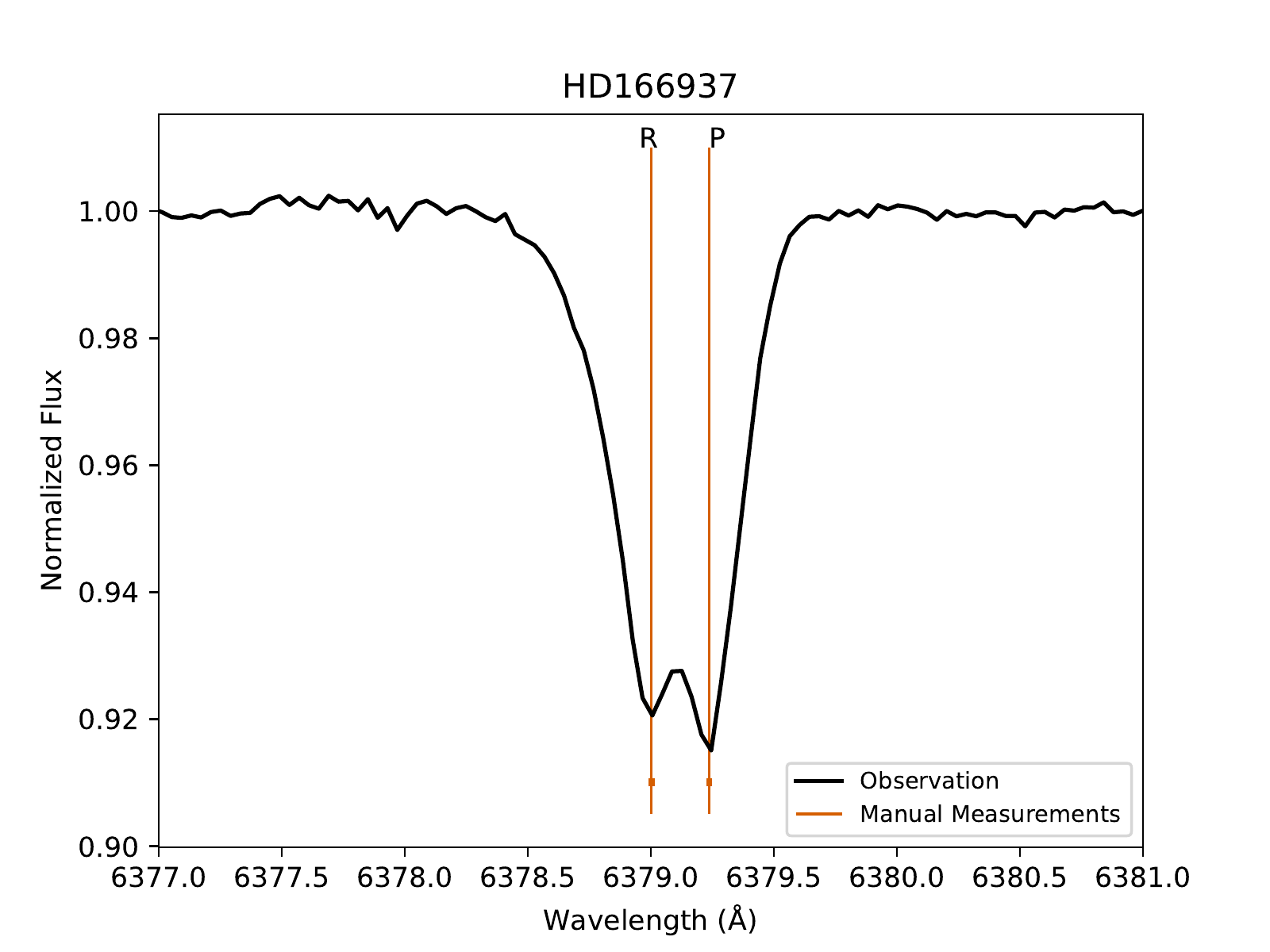}
\includegraphics[width=\columnwidth]{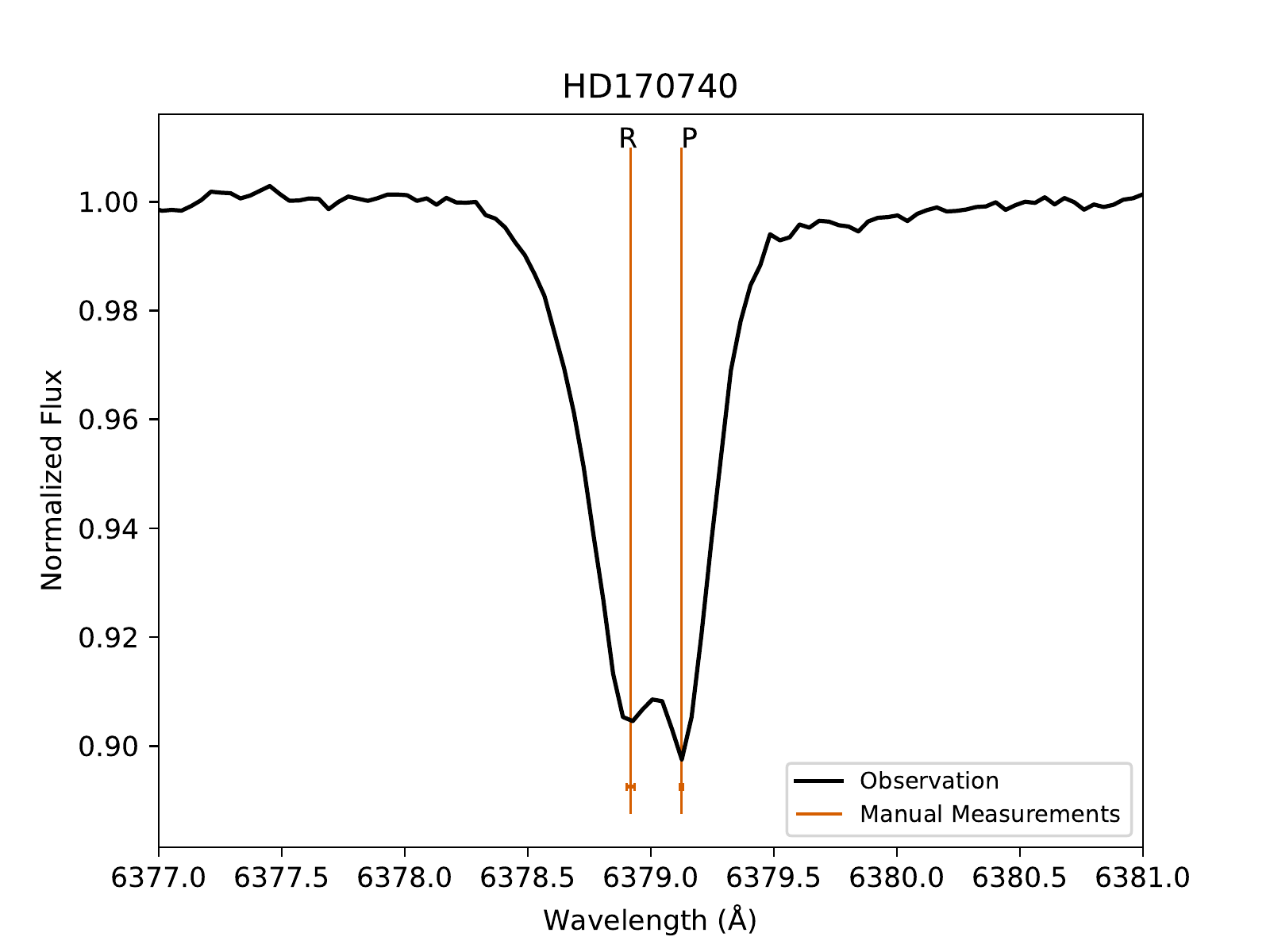}}
\caption{Same as Fig.~\ref{6614_fit_results_1} but for the 6379{\AA} DIB.}
\label{6379_fit_results_1}
\end{figure*}
\begin{figure*}
\centering

\resizebox{0.8\hsize}{!}{
\includegraphics[width=\columnwidth]{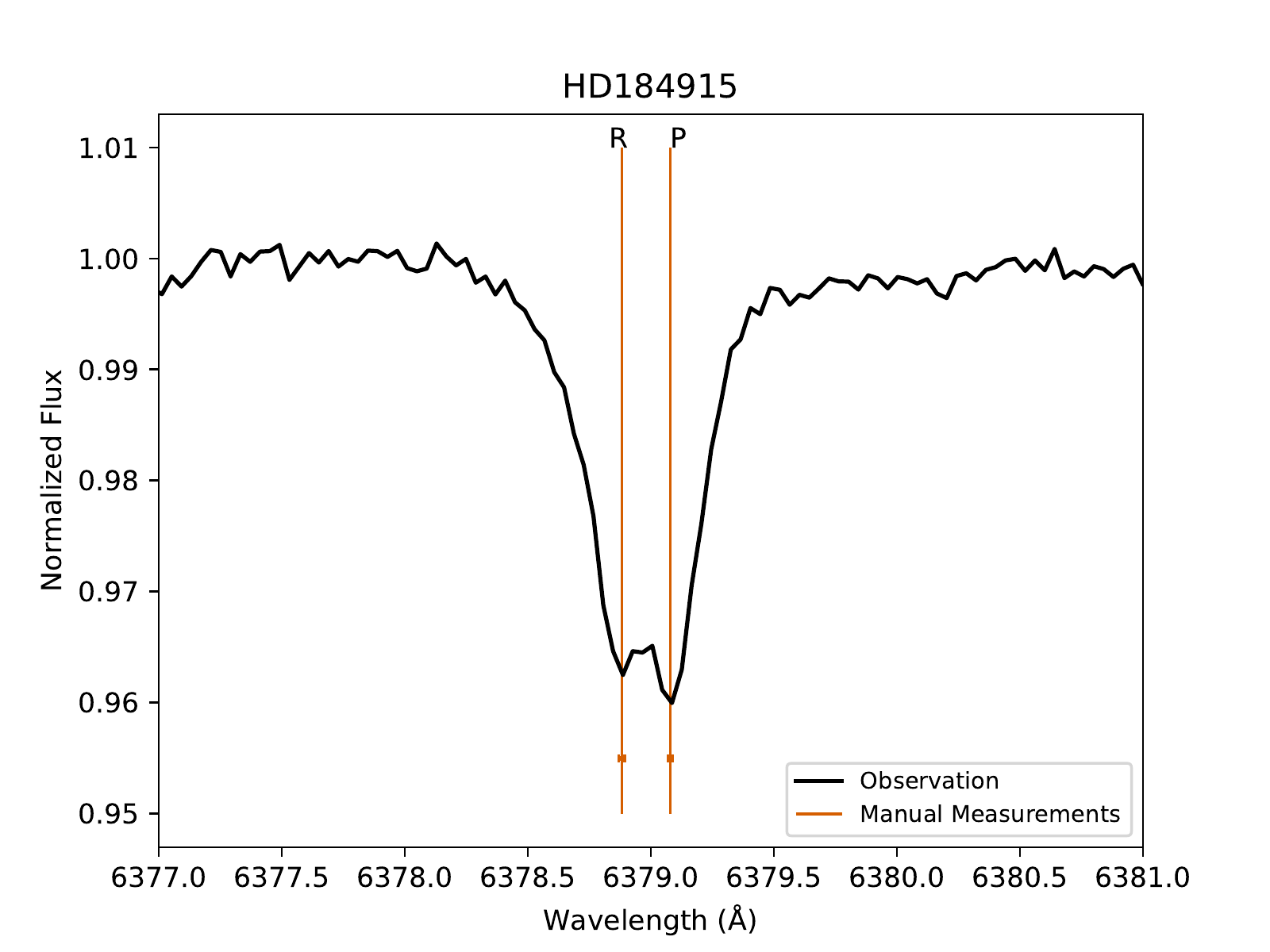}
\includegraphics[width=\columnwidth]{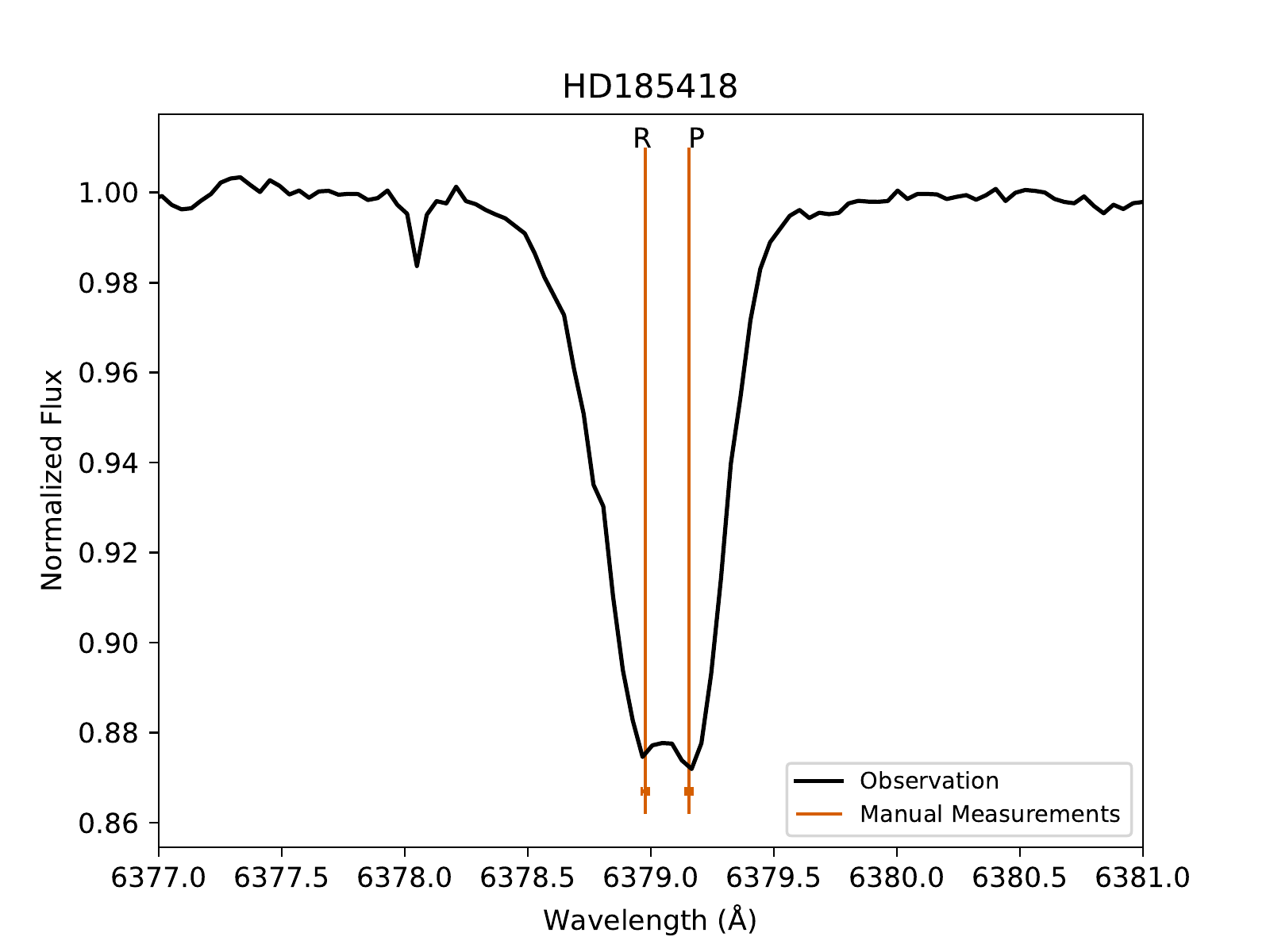}}
\resizebox{0.8\hsize}{!}{
\includegraphics[width=\columnwidth]{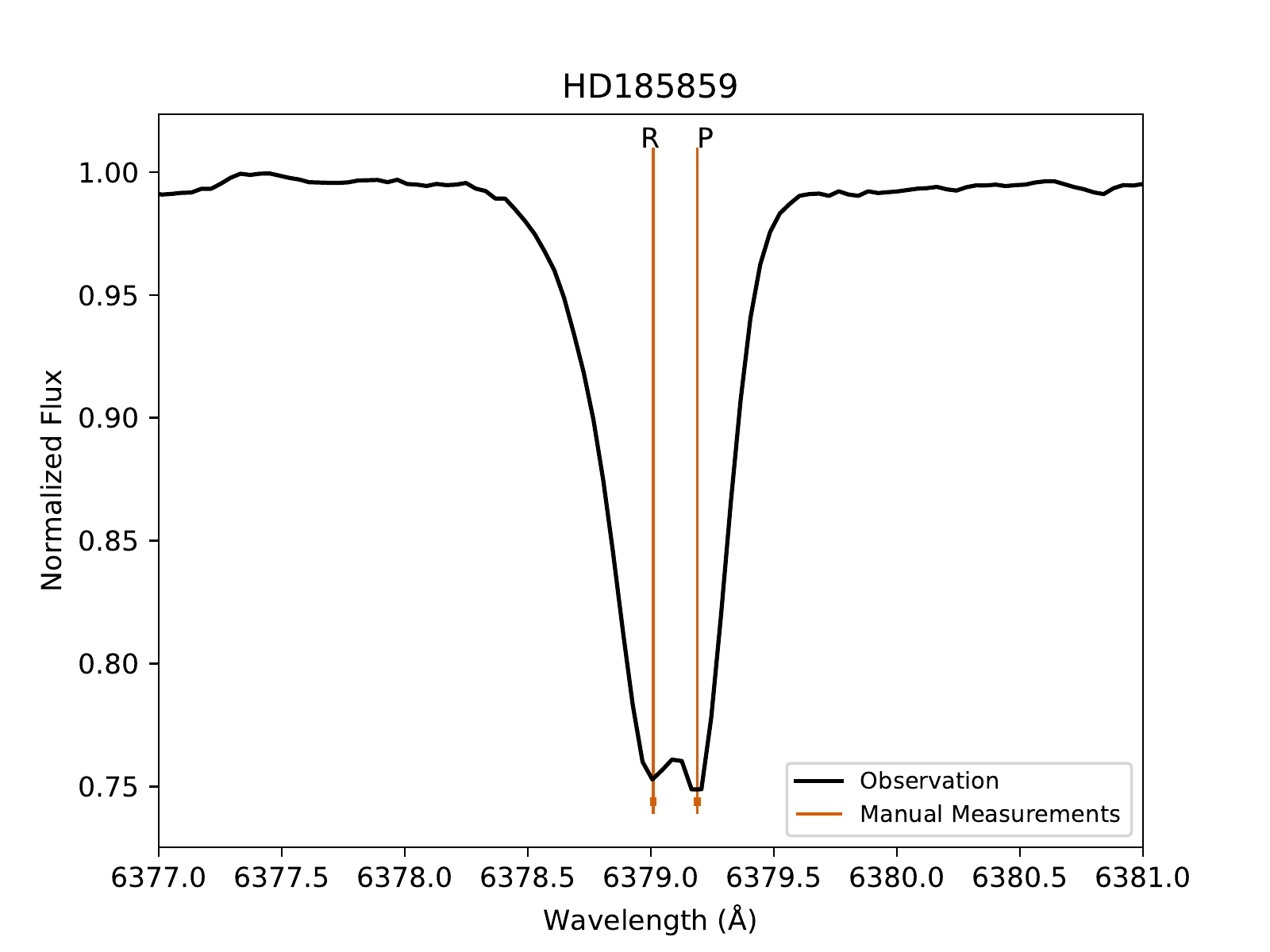}
\includegraphics[width=\columnwidth]{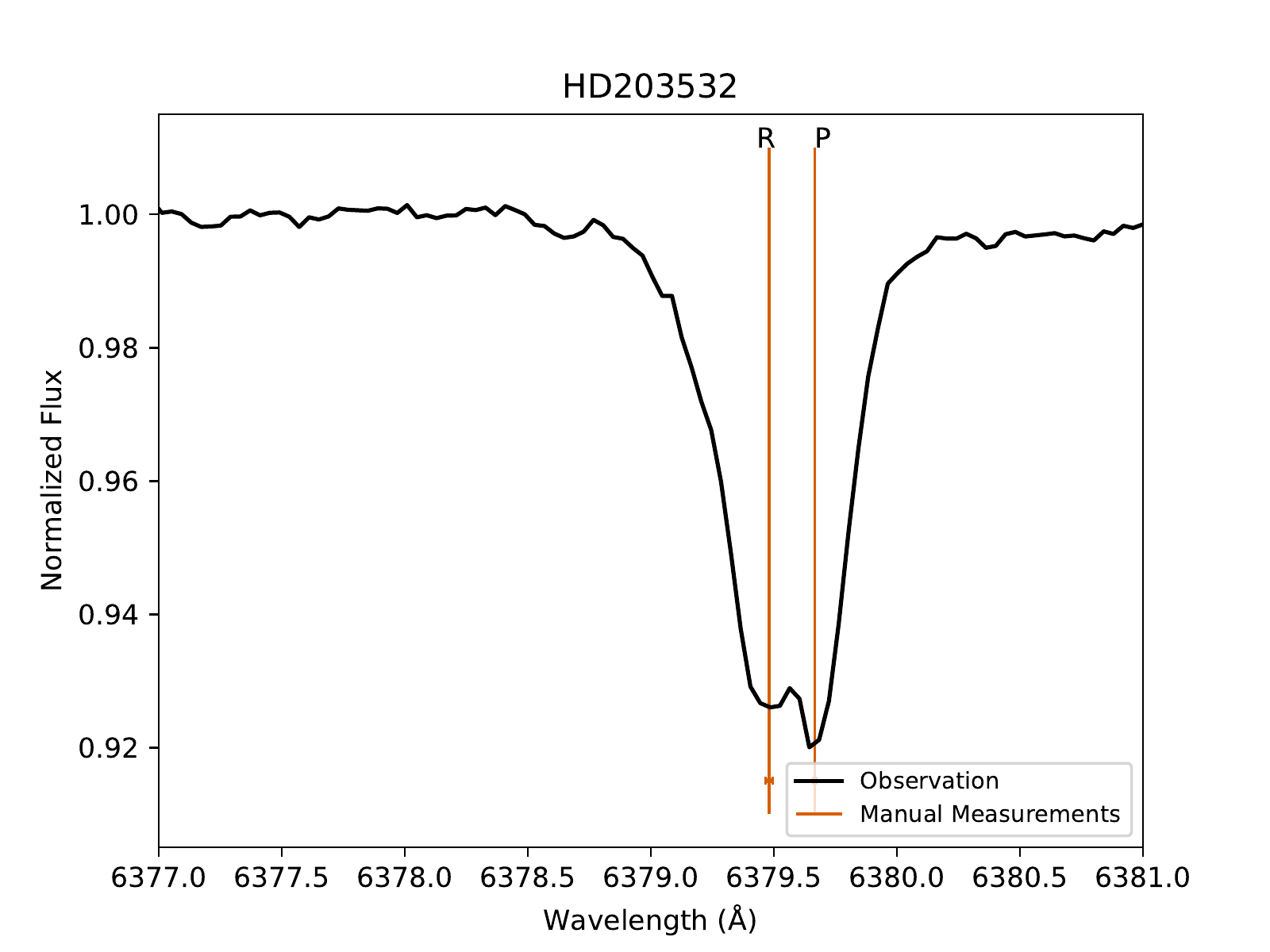}}
\resizebox{0.8\hsize}{!}{
\includegraphics[width=\columnwidth]{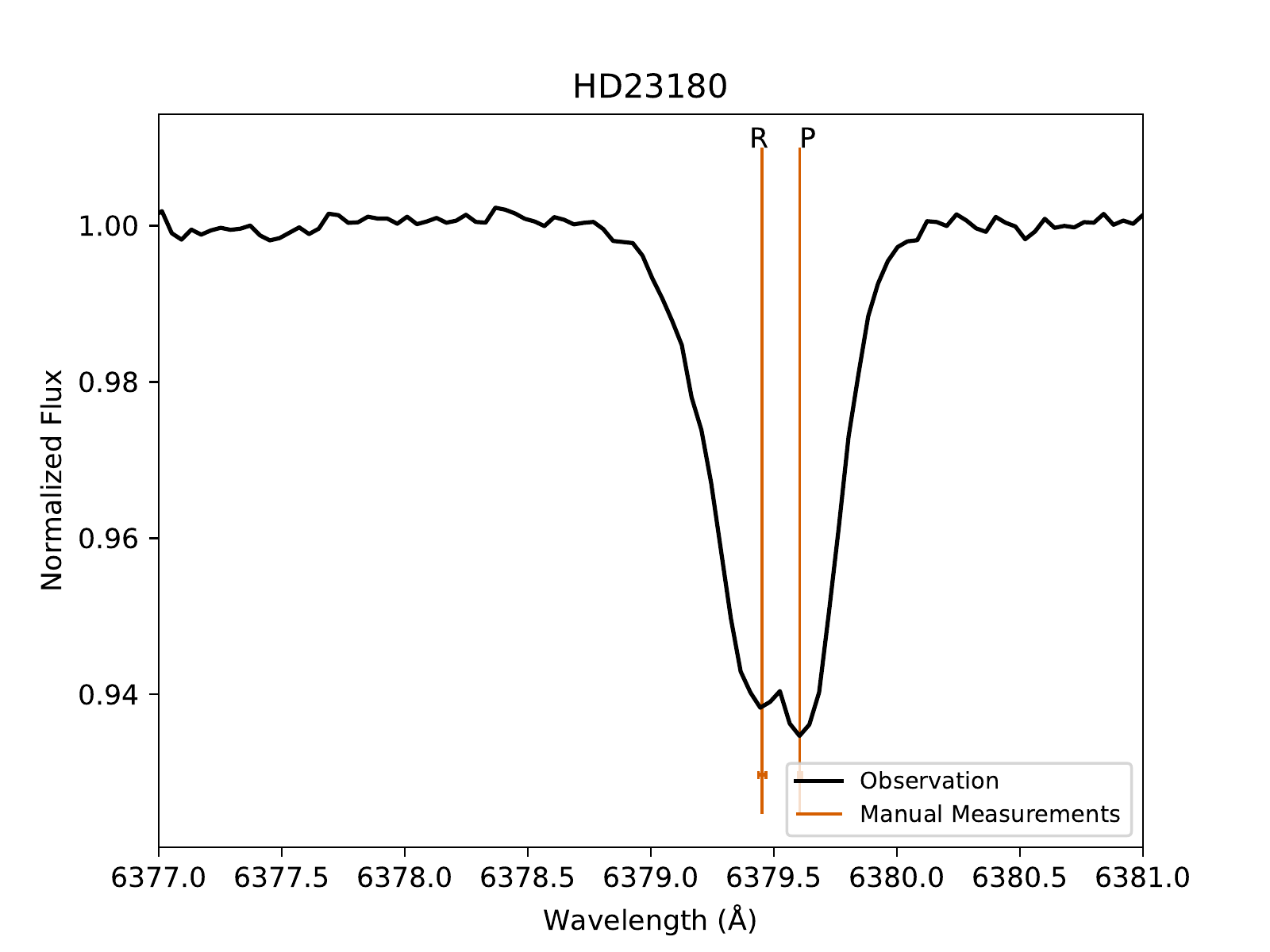}
\includegraphics[width=\columnwidth]{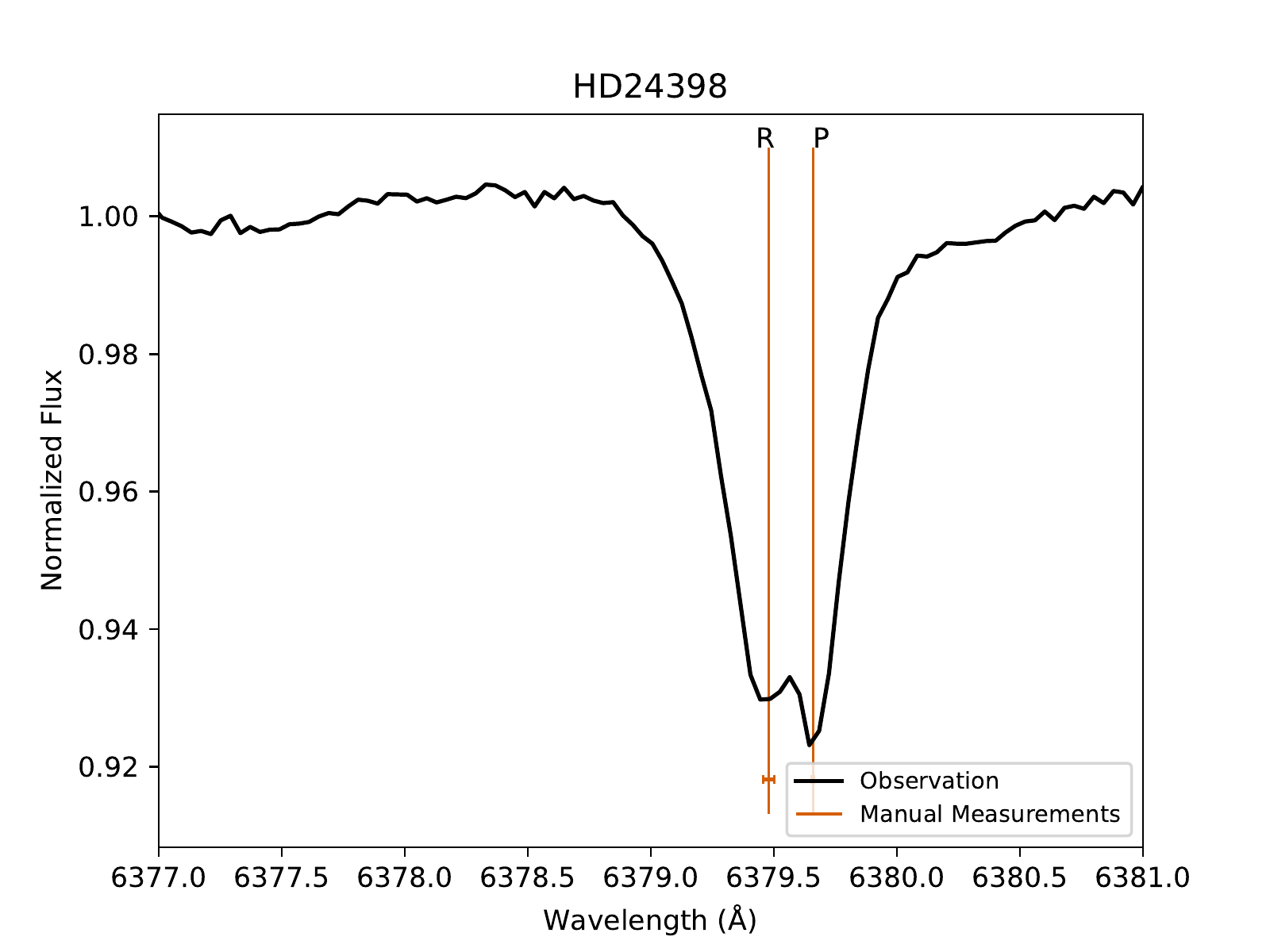}}
\caption{continued.}
\label{6379_fit_results_2}
\end{figure*}

\end{appendix}
\end{document}